\documentclass[fp]{jpsj3}
\usepackage{txfonts}
\usepackage{graphicx}
\usepackage{dcolumn}
\usepackage{bm}
\usepackage[dvipsnames]{xcolor}
\usepackage{braket}

\title{
Odd-Parity Crystalline Electric Field and Multipoles in Icosahedral Quasicrystal and Approximant: Magnetic Toroidal and Electric Dipoles in Ce-Based Systems
}

\author{Shinji Watanabe$^{1*}$, Syunta Furuya$^1$, Mitsuaki Kawamura$^2$}
\inst{$^1$Department of Basic Sciences, Kyushu Institute of Technology, Kitakyushu, Fukuoka 804-8550, Japan
\\
$^2$Faculty of Engineering, Yokohama National University, Yokohama, 240-8501, Kanagawa, Japan
}

\abst{
Quasicrystals (QCs) lack periodic arrangement of atoms but possess unique structural order distinct from periodic crystals and random systems. The electronic property is far from complete understanding. Here we have formulated not only even-parity but also odd-parity crystalline electric field (CEF) in the Ce-based icosahedral QC and approximant crystal (AC) on the basis of the point charge model. Under the odd-parity CEF due to local-inversion symmetry breaking at the Ce site, the hybrid wavefunction of 4f and 5d states at Ce is shown to give rise to the odd-parity multipoles such as electric dipole and magnetic toroidal (MT) dipole. We have found that the vortex order in the 1/1 AC where the magnetic moments on each icosahedron swirl is characterized by the cluster-type MT moment. The ferroic alignment of the cluster MT moments in the vortex order gives rise to the electro-magnetic effect. The cluster-MT moment is shown to be resonated with the MT dipole moments in each icosahedron by tuning the 4f energy level and 5d energy level so as to approach. We have revealed a noncollinear magnetic structure is stabilized in the ground state of the icosahedral QC cluster, behind which there exists underlying mechanism to prefer the ferroic interaction for the net MT dipole moments between the icosahedrons. We point out that the 4f and 5d energy levels can be made approach at the valence transition or the valence crossover, which is ubiquitous in the rare-earth based ACs and QCs. 
}

\begin{document}

\def\gsim{\mathop {\vtop {\ialign {##\crcr 
$\hfil \displaystyle {>}\hfil $\crcr \noalign {\kern1pt \nointerlineskip } 
$\,\sim$ \crcr \noalign {\kern1pt}}}}\limits}
\def\lsim{\mathop {\vtop {\ialign {##\crcr 
$\hfil \displaystyle {<}\hfil $\crcr \noalign {\kern1pt \nointerlineskip } 
$\,\,\sim$ \crcr \noalign {\kern1pt}}}}\limits}

\maketitle

\section{Introduction}
Quasicrystals (QCs) are solids without three dimensional periodicity of atoms but posses the structural order, which are distinct from periodic crystals and random systems. 
Since the discovery by D. Schechtman in 1984~\cite{Shechtman}, 
the understanding of the atomic structure of QCs has proceeded~\cite{Tsai,Takakura}. However, the electronic states in QCs and the physical properties are far from complete understanding because the Bloch theorem based on the translation symmetry can no longer be applied. 
For a long time, any long-range orders of electron-degrees of freedom have never been observed in QCs. Hence, it has been the dominant paradigm that electrons might not exhibit any ordering in QCs. 

This belief has,  however, been disproved by discoveries of quantum critical phenomena of valence transition in the QC Au-Al-Yb~\cite{Deguchi,Imura2020} and the superconductivity in the QC Al-Zn-Mg~\cite{Kabeya}. 
Recent discoveries of nonreciprocal and energy-asymmetric phonons in the QC Al-Pd-Mn~\cite{Matsuura}, atomic diffusion in Al-Ru-Pd~\cite{Nagai} and superconductivity in van der Waals layered QC Ta-Te~\cite{Tokumoto} have brought about a paradigm shift of the fundamental concept of physics established in periodic crystals, which have attracted much attention. 
One of the remaining significant questions about QCs has been whether the magnetic long-range order is realized in real QCs. 
To resolve this issue, considerable efforts have been devoted 
experimentally~\cite{Goldman,Suzuki2021}. 

In approximant crystals (ACs) which have the atomic configurations common to those in QCs with periodicity, ferromagnetic (FM) long-range order and antiferromagnetic (AF) long-range order have been observed
in Cd$_6$R (R=Pr, Nd, Sm, Gd, Tb, Dy, Ho, Eu, and Tm)~\cite{Tamura2010,Mori} and A-SM-R (A=Au and Ag; SM=Si, Al, Ge, and Sn; R=Ce, Gd, Tb, Dy, and Ho)~\cite{Hiroto2013,Hiroto2014,Imura,Sawano,Muro}. 
The magnetic phase transitions have been detected by the bulk measurements such as magnetic susceptibility and specific heat. 
Among them, the magnetic structures have been started to be identified in Tb-based compounds and Ho-based compound by neutron measurements~\cite{Das,Hiroto,Sato2019,Gebresenbut,Nawa2023,Labib}. 
These systems exhibit positive Curie-Weiss temperatures. 

Interestingly, negative Curie-Weiss temperature was observed in the Ce-based AC~\cite{Muro}. 
In the 1/1 AC Ag-In-Ce and Au-Al-Ce, a peak anomaly in magnetic susceptibility as well as specific heat was observed below 1~K~\cite{Imura,Sawano,Muro}. The origin of this anomaly is obscure although spin-grass like short-range magnetic order was discussed~\cite{Imura,Sawano}, which calls for theoretical study. 

As for the QC, the spin-grass behavior was observed in the QC Cd-R (R=Gd, Tb, Dy, Ho, Er, and Tm)~\cite{Goldman}.
Recently, the FM long-range order has been discovered in the rare-earth based QCs Au-Ga-R (R=Tb, Gd, and Dy)~\cite{Tamura2021,Takeuchi2023}. 
Recent discovery of the AFM long-range order in the QC Au-In-Eu~\cite{Tamura2025} has also attracted much attention. 
The 4f electrons at the rare-earth atoms are responsible for the magnetism.  
Since the wave function of the 4f electron is well localized in the vicinity of the rare-earth atom, strong Coulomb repulsion works between the 4f electrons.
Hence, the clarification of the electronic state in the rare-earth-based QC and AC not only advances the frontier of the condensed matter physics but also explores the new research field of the strongly correlated electrons~\cite{NKSato2022}.  
Recent breakthrough in the magnetism of the rare-earth-based QCs and ACs has opened a new stage of the research~\cite{WYTF2025}. 
Moreover, electric quadrupole in icosahedral quasicrystal has been theoretically discussed recently~\cite{Jeon2024}, which has attracted much attention.

The rare-earth atom in the icosahedral QCs and ACs is surrounded by ligand atoms with five-fold rotational symmetry [see Fig.~\ref{fig:Tsai_cluster}(f)]. Since five-fold symmetry is not compatible with translational symmetry of crystals, the conventional CEF theory based on crystallographic point group is not applicable to the QCs and ACs. Hence, lack of the CEF theory has prevented us from understanding the electronic state of QCs and ACs.  

Recently, the theory of the CEF in the rare-earth based QC and AC has been formulated on the basis of the point charge model~\cite{WK2021}. 
Then, the CEF in the quantum critical QC Au-Al-Yb~\cite{WK2021}, Au-Ga-Tb~\cite{WPNAS}, and Au-Ga-Dy~\cite{WI2023} was analyzed theoretically.   
Interestingly, uniaxial magnetic anisotropy was shown to arise from the CEF ground state of 4f electrons, giving rise to noncollinear magnetic structures on the icosahedral AC and QC~\cite{WPNAS,WSR,WI2025}. 
As noted above, positive Curie-Weiss temperatures were observed in the Tb- and Ho-based ACs where the FM interaction works between the 4f magnetic moments. In the Ce-based AC, where negative Curie-Weiss temperature was observed, the AFM interactions are dominant. Hence, it is interesting to clarify the effect of the magnetic frustration due to triangular geometry of the icosahedron [see Fig.~\ref{fig:Tsai_cluster}(c)]. 

An interesting feature of the rare-earth based QCs and ACs is that the local inversion symmetry is broken at the rare-earth site. Therefore, the odd-parity CEF, which is absent in centrosymmetric systems, exists. 
In centrosymmetric systems, even-parity CEF exists where the onsite hybridization between the states with different angular momentum is forbidden. However, in the non-centrosymmetric systems, odd-parity CEF arises, where parity mixing such as 4f-5d states occurs. 
The odd parity CEF based on the crystallographic point group was investigated theoretically~\cite{Hayami2018}. However, the odd-parity CEF in the QCs and ACs has not been studied to date. 
In this paper, we theoretically clarify not only the even-parity but also the odd-parity CEF in the Ce-based QC and AC. By constructing the magnetic model taking into account the magnetic anisotropy arising from the CEF, we reveal that unique magnetic states with noncollinear alignments of the magnetic moments appear in the ground-state phase diagrams in the 1/1 AC and QC. 
Under the odd-parity CEF, the 4f-5d hybrid wavefunction gives rise to the electric dipole as well as the magnetic toroidal (MT) dipole. We find that the vortex order in the 1/1 AC is characterized by the cluster-type MT moment, which can be resonated with the odd-parity CEF induced MT dipole. We reveal that the new type of magnetic order with noncollinear alignments of magnetic moments is stabilized in the icosahedral QC behind which there exists underlying mechanism to prefer the FM interaction for the MT dipole moments between the icosahedrons (ICs).

\section{\bf CEF for 4f electron in Au-SM-Ce} 
\label{sec:CEF_4f}

\begin{figure}[tb]
\centering
\includegraphics[width=14cm]{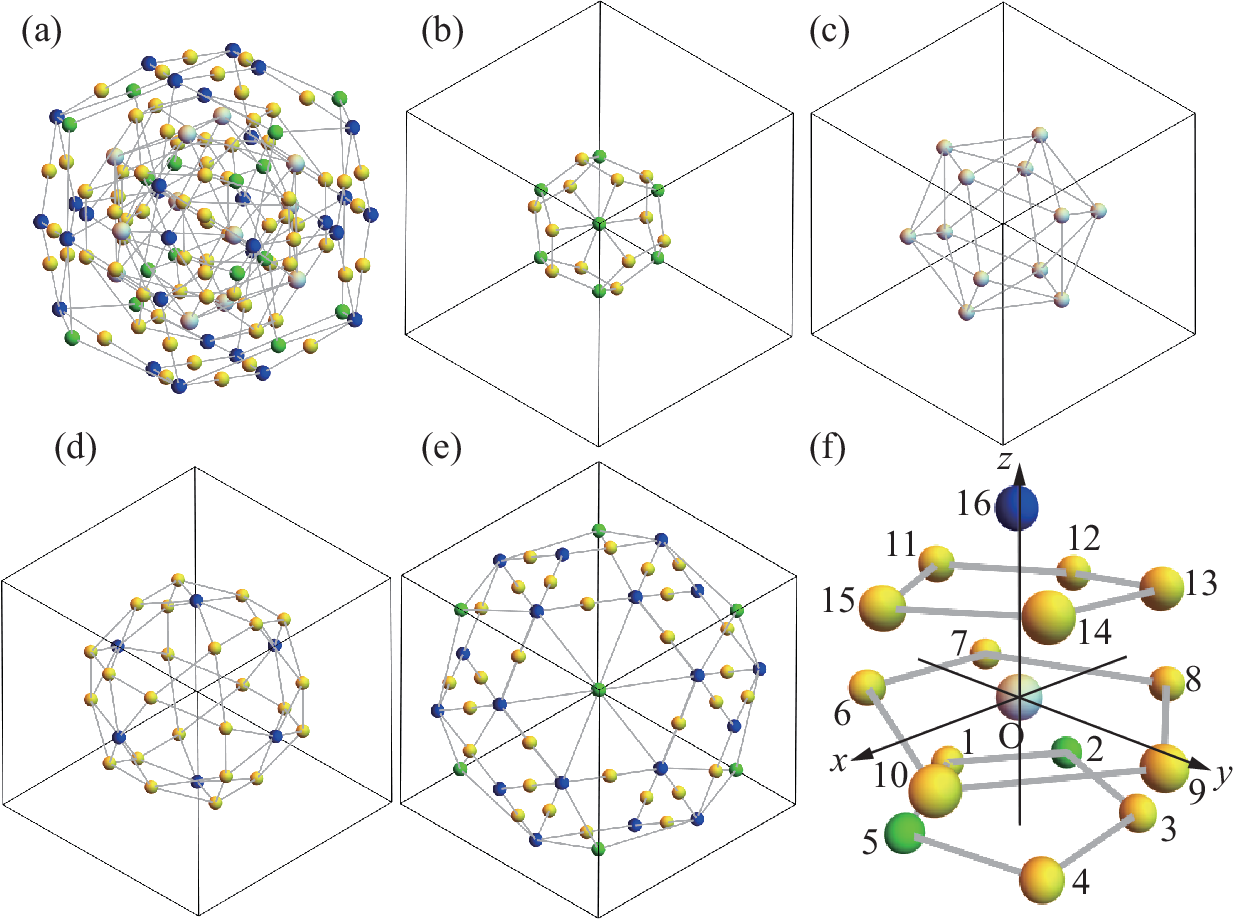}
\caption{(Color online) 
(a) Tsai-type cluster, (b) dodecahedron, (c) IC, (d) icosidodecahedron, and (e) defect rhombic triacontahedron with Ce (gray), SM (blue), Au (yellow), SM/Au mixed site (green). 
In (b)-(e), the frame box denotes the (expanded) bcc unit cell of the 1/1 AC.  The edge of the cube is the lattice constant $a$ of the 1/1 AC. 
(f) Local configuration of atoms around Ce. Surrounding Au and SM are labeled with numbers. 
}
\label{fig:Tsai_cluster}
\end{figure}

The rare-earth based QC and AC are composed of the Tsai-type cluster shown in Fig.~\ref{fig:Tsai_cluster}(a), which consists of the concentric shell structures of atoms shown in Figs.~\ref{fig:Tsai_cluster}(b)-(e).
Here we show the lattice structure experimentally identified in the 1/1 AC Au$_{65}$Sn$_{20}$Ce$_{15}$ with the lattice constant $a=15.190~{\rm \AA}$~\cite{Kenzari} as the prototype of the Au-SM-Ce (SM=Al, Sn, In).  
The rare-earth-based 1/1 AC is space group No. 204 $Im\bar{3}$ and site symmetry of the rare-earth site is $m..$. 
The rare-earth atoms are located at 12 vertices of the IC. 
The local configurations of atoms around the rare-earth atom is shown in Fig.~\ref{fig:Tsai_cluster}(f). 
We see that the (pseudo) five-fold rotation symmetry exists at the rare-earth site. 


By applying the formulation of the CEF theory developed in Ref.~\cite{WK2021} to the Ce-based QC and AC, we have analyzed the CEF for the Ce$^{3+}$ ion with $4f^1$ electron configuration. 

Let us outline the general formulation developed in Ref.~\cite{WK2021}. 
We consider the local coordinate shown in Fig.~\ref{fig:Tsai_cluster}(f), whose origin is set to be the Ce site.  
The electrostatic interaction between the electron located at ${\bm r}$ with the charge $q$ and the ligand ions located at ${\bm R}_i$ with the charge $q_i$ is expressed as
\begin{eqnarray}
H_{\rm CEF}=q\sum_{i=1}^{16}\frac{q_i}{|{\bm R}_i-{\bm r}|}, 
\label{eq:HCEF}
\end{eqnarray}
where the summation is taken over the Au and SM atoms labeled by $i=1$-16 surrounding the Ce atom in Fig.~\ref{fig:Tsai_cluster}(f). 
Here, $q_i=Z_{\rm SM}|e|$ for the SM ion and $q_i=Z_{\rm Au}|e|$ for the Au ion, where $Z_{\rm SM}$ and $Z_{\rm Au}$ are the effective valences of the SM ions and the Au ions, respectively.
As a first step of analysis, here we consider the case that SM is occupied $100\%$ at the SM/Au mixed site.
It is noted that the CEF in the 1/1 AC Au-SM-Tb (SM=Si, Al) was analyzed theoretically with similar configuration of ligand ions shown in Fig.~\ref{fig:Tsai_cluster}(f)~\cite{WPNAS}. The result was shown to explain the measured noncollinear magnetic structures, which was also endorsed by the CEF calculation considering farer ligand ions located at up to 30~${\rm \AA}$ (5000 sites)~\cite{Hiroto}. Hence, in this study we analyze the CEF considering the ligand ions shown in Fig.~\ref{fig:Tsai_cluster}(f). 

In Eq.~(\ref{eq:HCEF}), $|{\bm R}_i-{\bm r}|^{-1}$ can be expanded as
\begin{eqnarray}
\frac{1}{|{\bm R}_i-{\bm r}|}=\frac{1}{R_i}\sum_{\ell=0}^{\infty}\left(\frac{r}{R_i}\right)^{\ell}P_{\ell}(\cos{w_i}), 
\label{eq:R_r}
\end{eqnarray}
where $P_{\ell}(\cos{w_i})$ is the Legendre polynomial with $w_i$ being the angle between the vectors ${\bm R}_i$ and ${\bm r}$. 
Here we note that the addition theorem of the spherical harmonics holds as 
\begin{eqnarray}
P_{\ell}(\cos{w_i})=\frac{4\pi}{2\ell+1}\sum_{m=-\ell}^{\ell}Y_{\ell}^{m}(\theta_i,\varphi_i){Y_{\ell}^{m}}^{*}(\theta,\varphi), 
\label{eq:P_Y}
\end{eqnarray}
where ${\bm R}_i=(R_i,\theta_i,\varphi_i)$ is the position vector of ligand ions at the $i$th site and ${\bm r}$ is expressed as ${\bm r}=(r,\theta,\varphi)$ in the polar coordinate, where $\theta$ is the polar angle and $\varphi$ is the azimuth angle of the vector ${\bm r}$. By substituting Eq.~(\ref{eq:P_Y}) into Eq.~(\ref{eq:R_r}), we obtain
\begin{eqnarray}
H_{\rm CEF}=\sum_{i=1}^{16}\sum_{\ell=0}^{\infty}\sum_{m=-\ell}^{\ell}\frac{qq_i}{R_i}
\left(\frac{r}{R_i}\right)^{\ell}
\frac{4\pi(-1)^m}{2\ell+1}Y_{\ell}^{m}(\theta_i,\varphi_i)Y_{\ell}^{-m}(\theta,\varphi).
\nonumber
\\
\label{eq:HCEF2}
\end{eqnarray}
Here we have used the relation ${Y_{\ell}^{m}}^{*}(\theta,\varphi)=(-1)^mY_{\ell}^{-m}(\theta,\varphi)$. 

Then, by using the Wigner-Eckart theorem~\cite{WEtheorem}, $H_{\rm CEF}$ for 4f electron at Ce can be expressed as~\cite{WK2021} 
\begin{eqnarray}
H_{\rm CEF}^{4f}=\sum_{\ell=2,4,6}\left[
B_{\ell}^{0}(c)O_{\ell}^{0}(c)+\sum_{\lambda=c,s}\sum_{m=1}^{\ell}B_{\ell}^{m}(\lambda)O_{\ell}^{m}(\lambda)
\right],
\nonumber
\\
\label{eq:HCEF4f}
\end{eqnarray}
where $O_{\ell}^{m}(c)$ and  $O_{\ell}^{m}(s)$ are the total-angular-momentum operators called the Stevens operators~\cite{Stevens,Hutchings} and $B_{\ell}^{m}(c)$ and $B_{\ell}^{m}(s)$ are the coefficients. 
The definition of $B_{\ell}^{m}(\lambda)$ is given by 
\begin{eqnarray}
B_{\ell}^{m}(\lambda)=-|e|C_{\ell}^{m}\langle r^{\ell}\rangle\alpha_{\ell}h_{\ell}^{m}(\lambda), 
\label{eq:B_def}
\end{eqnarray}
where the charge $q$ in for $4f^1$ configuration at Ce Eq.~(\ref{eq:HCEF}) is set to be $-|e|$ with $e$ is the elementary charge. 

The Hund's rule tells us that the ground multiplet is given by the total angular momentum $J=5/2$. 
The Stevens factors of the $4f^1$ electron at the Ce$^{3+}$ ion for $J=5/2$ are given by $\alpha_2=-2/35$, $\alpha_4=2/315$, and $\alpha_6=0$~\cite{Stevens}. 
The expectation values by the radial part of 4f wave function have been evaluated for an isolated Ce$^{3+}$ ion~\cite{Giannozzi} by the density functional calculation using exchange correlation functional B3LYP~\cite{Becke} as $\langle r^2\rangle=0.4016~{\rm \AA}^2$ and $\langle r^4\rangle=0.3989~{\rm \AA}^4$ for the $J=5/2$ state (for detail, see Appendix~\ref{sec:DFT}).

The explicit forms of the coefficients $C_{\ell}^{m}$ are given in Ref.~\citen{WK2021} and  
$h_{\ell}^{m}(\lambda)$ are given as~\cite{WK2021}
\begin{eqnarray}
h_{\ell}^{m}(c)&=&{\rm Re}\left[
\sum_{i=1}^{16}\frac{q_i}{R_i^{\ell+1}}(-1)^mY_{\ell}^{m}(\theta_i,\varphi_i)
\right], 
\label{eq:hlm}
\\
h_{\ell}^{m}(s)&=&{\rm Im}\left[
\sum_{i=1}^{16}\frac{q_i}{R_i^{\ell+1}}(-1)^mY_{\ell}^{m}(\theta_i,\varphi_i)
\right]. 
\label{eq:hlm_s}
\end{eqnarray}
Here we note that Eqs. (7) and (8) themselves are not restricted to $m\ge 0$ but hold for $m<0$.
The Stevens operators $O_{\ell}^{m}(\lambda)$ for $\lambda=c,s$ are given in Ref.~\citen{rotter}. 
 
\begin{figure}[tb]
\centering
\includegraphics[width=14cm]{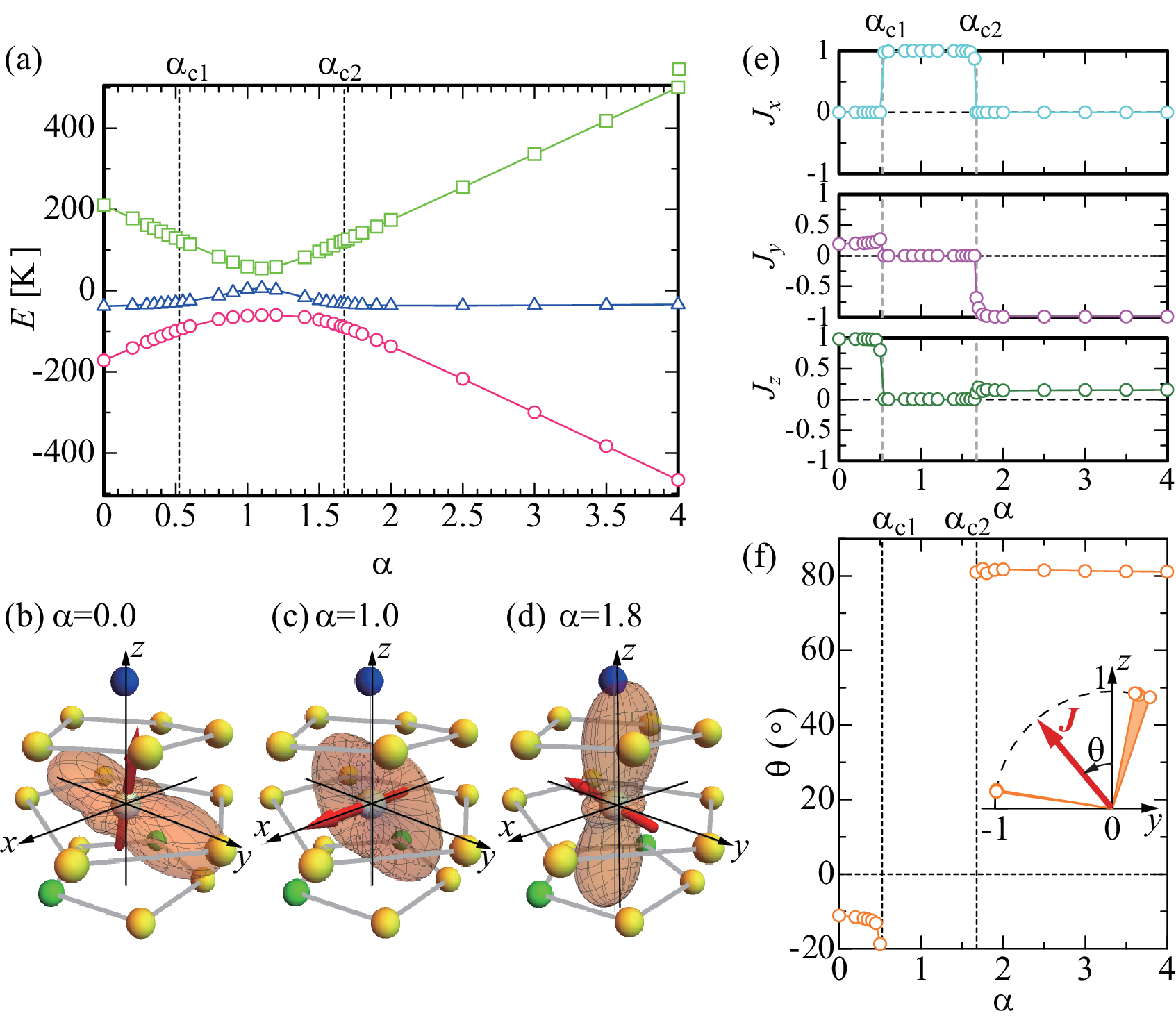}
\caption{(Color online) 
(a) The $\alpha$ dependence of the CEF energy of 4f electrons at the Ce site.  
Spherical part of the 4f wave function in the CEF ground state 
$|\psi_{0+}^{4f}(\hat{\bm r})|^2$ in the local coordinate with Ce (gray), Au (yellow), and Sn (blue) for (b) $\alpha=0.0$, (c) $\alpha=1.0$, and (d) $\alpha=1.8$. 
Red arrows indicate the magnetic easy axes. 
(e) The $\alpha$ dependence of the $x$ (top), $y$ (middle), $z$ (bottom) components of the largest magnetic moment in the CEF ground state. 
(f) The $\alpha$ dependence of the magnetic-moment ${\bm J}$ direction defined by the angle from the 5-fold $(z)$ axis in the mirror $(yz)$ plane (inset). 
In inset, the orange circles with the orange-colored regions correspond to the data indicated by ${\bm J}$ for $\alpha<\alpha_{\rm c1}$ and $\alpha>\alpha_{\rm c2}$. 
In (a), (e), and (f), the vertical dashed lines indicate $\alpha_{{\rm c}1}$ and $\alpha_{{\rm c}2}$. 
}
\label{fig:E4f_CEF}
\end{figure}

By diagonalizing the $6\times 6$ Hamiltonian matrix of Eq.~(\ref{eq:HCEF4f}) for the basis set of $|J=5/2, J_z\rangle_{4f}$ with $J_z=-5/2, -3/2, \cdots, 5/2$, we have obtained the CEF energy $E_n$~$(n=0,1,2)$ and the eigenstate 
\begin{eqnarray}
|\psi^{4f}_{n\eta}\rangle=\sum_{J_z=-5/2}^{5/2}a_{n\eta,J_z}|J=5/2,J_z\rangle_{4f}, 
\label{eq:WF_4f}
\end{eqnarray} 
where $\eta=\pm$ denotes the Kramers states. 
The results are shown in Fig.~\ref{fig:E4f_CEF}. 
Since the QC Au-SM-Ce and AC are metallic compounds, the valences of the Au ion and SM ion (e.g., Sn ion) are considered to be deviated from the formal valences +1 and +4, respectively~\cite{Pearson} because of the screening effect by conduction electrons. Hence, we plot the CEF energy as a function of the ratio of the valences of the screened ligand ions $\alpha=Z_{\rm SM}/Z_{\rm Au}$ in Fig.~\ref{fig:E4f_CEF}(a). 
Here we show the result for $Z_{\rm Au}=0.223$. This value was identifiled by the inelastic neutron measurement of the CEF in the 1/1 AC Au-Si-Tb~\cite{Hiroto}, which we here employ as a typical value.   
The $J=5/2$ states are split into 3 energy levels, each of which has the Kramers degeneracy.  
The energy difference between the ground state and the first-excited state $E_1-E_0$ is larger than 30~K. This indicates that the ground Kramers states dominate the physics in the low-temperature regime.
 
The spherical part of the wave function in the CEF ground state is obtained by calculating $\psi_{0\eta}^{4f}(\hat{\bm r})\equiv\langle\hat{\bm r}|\psi_{0\eta}^{4f}\rangle$ with $\hat{\bm r}$ being the unit vector along the radial direction. 
The $\alpha$ dependence of $|\psi_{0+}^{4f}(\hat{\bm r})|^2$ is shown in Figs.~\ref{fig:E4f_CEF}(b)-(d).  
At $\alpha=0.0$, $|\psi_{0+}^{4f}(\hat{\bm r})|^2$ has the elongated shape slightly tilted from the $y$ axis as shown in Fig.~\ref{fig:E4f_CEF}(b).   
At $\alpha=1.0$, the wave function has the flat shape lying in the $yz$ plane which is the mirror plane [Fig.~\ref{fig:E4f_CEF}(c)]. 
As $\alpha$ increases, the wave function is further deformed as the elongated shape along the diagonal direction in the $yz$ plane [Fig.~\ref{fig:E4f_CEF}(d)] at $\alpha=1.8$.

To clarify the magnetic anisotropy of the CEF ground state, we have calculated 
\begin{eqnarray}
\langle\psi_{0\eta}^{4f}|\hat{J}_{\mu}\hat{J}_{\nu}|\psi_{0\eta}^{4f}\rangle~(\mu, \nu=x, y, z), 
\label{eq:M_mat} 
\end{eqnarray}
where $\hat{J}_{\mu}$ is the operator of the total angular momentum. By diagonalizing the $3\times 3$ matrix, we have obtained the normalized eigenvector ${\bm J}$ for the largest eigenvalue, which gives the principal axis of the magnetic moment for the CEF ground state. The $\alpha$ dependence of each component of ${\bm J}$ is shown in Fig.~\ref{fig:E4f_CEF}(e). 
For $\alpha_{{\rm c}1}=0.525<\alpha<\alpha_{{\rm c}2}=1.675$, 
the results show ${\bm J}=(-1,0,0)$, which indicates that the magnetic easy axis is perpendicular to the mirror $(yz)$ plane. 
For $\alpha<\alpha_{{\rm c}1}$ and $\alpha>\alpha_{{\rm c}2}$, the results show $\langle J_x\rangle=0$ and finite $\langle J_y\rangle$ and $\langle J_z\rangle$ indicating that the magnetic easy axis is lying in the mirror plane.  
Namely, as $\alpha$ increases, the magnetic easy axis rotates to anti-clockwise direction in the $yz$ plane as shown in Fig.~\ref{fig:E4f_CEF}(f). 
Here we define the anisotropy angle $\theta$ between the magnetic easy axis and the (pseudo) five-fold axis ($z$ axis) in the mirror plane ($yz$ plane), as in the inset of Fig.~\ref{fig:E4f_CEF}(f). We note here the correspondence between $\theta$ and $\alpha$ 
$\theta=-11.1^{\circ}$ for $\alpha=0.0$, $\theta=-12.4^{\circ}$ for $\alpha=0.4$ as typical values. 
$\theta=77.0^{\circ}$ for $\alpha=1.7$ and $\theta=81.0^{\circ}$ for $\alpha=4.0$.
The magnetic easy axis is illustrated by red arrow in Figs.~\ref{fig:E4f_CEF}(b)-(d), which is directed to the perpendicular direction of the flat shape of $|\psi^{4f}_{0\eta}(\hat{\bm r})|^2$, as expected intuitively from the definition of the angular momentum.  

Notable is that the drastic change in the magnetic easy axis toward three different directions has not been reported in the CEF analyses of the other rare-earth systems such as Yb-, Tb-, and Dy-based QC and AC~\cite{WK2021,WPNAS,WI2023}, which indicates a unique feature in the Ce-based QC and AC.  
This is considered to be due to the lattice structure used in the present study, i.e., the 1/1 AC Au$_{65}$Sn$_{20}$Ce$_{15}$~\cite{Kenzari}. Namely, the number of the SM sites and their positions in the local coordinate around the rare-earth site [see Fig.~\ref{fig:Tsai_cluster}(f)] are different from those in previous studies~\cite{WK2021,WPNAS,WI2023}, which are considered to be the main reason of the emergence of the $\alpha<\alpha_{\rm c1}$ region. 


\section{Ground state of magnetic model in 1/1 approximant crystal}
\label{sec:PD_AC}

Next, let us analyze the magnetic structure in the 1/1 AC Au-SM-Ce with the IC, the 12 vertices of which are composed of the Ce atoms, located at the center and corner of the body-center-cubic (bcc) lattice.
We consider the effective model taking into account the magnetic anisotropy arising from the CEF
\begin{eqnarray}
H=-\sum_{\langle i,j\rangle}J_{ij}\hat{\bm J}_{i}\cdot\hat{\bm J}_{j},
\label{eq:H}
\end{eqnarray}
where $\hat{\bm J}_i$ is the unit vector operator parallel or anti-parallel to the magnetic easy axis at the $i$th Ce site and $J_{ij}$ is the interaction. 
In Eq.~(\ref{eq:H}), $J_{ij}$ is taken as the nearest-neighbor (N.N.) interaction $J_1$ and the next-nearest-neighbor (N.N.N.) interaction $J_2$ not only for the intra IC but also for the inter ICs. 
We set the N.N. bonds and the N.N.N. bonds similarly to our recent study for the Tb-based 1/1 AC~\cite{WI2025}. The readers refer to Ref.~\citen{WI2025} for the detail.
Here we consider the AF interaction $J_1<0$ and $J_2<0$ since negative Curie-Weiss temperature was observed in the 1/1 AC Au-Al-Ce~\cite{Muro}. The energy unit is taken as $J_1$ throughout this paper, i.e., $J_1=-1$. 

By performing the numerical calculation under the periodic boundary condition, we have determined the ground-state phase diagram of the effective model (\ref{eq:H}) in the 1/1 AC. Namely, we have calculated all the energies of $H$ for all eigenstates numerically and have identified the ground state with degeneracy. It is noted that the method of the numerical calculation is equivalent to the exact diagonalization.

\begin{figure}[h]
\centering
\includegraphics[width=14cm]{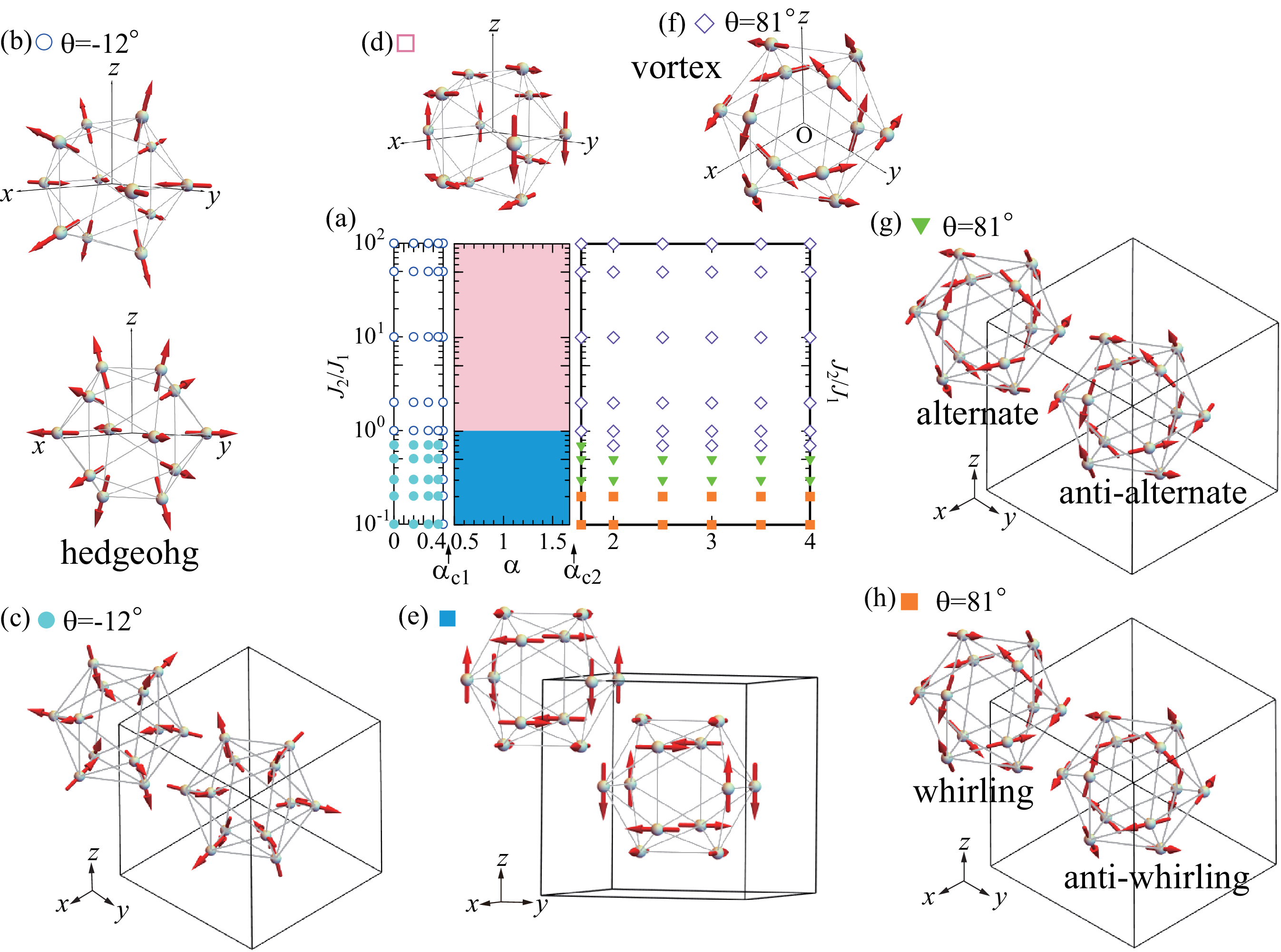}
\caption{(Color online) 
(a) Ground-state phase diagram for $J_2/J_1$ $(J_1<0, J_2<0)$ and $\alpha$. 
(b) and (c) magnetic states for $\theta=-12^{\circ}$. 
(d) and (e) Magnetic states for $\alpha_{{\rm c}1}<\alpha<\alpha_{{\rm c}2}$. 
(f) Vortex state seen from the (111) direction, (g) Alternate-anti-alternate state, and (h) whirling-anti-whirling state for $\theta=81^{\circ}$.  
In (c), (e), (g), and (h), the frame box denotes the expanded unit cell of the bcc lattice. 
}
\label{fig:PD_AC}
\end{figure}

The result is shown in Fig.~\ref{fig:PD_AC}(a) in the plane of $J_2/J_1$ and $\alpha$ corresponding to Fig.~\ref{fig:E4f_CEF}(a). 
The open symbol denotes the magnetic state on the IC, which is distributed uniformly, i.e., ferroic disdtribution, while the filled symbol denotes the magnetic state where the alignment of the magnetic moments at the 12 vertices of the IC located at the corner of the bcc unit cell are obtained by all inversing their directions at the 12 vertices of the IC located at the center of the bcc unit cell, i.e., antiferroic distribution.  
Namely, the magnetic propagation vector ${\bm q}=\frac{2\pi}{a}(hkl)$ with $a$ being a lattice constant for the former states and the latter states are $(hkl)=(000)$ and $(111)$, respectively (see Table~\ref{tb:mag_state}).  Hence, the latter states are referred to as antiferromagnetic states.


In the right panel of Fig.~\ref{fig:PD_AC}(a) for $\alpha>\alpha_{{\rm c}2}$, 
for $J_2/J_1\ge 0.89$, the magnetic state shown in Fig.~\ref{fig:PD_AC}(f) is realized at the center and corner ICs in the bcc unit cell. 
In this state, the magnetic moments on the IC swirl when we see from the (111) direction [Fig.~\ref{fig:PD_AC}(f)], which we call the vortex state [note that the vortex swirling is seen only from the (111) direction but not from the other diagonal axes of the bcc unit cell]. 
The vortex state has the 3-fold symmetry along the (111) axis, whose magnetic space group is identified to be $R\bar{3'}$ No. 148.3.1249 in the OG notation~\cite{OG} ($R\bar{3}'$ No. 148.19 in the BNS notation~\cite{BNS}).
Our numerical calculation has shown that the ground state has 8 degeneracy. This reflects the number of choosing one 3-fold axis among four diagonal directions i.e., $(\pm1,\pm1,1)$ axes and all magnetic moments are energetically degenerated with the all-moments-inverted state, which results in $4\times 2=8$ degeneracy.  

It is noted that the total magnetic moment on the IC in the vortex state
\begin{eqnarray}
{\bm J}_{\rm IC}=\sum_{i=1}^{12}{\bm J}_i,
\label{eq:J_IC}
\end{eqnarray}
is zero ${\bm J}_{\rm IC}={\bf 0}$. 

\begin{figure}[tb]
\centering
\includegraphics[width=5cm]{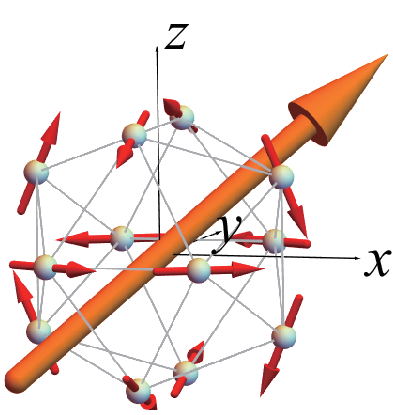}
\caption{(Color online) 
Side view of the vortex state with the magnetic toroidal moment ${\bm T}_{\rm IC}$ being illustrated as an orange arrow [see Eq.~(\ref{eq:MT})]. Red arrows denote the magnetic moments. 
}
\label{fig:vortex_MT}
\end{figure}

The vortex state where the magnetic moments on the IC swirl can be characterized by the MT moment
\begin{eqnarray}
{\bm T}_{\rm IC}=\sum_{i=1}^{12}{\bm r}_i\times{\bm J}_i, 
\label{eq:MT_cluster}
\end{eqnarray}
where ${\bm r}_i$ is the position vector of the Ce site measured from the center of the IC and ${\bm J}_i$ is the magnetic moment at the $i$th site in the ground state. 
For the vortex state shown in Fig.~\ref{fig:PD_AC}(f), the calculation of Eq.~(\ref{eq:MT_cluster}) results in ${\bm T}_{\rm IC}=a(t,t,t)\propto (111)$ where $a$ is the lattice constant of the 1/1 AC (see the vector ${\bm T}_{\rm IC}$ illustrated as an orange arrow in Fig.~\ref{fig:vortex_MT}). 

It is noted here that in Eq.~(\ref{eq:MT_cluster}), the origin of the position vector is set to be the origin of the IC, i.e., the center of the primitive unit cell. Exactly speaking, the choice of the origin to define the cluster MT moment may affect the quantitative value in the crystalline solids~\cite{King-Smith1993,Vanderbilt1993,SatoHayami2025,Vanderbilt_book}. However, since the vortex state is uniformly distributed in the 1/1 AC where each IC is isolated each other as shown in Fig.~\ref{fig:PD_AC}, the existence of the ferroic MT moment is considered to be at least qualitatively valid.

Since the vortex state is uniformly distributed at each IC, the ferroic alignment of the MT moments along the (111) direction is realized in the 1/1 AC. This causes the electro-magnetic effect where the magnetic moment ${\bm M}$ is induced by applying electric field ${\bm E}$ along the perpendicular direction ${\bm M}\perp{\bm E}$~\cite{Hayami2014,Hayami2024}.

To explore the topological property of the magnetic texture on the IC, we have calculated the topological charge $n$ defined on the IC, which has the physical meaning that the number covering the unit sphere by the magnetic moments at each vertex of the IC. 
The definition of $n$ is given by
\begin{eqnarray}
n\equiv\frac{\Omega}{4\pi},
\label{eq:TC}
\end{eqnarray} 
where $\Omega$ is the solid angle subtended by 12 magnetic moments on the IC and $4\pi$ is the surface area of the unit sphere~\cite{WPNAS}. The solid angle on the IC is given by 
\begin{eqnarray}
\Omega=\sum_{i,j,k\in{\rm IC}}\Omega_{ijk}, 
\label{eq:SA}
\end{eqnarray} 
where $\Omega_{ijk}$ is the solid angle for three magnetic moments ${\bm J}_i$, ${\bm J}_j$, and ${\bm J}_k$. Here, $\Omega_{ijk}$ can be calculated by~\cite{Eriksson}
\begin{eqnarray} 
\Omega_{ijk}=2\tan^{-1}\left[\frac{\chi_{ijk}}{1+{\bm J}_i\cdot{\bm J}_j+{\bm J}_j\cdot{\bm J}_k+{\bm J}_k\cdot{\bm J}_i}\right],
\label{eq:SA2}
\end{eqnarray}
where $\chi_{ijk}$ is the scalar chirality 
\begin{eqnarray}
\chi_{ijk}={\bm J}_{i}\cdot({\bm J}_j\times{\bm J}_k). 
\label{eq:SC}
\end{eqnarray}

We calculated the topological charge for each magnetic ground state shown in Fig.~\ref{fig:PD_AC}(a), among which the topological magnetic state characterized by non-zero $n$ is the whirling-antiwhirling state [Fig.~\ref{fig:PD_AC}(h)]. 
The whirling state located at the center IC in Fig.~\ref{fig:PD_AC}(i) is characterized by $n=+3$ and antiwhirling state located at the corner IC is characterized by $n=-3$~\cite{WSR}. 
The total magnetic moment per IC is zero ${\bm J}_{\rm IC}={\bf 0}$ in the whirling state and also in the anti-whirling state. 
We note that the whirling-antiwhirling state also appears in the ground-state phase diagram of the 1/1 AC for the FM interactions $J_1$ and $J_2$ for large $J_2/J_1$ regime~\cite{WPNAS,WI2025}, which was observed by neutron measurement in the 1/1 AC Au-Al-Tb~\cite{Sato2019,Labib2023}. As reported theoretically~\cite{WI2025} and experimentally~\cite{Labib2023}, the magnetic space group of the  whirling-anti-whirling state is $I_{\rm P}m'\bar{3'}$ No. 204.5.1534 in the OG notation ($PIn\bar{3}$, No. 201.21 in the BNS notation). 

It is known that the noncollinear alignment of magnetic moments give rise to the emergent fictious magnetic field~\cite{Aoyama2021}. We define the total chirality~\cite{WPNAS,Aoyama2021,WI2025}
\begin{eqnarray}
{\bm \chi}^{\rm T}=\sum_{\langle{i,j,k}\rangle}\chi_{ijk}\hat{n}_{ijk},
\label{eq:chi_T}
\end{eqnarray}
where the summation is taken over all triangular surfaces of the IC formed by vertices $i$, $j$ and $k$ with $\hat{n}_{ijk}$ being the surface normal~\cite{WPNAS}. 
When the total chirality is not zero, the topological Hall effect emerges as~\cite{Tatara}
\begin{eqnarray}
\sigma^{\rm T}_{\mu\nu}\propto\epsilon_{\mu\nu\rho}\chi_{\rho}^{\rm T},
\label{eq:sigma_T}
\end{eqnarray}
where $\sigma_{\mu\nu}^{\rm T}$ is the electric conductivity arising from the emergent fictious magnetic field and $\epsilon_{\mu\nu\rho}$ is the Levi-Civita symbol. 

We have calculated the total chirality ${\bm \chi}^{\rm T}$ for each ground state shown in Fig.~\ref{fig:PD_AC}(a) and it has turned out that the magnetic states shown in Fig.~\ref{fig:PD_AC}(c) and \ref{fig:PD_AC}(g) exhibit non-zero ${\bm \chi}^{\rm T}$. Namely, in the magnetic state located at center (corner) IC in the bcc unit cell shown in Fig.~\ref{fig:PD_AC}(c), the total chirality vector points to the (111) [$-(111)$] direction. 
Note that this ground state has eight degeneracy where ${\bm \chi}^{\rm T}$ of each state points to the diagonal-axis direction $(\pm1,\pm1,\pm1)$ (see Table~\ref{tb:mag_state}). 
In the alternate (anti-alternate) state located at center (corner) IC in the bcc unit cell shown in Fig.~\ref{fig:PD_AC}(g), the total chirality vector points to the $-(111)$ [(111)] direction.  
Note that this ground state has eight degeneracy where ${\bm \chi}^{\rm T}$ of each state points to the diagonal-axis direction $(\mp1,\mp1,\mp1)$. 
In these states, the total chirality per unit cell is cancelled in total.

The total magnetic moment ${\bm J}_{\rm IC}$ for the anti-alternate state (alternate state) located at the center (corner) IC of the frame box in Fig.~\ref{fig:PD_AC}(g) is directed along the (111) [$-(111)$] direction. 
In the magnetic state located at the center (corner) IC shown in Fig.~\ref{fig:PD_AC}(c), the total magnetic moment ${\bm J}_{\rm IC}$ points to the $-(111)$ [(111)] direction. 
As noted above, these states have eight degeneracy, where ${\bm J}_{\rm IC}$ of each state points to the diagonal-axis direction  $(\pm1,\pm1,\pm1)$ (see Table~\ref{tb:mag_state}).  

The magnetic space group of the magnetic state shown in Fig.~\ref{fig:PD_AC}(c) and the alternate-anti-alternate state shown in Fig.~\ref{fig:PD_AC}(g) is $R_{\rm R}\bar{3}$ No.148.4.1250 in the OG notation ($RI\bar{3}$ No. 148.20 in the BNS notation). 
 
As for the magnetic sates realized for $\alpha_{\rm c1}\le\alpha\le\alpha_{\rm c2}$ [see Figs.~\ref{fig:PD_AC}(d) and \ref{fig:PD_AC}(e)], collinear alignment of the magnetic moments for the N.N. sites along the edge of the IC appears, due to which scalar chirality vanishes [see Eq.~(\ref{eq:SC})] so that the solid angle $\Omega_{ijk}$ cannot be defined as noted in Ref.~\cite{WPNAS}. Hence, the topological charge $n$ and the total chirality ${\bm \chi}^{\rm T}$ are indeterminate. The total magnetic moment ${\bm J}_{\rm IC}$ is zero in these states. 

The magnetic space group of the magnetic state shown in Fig.~\ref{fig:PD_AC}(d) is $R\bar{3}'$ No. 148.3.1249 in the OG notation ($R\bar{3}'$ No. 148.19 in the BNS notation). The magnetic space group of the magnetic state shown in Fig.~\ref{fig:PD_AC}(e) is $I_{\rm P}23$ No. 197.3.1510 in the OG notation ($P_{\rm I}23$ No. 195.3 in the BNS notation).

In the noncollinear magnetic state shown in Fig.~\ref{fig:PD_AC}(b), the total magnetic moment is zero ${\bm J}_{\rm IC}={\bm 0}$ and the total chirality is zero ${\bm \chi}^{\rm T}={\bm 0}$. 
The magnetic space groups of these states are $Im'm'm'$ No. 71.5.625 in the OG notation ($Im'm'm'$ No. 71.537 in the BNS notation).
 
The results of ${\bm J}_{\rm IC}$, ${\bm \chi}^{\rm T}$, $n$ as well as the degeneracy for each magnetic ground state shown in Fig.~\ref{fig:PD_AC}(a) are summarized in Table.~\ref{tb:mag_state}. Note that for the antiferromagnetic states on the center IC and corner IC, ${\bm J}_{\rm IC}$ and ${\bm \chi}^{\rm T}$ for the center IC are listed in Table~\ref{tb:mag_state}. Note that all the states at least have the degenerated state where all the magnetic moments are inverted reflecting time-reversal symmetry in the finite-size calculation with periodic boundary condition so that the even number of degeneracy appears.

\begin{table}
\caption{
Characteristics of each magnetic ground state in the 1/1 AC.
The symbol in the first column denotes the magnetic state shown in Fig.~\ref{fig:PD_AC}.  
The second column denotes magnetic space group in OG notation.
The third column denotes the magnetic propagation vector ${\bm q}=\frac{2\pi}{a}(hkl)$ with $a$ being a lattice constant.
The fourth column denotes the degeneracy of the ground state. 
The fifth column and sixth column denotes the total magnetic moment ${\bm J}_{\rm IC}$ and total chirality ${\bm \chi}^{\rm T}$ per IC.  
The seventh column denotes the topological charge $n$. For the antiferromagnetic state, ${\bm J}_{\rm IC}$, ${\bm \chi}^{\rm T}$ and $n$ for the IC located at the center of the bcc unit cell shown in Fig.~\ref{fig:PD_AC} are denoted (see text). 
}
\label{tb:mag_state}
\renewcommand{\arraystretch}{1.2}
\begin{tabular}{lcccccc}
\hline
\textrm{state}&
mag. sp. gr. &
$(hkl)$ &
\textrm{dege.}& \textrm{${\bm J}_{\rm IC}$} & \textrm{${\bm \chi}^{\rm T}$} &
\textrm{$n$}\\
\hline
{\color{blue}${\circ}$} Fig.~\ref{fig:PD_AC}(b) & $Im'm'm'$ & (000) & 6 & {\bf 0} & {\bf 0} & 0 \\
\hline
{\color{pink}$\square$} Fig.~\ref{fig:PD_AC}(d) & $R\bar{3}'$ & (000) & 8 & {\bf 0} & - & - \\
\hline
{\color{purple}$\diamond$} Fig.~\ref{fig:PD_AC}(f) & $R\bar{3}'$ & (000) & 8 & {\bf 0} & {\bf 0} & 0 \\
\hline
{\color{SkyBlue}$\bullet$} Fig.~\ref{fig:PD_AC}(c) & $R_{\rm R}\bar{3}$ & (111) & 8 & $\propto-(111)$ & $\propto(111)$ & 0 \\
\hline
{\color{Cyan}$\blacksquare$} Fig.~\ref{fig:PD_AC}(e) & $I_{\rm P}23$ & (111) & 8 & {\bf 0} & - & - \\
\hline
{\color{Green}$\blacktriangledown$} Fig.~\ref{fig:PD_AC}(g) & $R_{\rm R}\bar{3}$ & (111) & 8 & $\propto(111)$ & $\propto-(111)$ & 0 \\
\hline
{\color{orange}$\blacksquare$} Fig.~\ref{fig:PD_AC}(h) & $I_{\rm P}m'\bar{3}'$ & (111) & 2 & {\bf 0} & {\bf 0} & 3 \\
\hline
\end{tabular}
\renewcommand{\arraystretch}{1}
\end{table}

\section{Quasicrystal cluster} 

To gain insight into magnetism in Ce-based icosahedral QC, we have theoretically studied the QC cluster. The Ce-based QC has the Cd$_{5.7}$Yb-type lattice structure~\cite{Takakura} composed of the Tsai-type cluster shown in Fig.~\ref{fig:Tsai_cluster}(a), where Yb is replaced with Ce. In this study, we have considered a QC cluster, which consists of 30 ICs located at the vertices of the icosidodecahedron, as a first step of analysis. This structure is repeatedly contained in the Cd$_{5.7}$Yb-type QC in the self-similar manner~\cite{Takakura}.

Here we have applied the effective model Eq.~(\ref{eq:H}) to the icosidodecahedron cluster with the 30 ICs.  
We consider the N.N. interaction $J_1$ and the N.N.N. interaction $J_2$ not only for the intra IC but also for the inter IC under the open boundary condition. In the present study, we have studied the case that the magnetic easy axis is directed around $\theta=80^{\circ}$ within the mirror $(yz)$ plane in the local coordinate [see Fig.~\ref{fig:Tsai_cluster}(f)] with the anisotropy angle $\theta$ being defined in the inset of Fig.~\ref{fig:E4f_CEF}(f), which corresponds to the $\alpha>\alpha_{\rm c2}$ regime in Fig.~\ref{fig:E4f_CEF}(a). 

By performing the numerical calculation of the 2-ICs cluster mean-field method (see Appendix~\ref{sec:2ICMF} for detail), we have explored the ground state.  Namely, by inputting several initial states on the ICs obtained as the ground state in the 1/1 AC [see Fig.~\ref{fig:PD_AC}(a)],  we have sought for the converged states for $J_2/J_1$ and the anisotropy angle $\theta$ [see the inset of Fig.~\ref{fig:E4f_CEF}(f)], among which the lowest-energy state is identified. 

\begin{figure}[tb]
\centering
\includegraphics[width=14cm]{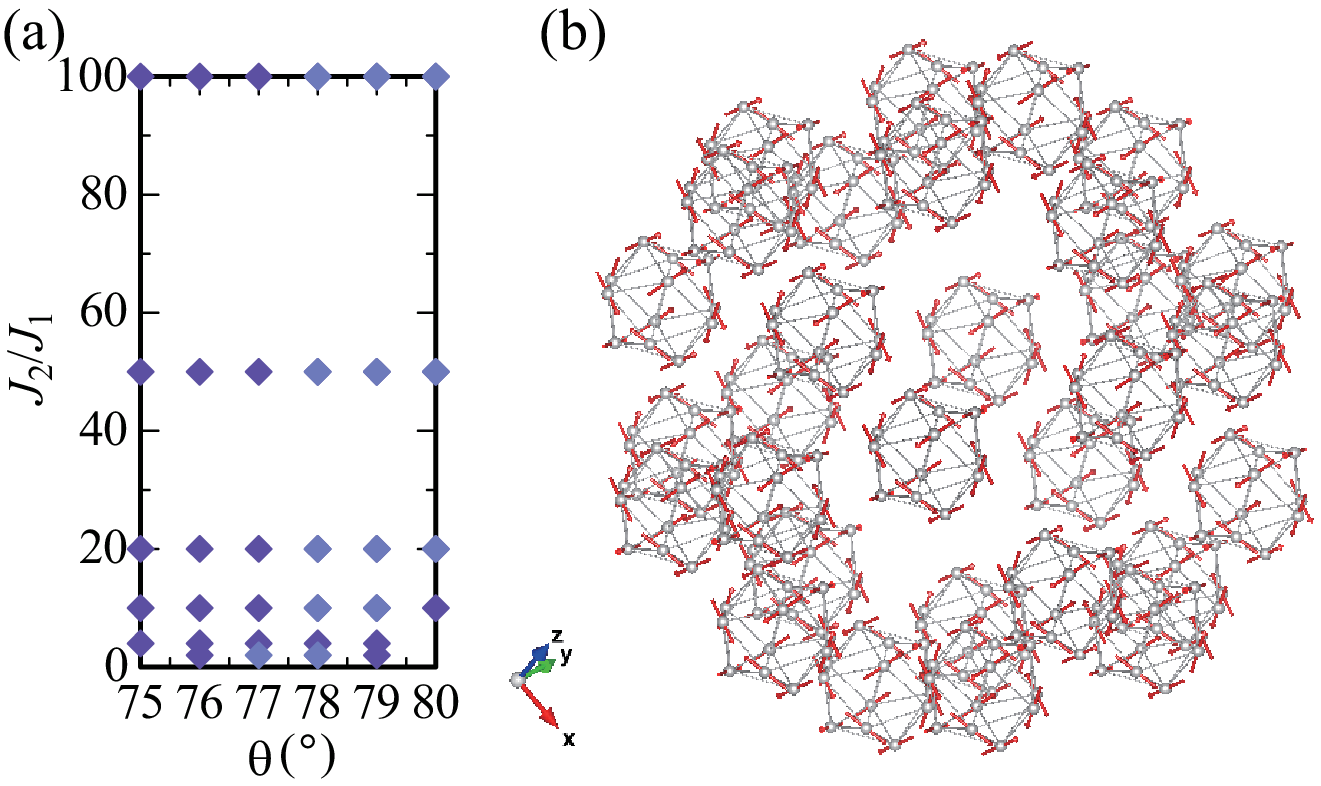}
\caption{(Color online) 
(a) Ground-state phase diagram for $J_2/J_1$ with $J_1<0$ and $\alpha$. 
(b) The magnetic moments for $\theta=77^{\circ}$ at $J_1=-1$ and $J_2=-1$.  
(c) Total magnetic toroidal moment per IC ${\bm t}_{\rm tot}$ in the magnetic order denoted by the purple diamonds in (a). 
}
\label{fig:PD_QC}
\end{figure}

The results are summarized in the phase diagram shown in Fig.~\ref{fig:PD_QC}(a), where the converged states are obtained at the symbols denoted by purple diamonds. The magnetic structure for $J_2/J_1=10$ at $\theta=77^{\circ}$ is shown in Fig.~\ref{fig:PD_QC}(b). 
The non-collinear magnetic moments are aligned on the IC, which are different in IC by IC. 
We have calculated the total magnetic moment in the $j$th IC 
\begin{eqnarray}
{\bm J}^{\rm IC}_{j}=\sum_{i=1}^{12}\langle\hat{\bm J}_{j,i}\rangle, 
\label{eq:J_IC}
\end{eqnarray}
where $\hat{\bm J}_{j,i}$ is the unit vector operator parallel or anti-parallel to the magnetic easy axis direction. The magnitude of ${\bm J}^{\rm IC}_{j}$ and the direction are different in IC by IC. 
The total magnetic moment per IC defined by 
\begin{eqnarray}
{\bm J}_{\rm tot}=\frac{1}{30}\sum_{j=1}^{30}{\bm J}^{\rm IC}_{j} 
\label{eq:J_tot}
\end{eqnarray}
is finite as ${\bm J}_{\rm tot}=(-0.04,0.19,-0.02)$. 
After the summation of the non-collinear magnetic moments the total magnetic moments are almost cancelled so that small amount remains. 
The ground states at purple diamonds in Fig.~\ref{fig:PD_QC}(a) have similar features. 
In Sect.~\ref{sec:FM_MT_QC}, it will be shown that these states are characterized by the FM alignment of the magnetic toroidal (MT) dipoles [see Fig.~\ref {fig:t_tot}(b)]. 
We note that this state is different from the FM order in the QC where the net magnetic moments of the noncollinear ferrimagnetic state in all the ICs are directed to the 3-fold axis direction reported in Ref.~\cite{WPNAS} nor the hedgehog order where the net magnetic moment in each IC is zero  reported in Ref.~\cite{WSR}. 


\hspace{0.2cm}

\section{CEF for 4f and 5d electrons}

At the Ce site in the Tsai-type cluster, the 5-fold symmetry exists in the local coordinate as shown in Fig.~\ref{fig:Tsai_cluster}(f). The important point is that the spatial inversion symmetry is broken locally at the Ce site, which gives rise to the odd-parity CEF.  In the Ce-based compounds, at most one 5d electron exists at the Ce site. Hence, here we extend the CEF theory~\cite{WK2021} not only for 4f electron but also for 5d electron on the basis of the point charge model. 

\subsection{CEF for 5d electron}

Since the CEF theory developed in Ref.~\cite{WK2021} was formulated for general values of the total angular momentum $J$, 
it can be applied to the 5d state in the Ce$^{3+}$ ion. Namely, the CEF Hamiltonian for 5d electron $H_{\rm CEF}^{5d}$ is expressed by taking the summation for $\ell=2, 4$ in Eq.~(\ref{eq:HCEF4f}). 
The Hund's rule tells us that the ground multiplet of the $5d^1$ configuration is given by $J=3/2$. 

For 5d electrons, the calculation by the density functional theory~\cite{Giannozzi} using the B3LYP hybrid functional~\cite{Becke} gives $\langle r^2\rangle_{5d}=1.767~{\rm \AA}^2$ for the 5d state with $J=3/2$ (for detail, see Appendix~\ref{sec:DFT}).  
The Stevens factors of the $5d^1$ electron for $J=3/2$ is derived as $\alpha_{2}^{5d}=-2/15$~\cite{WSSC2021}, $\alpha_{4}^{5d}=0$, and $\alpha_{6}^{5d}=0$.

Then, the $4\times 4$ Hamiltonian matrix $H_{\rm CEF}^{5d}$ is obtained for the basis $|J=3/2, J_z\rangle_{5d}$ with $J_z=-3/2, \cdot\cdot, 3/2$. 

\subsection{Odd-parity CEF}
\label{sec:op_CEF}

The odd parity CEF is formulated as the off-diagonal matrix element of the CEF Hamiltonian Eq.~(\ref{eq:HCEF}) as 
%
\begin{eqnarray}
_{4f}\Bra{J=\frac{5}{2},J_z}H_{\rm CEF}\Ket{J'=\frac{3}{2},J'_z}_{5d} . 
\label{eq:HCEF4f5d}
\end{eqnarray}
Note that this term is usually zero in the centrosymmetric systems but in the icosahedral QC and AC, this term can be finite because the spatial inversion symmetry is broken locally at the rare-earth site as shown in Fig.~\ref{fig:Tsai_cluster}(f). 

The spherical part of the 5d wave function is defined by $\psi^{5d}_{J',J_z'}(\hat{\bm r})\equiv\langle\hat{\bm r}|J'=3/2, J_z'\rangle_{5d}$. 
The explicit form of the basis function  $\psi^{5d}_{J',J_z'}\equiv\langle\hat{\bm r}|J',J_z'\rangle_{5d}$ as well as $\psi^{4f}_{J,J_z}\equiv\langle\hat{\bm r}|J,J_z\rangle_{4f}$ is given in Appendix~\ref{sec:psi_5d}. 
As shown in Ref.~\cite{WK2021} and also in Sec.~\ref{sec:CEF_4f}, $H_{\rm CEF}$ can be expressed by the spherical harmonics as Eq.~(\ref{eq:HCEF2}). 
Then, by substituting $\psi^{4f}_{J,J_z}(\hat{\bm r})$ and $\psi^{5d}_{J',J_z'}(\hat{\bm r})$ into Eq.~(\ref{eq:HCEF4f5d}) and performing the integral with respect to the spherical part
\begin{eqnarray}
\int{Y_{{\ell}_1}^{m_{1}}}^{*}(\theta,\varphi)Y_{\ell}^{m}(\theta,\varphi)Y_{\ell_{2}}^{m_{2}}(\theta,\varphi)\sin\theta{d\theta}{d\varphi},   
\label{eq:YYY}
\end{eqnarray}
we obtain the $6\times 4$ matrix elements of Eq.~(\ref{eq:HCEF4f5d}). It is noted here that 
Eq.~(\ref{eq:YYY}) is only non zero when $m=m_1-m_2$ and $\ell_1+\ell_2$ is the even number for $|\ell_1-\ell_2|\le\ell\le\ell_1+\ell_2$.  
Namely, the odd-parity CEF Hamiltonian is expressed as
\begin{eqnarray}
H_{\rm CEF}^{4f-5d}=q\sum_{\ell=1,3,5}\sum_{m=-\ell}^{\ell}\frac{4\pi{r}^{\ell}}{2\ell+1}\left[h_{\ell}^{m}(c)+ih_{\ell}^{m}(s)\right]Y_{\ell}^{-m}(\theta,\varphi),
\nonumber
\\
\label{eq:CEF_4f-5d}
\end{eqnarray}
where $q$ is the charge of ${\rm R}^{3+}$ ion of the rare-earth atom R and 
$h_{\ell}^{m}(c)$ and $h_{\ell}^{m}(s)$ are given in Eq.~(\ref{eq:hlm}) and Eq.~(\ref{eq:hlm_s}), respectively. 
For Ce$^{3+}$ ion with $4f^1$ configuration, $q$ is set to as $q=-|e|$.

Then, the off-diagonal matrix element Eq.~(\ref{eq:HCEF4f5d}) is calculated for example as
%
\begin{eqnarray}
_{4f}\Bra{\frac{5}{2},\frac{5}{2}}H_{\rm CEF}\Ket{\frac{3}{2},\frac{3}{2}}_{5d}=
_{4f}\Bra{\frac{5}{2},\frac{5}{2}}H_{\rm CEF}^{4f-5d}\Ket{\frac{3}{2},\frac{3}{2}}_{5d}
\nonumber
\\
=\sqrt{\frac{1}{7}}\sqrt{\frac{1}{5}}\sum_{\ell=1,3,5}\left[h_{\ell}^{-1}+ih_{\ell}^{-1}(s)\right]\sqrt{\frac{4\pi}{2\ell+1}}\langle r^{\ell}\rangle
c^{\ell}(3,2,2,1)
\nonumber
\\
+\sqrt{\frac{6}{7}}\sqrt{\frac{4}{5}}\sum_{\ell=1,3,5}\left[h_{\ell}^{-1}+ih_{\ell}^{-1}(s)\right]\sqrt{\frac{4\pi}{2\ell+1}}\langle r^{\ell}\rangle
c^{\ell}(3,3,2,2), 
\label{eq:72_32_b}
\end{eqnarray}
where $c^{k}({\ell},m,{\ell}',m')$ is defined by
\begin{eqnarray}
c^{k}({\ell},m,{\ell}',m')=\sqrt{\frac{4\pi}{2k+1}}\int{Y_{\ell}^{m}}^{*}(\theta,\varphi)Y_{k}^{m-m'}(\theta,\varphi)Y_{\ell'}^{m'}(\theta,\varphi)\sin\theta{d\theta}d\varphi
\label{eq:c_func}
\end{eqnarray}
and $h_{\ell}^{m}$ and $h_{\ell}^{m}(s)$ are defined by Eq.~(\ref{eq:hlm}) and Eq.~(\ref{eq:hlm_s}), respectively. 
%
It is noted that the spherical integral in the form of Eq.~(\ref{eq:c_func}) is computable by performing the double integral with respect to $\theta$ and $\varphi$.  The list of the values of $c^{k}({\ell},m,{\ell}',m')$ can be found in literatures~\cite{ITO_Group_Theory}.   
As shown in Eq.~(\ref{eq:72_32_b}), the expectation values of the radial parts $\langle r\rangle$, $\langle r^3\rangle$, and $\langle r^5\rangle$ appear.  
By performing the calculation for an isolated Ce$^{3+}$ ion by the density functional theory~\cite{Giannozzi} using the B3LYP hybrid functional~\cite{Becke}, we have obtained $\langle r\rangle=0.5593~{\rm \AA}$, $\langle r^3\rangle=0.3622~{\rm \AA}^3$, and $\langle r^5\rangle=0.5525~{\rm \AA}^5$. We have calculated $\langle r^n\rangle$ for $n=1, 2,\dots,6$ for the $4f$ state and $5d$ state not only for $R^{3+}$ but also for $R^{2+}$ ($R$=Ce, Pr, Nd, Pm, Sm, Eu, Gd, Tb, Dy, Ho, Er, Tm, and Yb). For detail, readers refer to Appendix~\ref{sec:DFT}.  
With use of these values, we have calculated all the off-diagonal matrix elements of Eq.~(\ref{eq:CEF_4f-5d}) for the $4f$ state with $J=5/2$ and $J_z=5/2, 3/2, 1/2, -1/2, -3/2, -5/2$ and the $5d$ state with $J_z'=3/2$ and $J_z'=3/2, 1/2, -1/2, 3/2$.  
In Appendix~\ref{sec:H_CEF_odd}, the explicit forms of the off-diagonal matrix element Eq.~(\ref{eq:HCEF4f5d}) are given. 

Then, we have obtained the $10\times 10$ total matrix of $H_{\rm CEF}$. 
By diagonalizing it, we have obtained the CEF energies $E_{n}^{\rm tot}$ $(n=0,1,2,3,4)$ and the eigenstates  
\begin{eqnarray}
|\psi_{n\eta}^{\rm tot}\rangle&=&\sum_{J_z=-5/2}^{5/2} a_{n\eta,J_z}^{\rm tot}\Ket{J=\frac{5}{2},J_z}_{4f}
\nonumber
\\
&+&\sum_{J_z=-3/2}^{3/2}b_{n\eta,J_z}^{\rm tot}\Ket{J=\frac{3}{2},J_z}_{5d},
\label{eq:WFtot}
\end{eqnarray}
where $\eta=\pm$ denotes the Kramers states and $a_{n\eta,J_z}^{\rm tot}$ and $b_{n\eta,J_z}^{\rm tot}$ are the coefficients which satisfy $\sum_{J_z=-5/2}^{5/2}|a_{n\eta,J_z}^{\rm tot}|^2+\sum_{J_z=-3/2}^{3/2}|b_{n\eta,J_z}^{\rm tot}|^2=1$. 
We have also calculated the spherical part of the wave function of the CEF ground state $\psi^{\rm tot}_{0\eta}(\hat{\bm r})\equiv\langle\hat{\bm r}|\psi_{0\eta}^{\rm tot}\rangle$. 

\begin{figure}[tb]
\centering
\includegraphics[width=8cm]{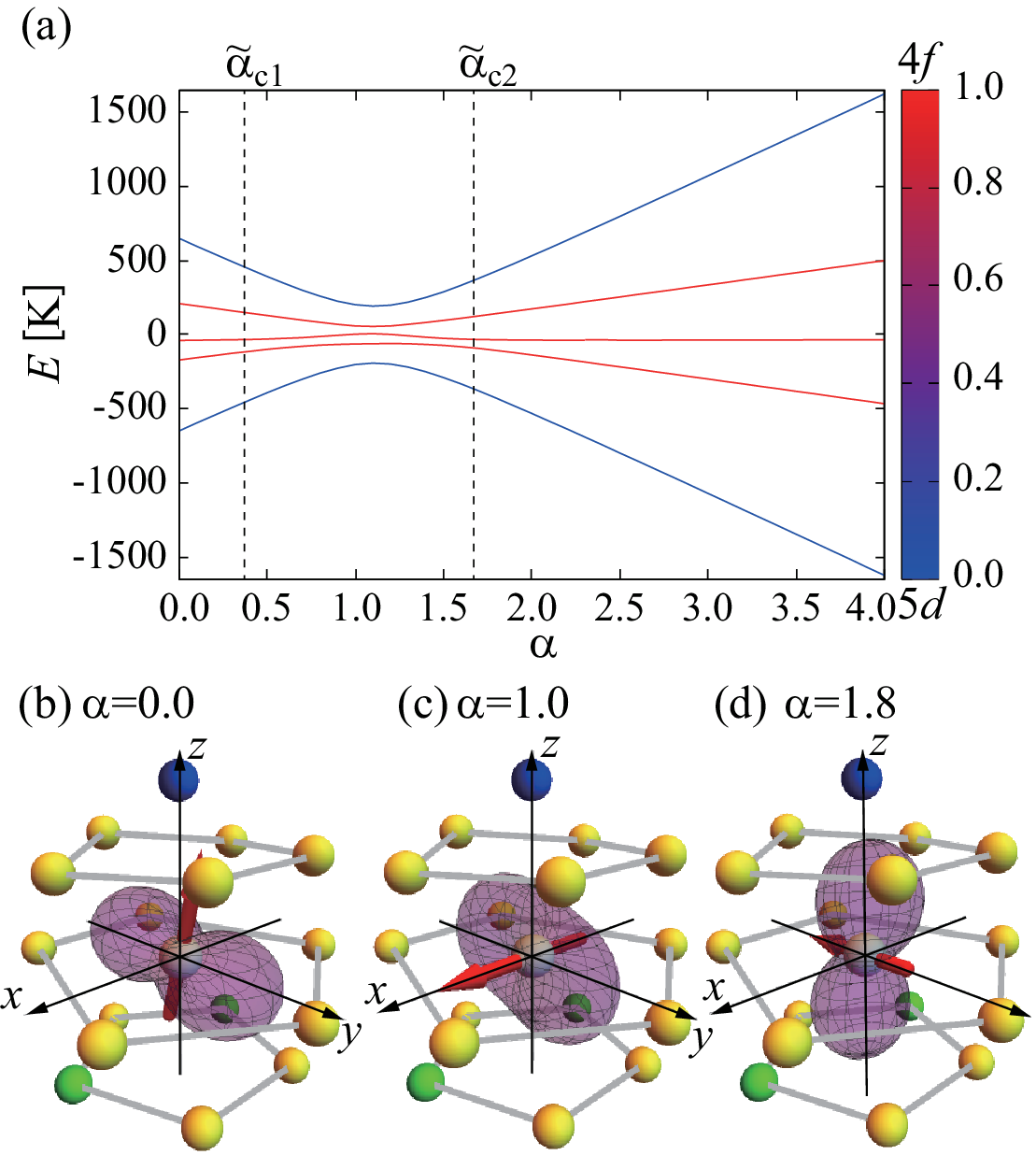}
\caption{(Color online) 
(a) The CEF energy $E_n^{\rm tot}$ vs $\alpha=Z_{\rm SM}/Z_{\rm Au}$ for $Z_{\rm Au}=0.223$. The color of the energy level represents the ratio of the 4f component ranging 0 (blue) to 1 (red). 
The vertical dashed lines indicate $\tilde{\alpha}_{\rm c1}$ and $\tilde{\alpha}_{\rm c2}$. 
$|\psi_{0+}^{\rm tot}(\hat{\bm r})|^2$ for (b) $\alpha=0.0$,
(c)  $\alpha=1.0$, and 
(d)  $\alpha=1.8$.  
In (b)-(d), the red arrow denotes the magnetic easy axis of the CEF ground state. 
}
\label{fig:E_CEF_tot}
\end{figure}

We show the CEF energies $E_n^{\rm tot}$ for $Z_{\rm Au}=0.223$ as a function of $\alpha=Z_{\rm SM}/Z_{\rm Au}$ in Fig.~\ref{fig:E_CEF_tot}(a). 
Each energy level is plotted with color ranging from blue to red, which reflect the ratio of the 4f component $\sum_{J_z=-5/2}^{5/2}|a_{n\eta,J_z}^{\rm tot}|^2=0$ to $1$ for the eigenstate $\Ket{\psi_{n\eta}^{\rm tot}}$.  
From the results, it has turned out that the ground state $|\psi_{0\eta}^{\rm tot}\rangle$ and the 4th exscited state $|\psi_{4\eta}^{\rm tot}\rangle$ consist of the 5d component $|\psi_{n\eta}^{\rm tot}\rangle\approx |\psi_{n\eta}^{5d}\rangle$ for $n=0$ and $4$ and the 4f component $|\psi_{n\eta}^{\rm tot}\rangle\approx |\psi_{n\eta}^{4f}\rangle$  for $n=1$, 2, and 3 in all the range of $\alpha$ presented in Fig.~\ref{fig:E_CEF_tot}(a). 
The spherical part of the ground-state wave function $|\psi_{0+}^{\rm tot}(\hat{\bm r})|^2\approx|\psi_{0+}^{5d}({\bm \hat{\bm r}})|^2$ are shown in Figs.~\ref{fig:E_CEF_tot}(b)-(d).  At $\alpha=0.0$, $|\psi_{0+}^{5d}({\bm \hat{\bm r}})|^2$ has a shape elongated along the $y$-axis direction. At $\alpha=1.0$, the wave function has a flat shape lying in the $yz$ plane. At $\alpha=1.8$, the wave function is elongated along slightly tilded $z$-axis direction. 

Here, we also plot the magnetic easy axis as a red arrow, which was obtained as the normalized eigenvector of the largest eigenvalue of the $3\times 3$ matrix $\langle\psi_{0\eta}^{5d}|\hat{J}_{\mu}\hat{J}_{\nu}|\psi_{0\eta}^{5d}\rangle$ similarly in Eq.~(\ref{eq:M_mat}). At $\alpha=0.0$, the magnetic easy axis is directed slightly tilted from the $z$ axis [Fig.~\ref{fig:E_CEF_tot}(a)]. At $\alpha=1.0$, the magnetic easy axis is pointed perpendicularly to the mirror $(yz)$ plane [Fig.~\ref{fig:E_CEF_tot}(b)]. At $\alpha=1.8$, the magnetic easy axis is directed almost perpendicularly to the $z$ axis in the $yz$ plane. 
These results indicate that 
similarly to the 4f-electron case, the drastic change in the magnetic easy axis of the 5d ground state takes place. We have identified the threshold values of $\alpha$ as $\tilde{\alpha}_{\rm c1}=0.3725$ and $\tilde{\alpha}_{\rm c2}=1.6725$. For $\alpha<\tilde{\alpha}_{\rm c1}$ and $\alpha>\tilde{\alpha}_{\rm c2}$, the magnetic easy axis is lying in the $yz$ plane while for $\tilde{\alpha}_{\rm c1}<\alpha<\tilde{\alpha}_{\rm c2}$, the magnetic easy axis is perpendicular to the $yz$ plane. 

\subsection{Lattice constant dependences of the f level and d level}

The CEF energy levels shown in Fig.~\ref{fig:E_CEF_tot} are split into 5 levels, among which the lowest- and highest-energies states are the 5d states and the middle 3 states are the 4f states. In this subsection, we analyze the lattice constant dependence of the f level and d level. 

In the point charge model, the 4f level $\varepsilon_{\rm f}$ is expressed as~\cite{WK2021}
\begin{eqnarray}
\varepsilon_{\rm f}\propto B_{2}^{0}\langle O_{2}^{0}\rangle +B_{4}^{0}\langle O_4^0\rangle, 
\label{eq:Ef}
\end{eqnarray}
where $O_{\ell}^0$ is the Stevens oprators for $J=5/2$. 
The d level is expressed as
\begin{eqnarray}
\varepsilon_{\rm d}\propto {B}_{2}^{0}\langle O_{2}^{0}\rangle,
\label{eq:Ed}
\end{eqnarray}
where $O_2^0$ is the Stevens operator for $J=3/2$. 
As explicitly shown in Eq.~(\ref{eq:hlm}), the coefficient $B_{\ell}^{m}$ is expressed as $B_{\ell}^m\propto {R_{i}}^{-\ell-1}$, i.e., $B_2^0\propto {R_i}^{-3}$ and $B_4^0\propto {R_i}^{-5}$. Hence, the lattice-constant dependences of $\varepsilon_{\rm f}$ and $\varepsilon_{\rm d}$ are different within the point charge model~\cite{WM2018}. 

On the other hand, it is known that the effect of hybridization between the wave functions of the 4f and conduction electrons also contribute to the CEF~\cite{Takahashi1985,Otsuki2005,WM2018}. According to the linear combination of atomic orbitals (LCAO) method~\cite{Andersen1,Andersen2}, the transfer integral between the electrons with the orbital angular momentum $\ell$ and $\ell'$ has the distance dependence as $V_{\ell,\ell'}\propto r^{-\ell-\ell'-1}$ where $r$ is the distance between each orbital. For example, in the case that the 4f electron at the Ce site hybridizes with the 6s electron at the $i$th Au site in Fig.~\ref{fig:Tsai_cluster}(f), the transfer integral $V^{4f,6s}_i$ is expressed as $V^{4f,6s}_{i}\propto {R_i}^{-4}$. By employing the 2nd-order perturbation theory with respect to the hybridization, the distance dependence of $\varepsilon_{\rm f}$ is estimated as
\begin{eqnarray}
\varepsilon_{\rm f}\propto\frac{|V^{4f,3p}_i|^2}{\Delta}\propto {R_{i}}^{-8},
\label{eq:Ef2}
\end{eqnarray}
where $\Delta$ is the excitation energy to the $4f^2$ state or $4f^0$ state.  

According to the LCAO method~\cite{Andersen1,Andersen2}, 
the transfer integral $V_i^{5d,3p}$ between the 5d electron at the Ce site and the 6s electron at the $i$th Au site is expressed as $V_i^{5d,6s}\propto R_i^{-3}$. Then, the distance dependence of $\varepsilon_{\rm d}$ is estimated as
\begin{eqnarray}
\varepsilon_{\rm d}\propto\frac{|V^{5d,3p}_i|^2}{\Delta'}\propto {R_{i}}^{-6},
\label{eq:Ed2}
\end{eqnarray}
where $\Delta'$ is the excitation energy to the $5d^2$ state or $5d^0$ state.  

Above estimation assumed that the hybridization between the 4f and 5d electrons at the Ce site and the 6s electrons at the Au site is dominant. Even in the case that the hybridization between the 4f and 5d electrons at the Ce site with the other conduction electrons is assumed, the distance dependences of $\varepsilon_{\rm f}$ and $\varepsilon_{\rm d}$ are different, as demonstrated in Eq.~(\ref{eq:Ef2}) and Eq.~(\ref{eq:Ed2}).  

In real systems, the CEF is considered to be contributed from both effects of the point charge and the hybridization. Hence, applying pressure or chemical substitution (yielding negative pressure) is expected to cause their remarkable dependences of $\varepsilon_{\rm f}$ and $\varepsilon_{\rm d}$.

\subsection{Effect of level crossing of $\varepsilon_{\rm f}$ and $\varepsilon_{\rm d}$}

As analyzed in previous subsection, by applying positive or negative pressure to Au-SM-Ce, it is considered to be possible to make the f level $\varepsilon_{\rm f}$ and the 5d level $\varepsilon_{\rm d}$ approach. 
As noted above, the CEF in real materials is contributed from both the effects of the point charge and hybridization. On these grounds, the coefficients in the CEF Hamiltonian in Eq.~(\ref{eq:HCEF4f}) e.g., $B_{\ell}^m$ are often treated as parameters in the calculation of the CEF energies~\cite{Hayami2018}. Here,
 to explore 
the effect of the close 4f- and 5d-energy levels, 
we analyze the ground-state property by changing the coefficient $B_2^0$ in Eq. (\ref{eq:Ed}), to be precise, the value of $\langle r^2\rangle_{\rm 5d}$ as a parameter. 
When positive or negative pressure is applied to Ce-based QCs and ACs, the 4f-level and 5d-level positions are generally changed. Hence, in reality, $B_2^0$ and $B_4^0$ in Eq.~(\ref{eq:Ef}) as well as the coefficients in the off diagonal $H_{\rm CEF}^{4f-5d}$ in Eq.~(\ref{eq:CEF_4f-5d}), i.e., $h_{\ell}^m(c)+ih_{\ell}^m(s)$ are expected to change their values. Here, to demonstrate what happens when $\varepsilon_{\rm f}$ and $\varepsilon_{\rm d}$ approach, we just change $\varepsilon_{\rm d}$ from the parameters used for Fig.~\ref{fig:E_CEF_tot}. The underlying thought behind this is that the coefficients are to be regarded as parameters of the {\it effective} CEF model considering the effects of the point charge and hybridization in the following discussions of Sects. 5.4, 6.1 and 6.2. The results for this representative parameter which will be shown below such as the emergence of onsite hybrid 4f-5d orbital and odd-parity multipoles are considered to be at least qualitatively correct.

\begin{figure}[tb]
\centering
\includegraphics[width=8cm]{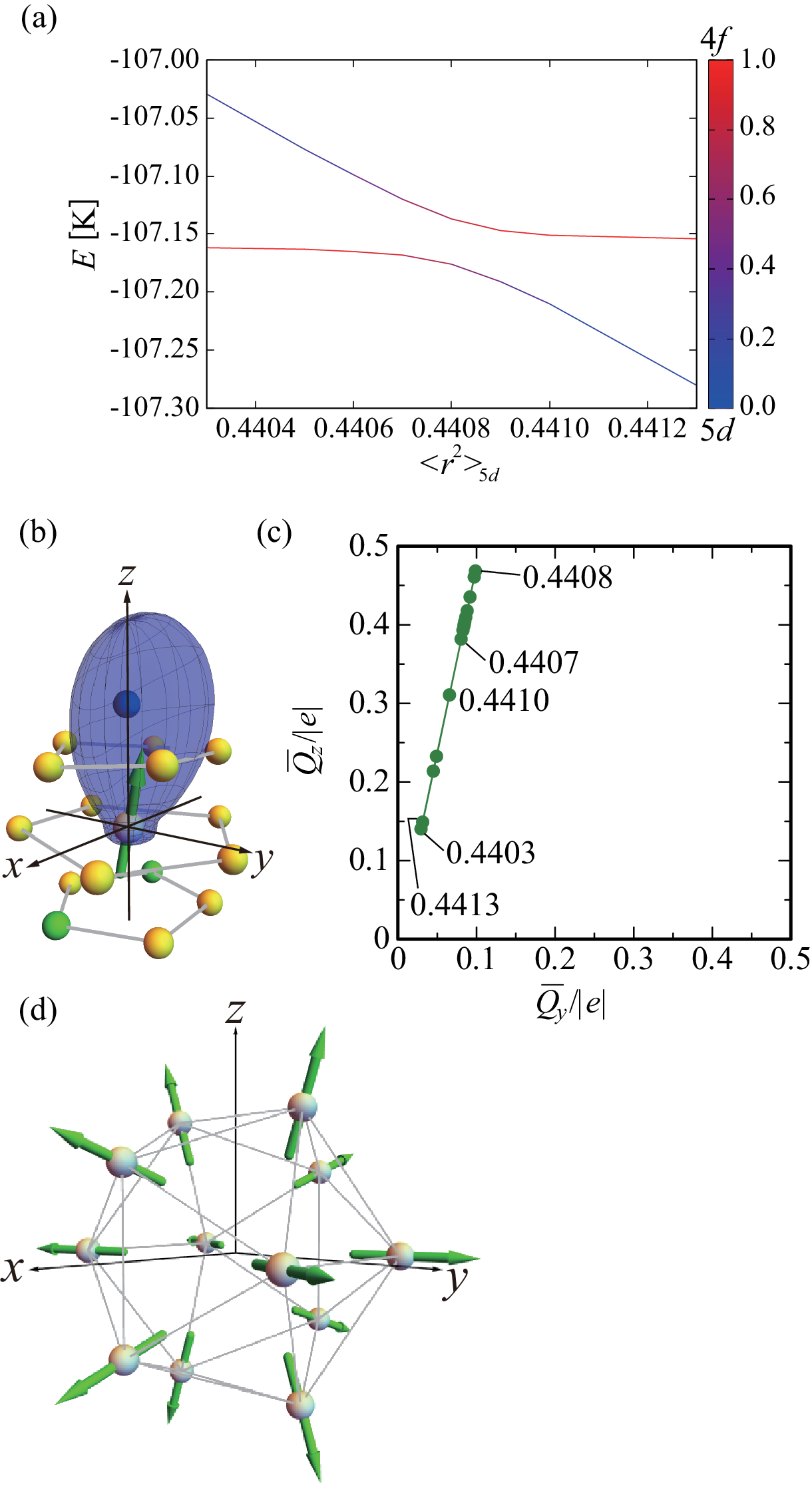}
\caption{(Color online) 
(a) The CEF energies $E_0^{\rm tot}$ and $E_1^{\rm tot}$ vs $\langle r^2\rangle_{\rm 5d}$ for $\alpha=1.8$ at $Z_{\rm Au}=0.223$. The color of the energy level represents the ratio of the 4f component ranging 0 (blue) to 1 (red). 
(b)  Spherical part of the ground state of the total CEF Hamiltonian $|\psi_{0+}^{\rm tot}(\hat{\bm r})|^2$ at $\langle r^2\rangle_{\rm 5d}=0.4408$. 
The green arrow denotes the electric dipole moment for  $\langle r^2\rangle_{\rm 5d}=0.4408$ at $\alpha=1.8$. 
(c)  $\bar{Q}_z/|e|$ vs $\bar{Q}_y/|e|$ for $0.4403\le \langle r^2\rangle_{\rm 5d}\le0.4413$ at $\alpha=0.18$. The number denotes the value of  $\langle r^2\rangle_{\rm 5d}$. 
(d)  Electric dipole moments in the IC for $\langle r^2\rangle_{\rm 5d}=0.4408$ at $\alpha=1.8$. 
}
\label{fig:E_r2_5d}
\end{figure}

Figure~\ref{fig:E_r2_5d}(a) shows the $\langle r^2\rangle_{\rm 5d}$ dependence of the CEF energies $E_n^{\rm tot}$ $(n=0, 1)$ for $\alpha=1.8$ at $Z_{\rm Au}=0.223$. 
Each energy level is plotted with color ranging from blue to red, which reflect the 4f component $\sum_{Jz=-5/2}^{5/2}|a_{n\eta,J_z}^{\rm tot}|^2=0$ to $1$ for the eigenstate $\Ket{\psi_{n\eta}^{\rm tot}}$. 
The dominant contribution to the CEF ground state changes from the 4f state at $\langle r^2\rangle_{\rm 5d}=0.4403$ to the 5d state at $\langle r^2\rangle_{\rm 5d}=0.4413$. 
In between, e.g., at $\langle r^2\rangle_{\rm 5d}=0.4408$, the 4f-5d hybrid state $|\psi_{0+}^{\rm tot}(\hat{\bm r})|^2$, which has the bulged shape along the +$zy$ direction, emerges as shown in Fig.~\ref{fig:E_r2_5d}(b). We confirmed that the Kramers pair state $|\psi_{0-}^{\rm tot}(\hat{\bm r})|^2$ has the same shape as  $|\psi_{0+}^{\rm tot}(\hat{\bm r})|^2$. 
This indicates that the onsite 4f-5d hybridization becomes prominent at the parameter where the 4f level and the 5d level approach mostly. 
This indicates that the electron polarization occurs at each Ce site.  
Namely, the electric dipole moment is active at the rare-earth site in the icosahedral QC and AC because of the local inversion symmetry breaking. Interestingly, the odd-parity multipole is active under the odd-parity CEF, which will be discussed in the next section.  

\section{Odd-parity multipoles}

In this section, we analyze the property of the odd-parity multipole in the QC and AC Au-SM-Ce, focusing on the electric dipole and the magnetic toroidal (MT) dipole.  

\subsection{Electric dipole}
\label{sec:ED}

To quantify the electron polarization, we have calculated the electron dipole moment 
\begin{eqnarray}
Q_{\mu}\equiv|e|\mu  \  \  \   (\mu=x, y, z)
\label{eq:ED}
\end{eqnarray} 
in the local coordinate [see Fig.~\ref{fig:Tsai_cluster}(f)].  
We have calculated the expectation value $\bar{Q}_{\mu}\equiv\langle\psi_{0\eta}^{\rm tot}|\frac{Q_{\mu}}{r}|\psi_{0\eta}^{\rm tot}\rangle$ with $r=\sqrt{x^2+y^2+z^2}$ for the CEF ground state Eq.~(\ref{eq:WFtot}).  
The explicit of form of the matrix element for the basis set of the 4f state $\Ket{J=5/2,J_z}$ and the 5d state $\Ket{J=3/2,J_z'}$ is given in Appendix~\ref{sec:EDapp}. 

The results show $\bar{Q}_{x}=0$ because of the mirror symmetry with respect to the $yz$ plane in the local coordinate. The results of $\bar{Q}_z/|e|$ vs $\bar{Q}_y/|e|$ calculated for the CEF ground state presented in Fig.~\ref{fig:E_CEF_tot}(a) are shown for $0.4403\le \langle r^2\rangle_{\rm 5d}\le0.4413$ in Fig.~\ref{fig:E_r2_5d}(c). 
We see that the electron polarization takes place along the $yz$ direction in the mirror plane. The magnitude of the electron polarization becomes maximum at $\langle r^2\rangle_{\rm 5d}=0.4408$, where $\bar{\bm Q}/|e|$ is given by $(0, 0.099, 0.468)$. 
We plot this polarization by distributing them at the 12 vertices of the IC, which are denoted as the $\bar{\bm Q}_i$ vector (green vector), in Fig.~\ref{fig:E_r2_5d}(d).
This result indicates that the electron polarization takes place at each site of the IC. Namely, when $\varepsilon_{\rm f}$ and $\varepsilon_{\rm d}$ most approach each other as Fig.~\ref{fig:E_r2_5d}(a), the bulged shape of $|\psi_{0\pm}^{\rm tot}(\hat{\bm r})|^2$ as well as the electron polarization at the Ce site of the IC emerges.  We note that the total electric polarization in the IC is zero $\sum_{i=1}^{12}\bar{\bm Q}_i={\bm 0}$ as seen in Fig.~\ref{fig:E_r2_5d}(d) where the electric dipole moments are directed oppositely each other at the diagonal positions with respect to the center of the IC.   

\subsection{Magnetic toroidal dipole}
\label{sec:MT}

\subsection{Magnetic toroidal moment}

Recently, the MT moment has been defined quantum mechanically~\cite{HK2018,Hayami2018,Kusunose2020,Hayami2024} as
\begin{eqnarray}
{\bm t}_{\ell}({\bm r})=\frac{{\bm r}}{\ell+1}\times\left(\frac{2{\bm \ell}}{\ell+2}+{\bm \sigma}\right), 
\label{eq:MT}
\end{eqnarray}
where ${\bm \ell}$ and ${\bm \sigma}$ are the orbital and spin angular-momentum operators respectively at the site ${\bm r}$. 
Here we calculate the MT dipole moment $\tilde{t}_{1\mu}$ $(\mu=x, y, z)$, where $\tilde{t}_{\ell\mu}$ is defined as the Hermite operator $\tilde{t}_{\ell\mu}\equiv \{t_{\ell\mu}({\bm r})+{t_{\ell\mu}}^{\dagger}({\bm r})\}/(2r)$~\cite{Kusunose2020} in the local coordinate [see Fig.~\ref{fig:Tsai_cluster}(f)]. Here we consider the MT dipole $\ell=1$, whose moment is given by $T_{1\mu}=-\mu_{\rm B}\tilde{t}_{1\mu}$, where $\mu_{\rm B}$ is the Bohr magneton~\cite{HK2018}. 

\begin{figure}[tb]
\centering
\includegraphics[width=9cm]{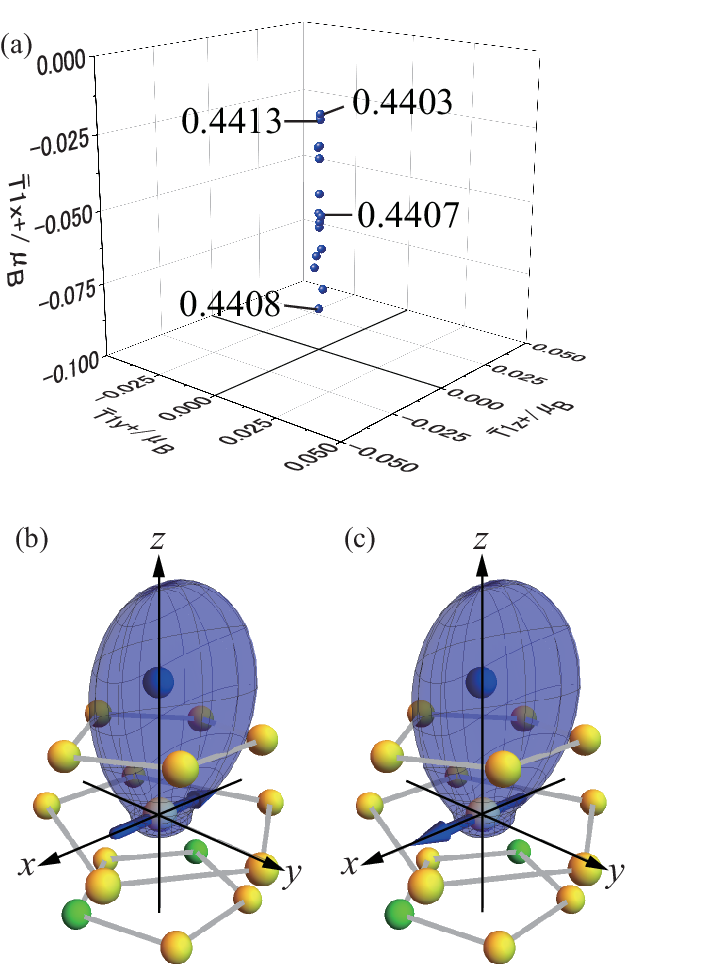}
\caption{(Color online) 
(a)  $\bar{\bm T}_{1+}/\mu_{\rm B}$ for $0.4403<\langle r^2\rangle_{\rm 5d}<0.4413$ at $\alpha=1.8$. The number indicates the value of $\langle r^2\rangle_{\rm 5d}$. 
(b)  Spherical part of the ground state of the total CEF Hamiltonian $|\psi_{0+}^{\rm tot}(\hat{\bm r})|^2$ at $\langle r^2\rangle_{\rm 5d}=0.4408$. 
The blue arrow denotes the magnetic toroidal dipole moment $\bar{\bm T}_{1+}$. 
(c)  Spherical part of the ground state of the total CEF Hamiltonian $|\psi_{0-}^{\rm tot}(\hat{\bm r})|^2$ at $\langle r^2\rangle_{\rm 5d}=0.4408$. 
The blue arrow denotes the magnetic toroidal dipole moment $\bar{\bm T}_{1-}$. 
}
\label{fig:MT}
\end{figure}

For the CEF ground state shown in Fig.~\ref{fig:E_CEF_tot}(a), we have calculated the expectation value  $\bar{t}_{1\mu\eta}\equiv\langle\psi_{0\eta}^{\rm tot}|\tilde{t}_{1\mu}|\psi_{0\eta}^{\rm tot}\rangle$ (see detail for the analytic form in Appendix~\ref{sec:MTapp}).
The results of $\bar{T}_{1\mu{+}}/\mu_{\rm B}=-\langle\psi_{0+}^{\rm tot}|\tilde{t}_{1\mu}|\psi_{0+}^{\rm tot}\rangle$ for $0.4403<\langle r^2\rangle_{\rm 5d}<0.4413$ are plotted in Fig.~\ref{fig:MT}(a). 
We see that $\bar{T}_{1y+}\approx 0$ and $\bar{T}_{1z+}\approx0$ within the numerical accuracy. This indicates that the MT dipole moment is directed perpendicularly to the mirror $(yz)$ plane in the local coordinate [see Fig.~\ref{fig:Tsai_cluster}(f)]. 
As  $\langle r^2\rangle_{\rm 5d}$ increases from 0.4403 to 0.4408, $\bar{T}_{1x+}$ decreses. As  $\langle r^2\rangle_{\rm 5d}$ further increases, $\bar{T}_{1x+}$ increases toward the data for $\langle r^2\rangle_{\rm 5d}=0.4413$.   
These results indicate that when $\varepsilon_{\rm f}$ approaches $\varepsilon_{\rm d}$ mostly, i.e., at $\langle r^2\rangle_{\rm 5d}=0.4408$, the magnitude of the MT dipole moment becomes maximum. 
The 4f-5d hybrid orbital $|\psi_{0+}^{\rm tot}(\hat{\bm r})|^2$ at $\langle r^2\rangle_{\rm 5d}=0.4408$ and the MT dipole moment $\bar{\bm T}_{1+}$ are shown in Fig.~\ref{fig:MT}(b).  

We have also calculated the magnetic dipole moment by the Kramers pair state $\bar{T}_{1\mu{-}}=-\mu_{\rm B}\langle\psi_{0-}^{\rm tot}|\tilde{t}_{1\mu}|\psi_{0-}^{\rm tot}\rangle$ and have confirmed that $\bar{\bm T}_{1-}=-\bar{\bm T}_{1+}$ holds. Namely, $\bar{\bm T}_{1-}$ is directed oppositely to  $\bar{\bm T}_{1+}$, as shown in Fig.~\ref{fig:MT}(c). Hence, the net MT dipole moment in the CEF ground state at each Ce site is zero in the paramagnetic phase because of the cancellation due to the degeneracy of the Kramers pair states. 

\subsection{Resonance of cluster MT and atomic MT dipole in 1/1 AC}

\begin{figure}[tb]
\centering
\includegraphics[width=9cm]{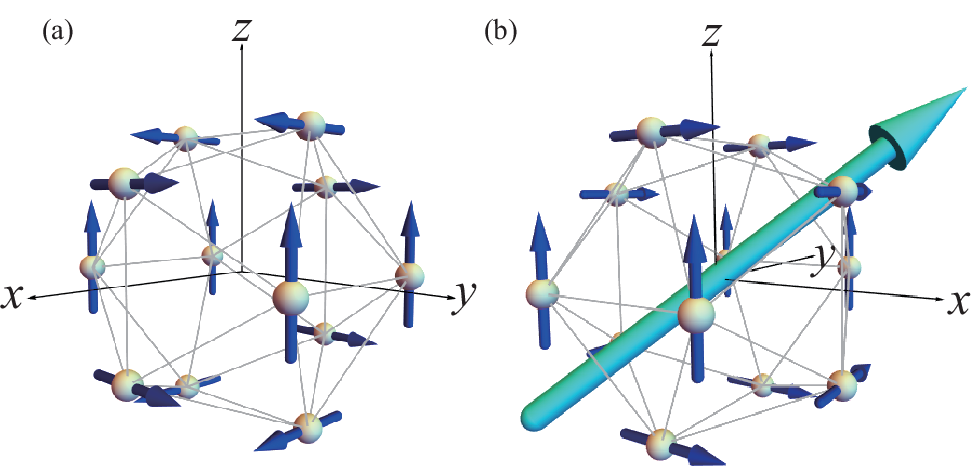}
\caption{(Color online) 
(a) Magnetic toroidal dipole moments for the vortex ordered phase in the 1/1 AC at $\alpha=1.8$ [see Fig.~\ref{fig:PD_AC}(a)]. 
(b) Side view of (d). A light-blue arrow denotes the total magnetic toroidal dipole moments per IC ${\bm t}_{\rm IC}$ [see Eq.~(\ref{eq:MT_op})]. 
}
\label{fig:MT_IC}
\end{figure}

When magnetic ordering occurs, the time-reversal symmetry is broken, the MT dipole moment arises at each Ce site. 
We have discovered that in the vortex order shown in Fig.~\ref{fig:PD_AC}(h), the MT dipole moments at each vertex of the IC appear as shown in Fig.~\ref{fig:MT_IC}(a). Notable is that the MT dipole moments are pointed to all the positive $x$, $y$, and $z$ directions in the global coordinate so that the total MT dipole moment in the IC
\begin{eqnarray}
{\bm t}_{\rm IC}=\sum_{i=1}^{12}\bar{\bm t}_{1i}
\label{eq:MT_op}
\end{eqnarray}
is directed along the 3-fold axis, i.e., the (111) direction ${\bm t}_{\rm IC}\propto (111)$ [see Fig.~\ref{fig:MT_IC}(b)].   

This gives rise to the cooperative phenomena between the different types of the MT moments: 
One is the cluster multipole shown in Eq.~(\ref{eq:MT_cluster}) and the other originates from the hybrid atomic wave function under the odd-parity CEF, both of which are arising from the vortex alignment of the magnetic moments [Fig.~\ref{fig:PD_AC}(f)]. 
This finding provides an interesting possibility of {\it resonance} of ${\bm T}_{\rm IC}$ and ${\bm t}_{\rm IC}$, i.e., ${\bm T}_{\rm IC}\parallel{\bm t}_{\rm IC}$. For instance, by applying positive or negative pressure to the vortex-ordered state in Fig.~\ref{fig:PD_AC}(f) and at the fine-tuned pressure where onsite 4f-5d hybridization becomes prominent, the sharp enhancement of the cross-correlation effect such as electro-magnetic effect due to the MT moments is expected to occur. As noted in Sect.~\ref{sec:PD_AC}, by applying electric field ${\bm E}$ to the vortex order, the magnetic moment ${\bm M}$ is induced  in the perpendicular direction ${\bm M}\perp{\bm E}$. By controlling the parameter of the system to tune so as $\varepsilon_{f}$ and $\varepsilon_{d}$ approach, a sharp enhancement of the electric-field-induced ${\bm M}$ is expected, which is the consequence of the {\it resonance} of ${\bm T}_{\rm IC}$ and ${\bm t}_{\rm IC}$. 
In Sec.~\ref{sec:discussion}, in order to realize such a situation, the effect of applying pressure or chemical substitution to apply negative pressure will be discussed. 

\subsection{Ferromagnetic alignment of MT dipoles in QC cluster}
\label{sec:FM_MT_QC}

\begin{figure}[b]
\centering
\includegraphics[width=8cm]{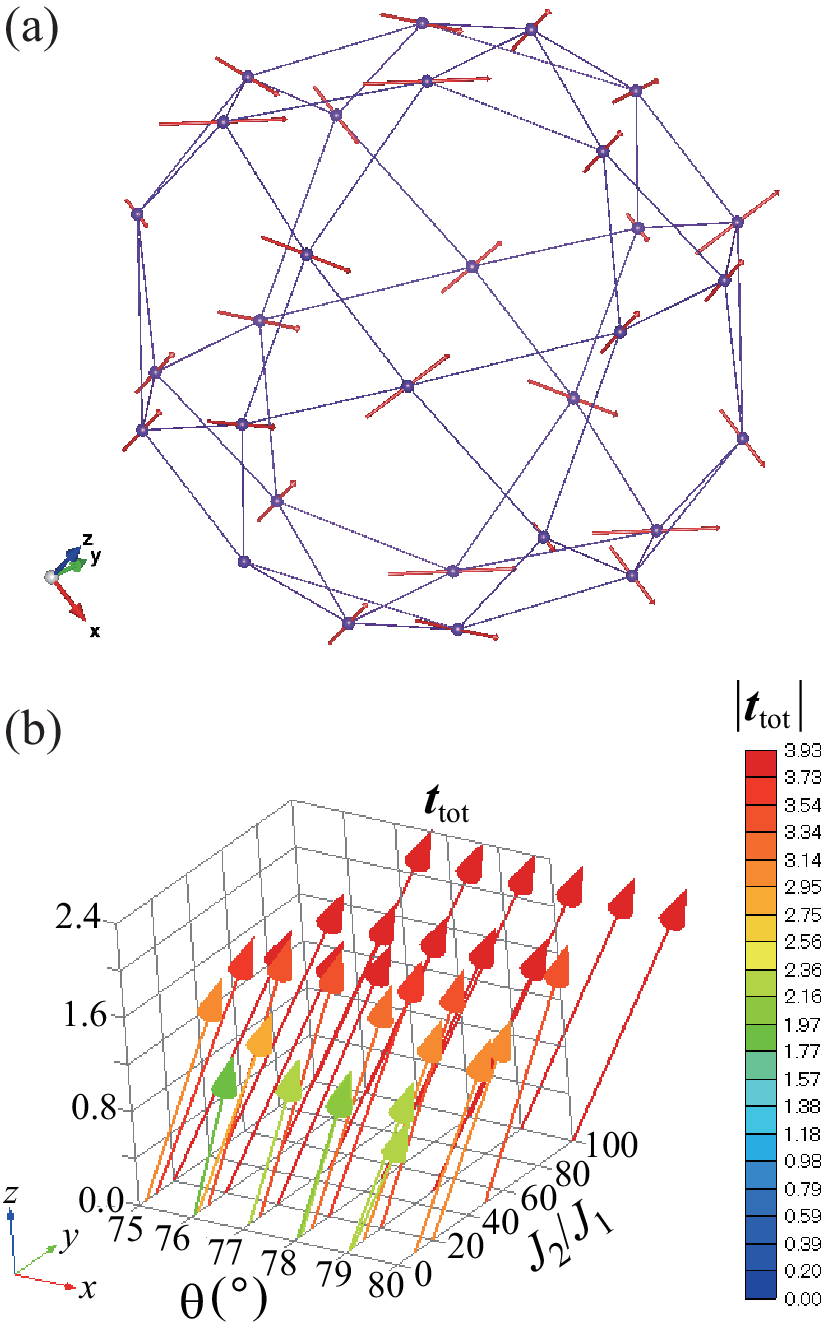}
\caption{(Color online)  
(a)  Distribution of the total magnetic toroidal dipole moment per each IC located at vertices of the icosidodecahedron when the magnetic order shown in Fig.~\ref{fig:PD_QC}(b) occurs. 
(b) Total magnetic toroidal moment per IC ${\bm t}_{\rm tot}$ in the magnetic order denoted by the purple diamonds in (a). 
}
\label{fig:t_tot}
\end{figure}

Interestingly, the positive alignments of the MT dipole moments are also revealed in the QC cluster. In the magnetic order shown in Fig.~\ref{fig:PD_QC}(b), we find that the MT dipole moments are pointed to all the positive directions of  $x$, $y$, and $z$ axes in the global coordinate as shown in Fig.~\ref{fig:PD_QC}(b). 
Here, we plot the total MT dipole moment per IC 
\begin{eqnarray}
{\bm t}_{{\rm IC},j}=\sum_{i=1}^{12}\langle\hat{\bm t}_{j,i}\rangle 
\label{eq:t_IC}
\end{eqnarray}
for the magnetic order in 
Fig.~\ref{fig:t_tot}(a), 
where   
$\hat{\bm t}_{j,i}$ is the unit vector operator of the MT dipole moment perpendicular to the mirror plane in the local coordinate at the $i$th Ce site.
We confirmed that each component of the net MT dipole moment in the IC ${\bm t}_{{\rm IC},j}$ is positive or zero in the magnetic order at the symbols of the purple diamond in the phase diagram of Fig.~\ref{fig:PD_QC}(a). 
At the symbols of the light purple diamonds in  Fig.~\ref{fig:PD_QC}(a), at only one IC, one component of ${\bm t}_{{\rm IC},j}$ has a negative value but in all the other ICs, all ${\bm t}_{\rm IC}$ have the positive (or zero) components. 

We calculate the total MT moment per IC defined by
\begin{eqnarray}
{\bm t}_{\rm tot}=\frac{1}{30}\sum_{j=1}^{30}{\bm t}_{{\rm IC},j}. 
\label{eq:t_tot}
\end{eqnarray}
We plot ${\bm t}_{\rm tot}$ in 
Fig.~\ref{fig:t_tot}(b),  
which is calculated for the magnetic order denoted by the purple diamond in Fig.~\ref{fig:PD_QC}(a). 
The results indicate that the considerable amount of the total MT dipole moment exist in the magnetic ordered phase denoted by purple diamonds in Fig.~\ref{fig:PD_QC}(a). 

This finding offers an interesting possibility that by applying positive or negative pressure to the magnetic order in the QC, the sharp enhancement of the cross-correlation effect due to the MT moments can be caused at the fine-tuned pressure which makes the onsite 4f-5d hybridization prominent. 

\section{Discussion}
\label{sec:discussion}

\subsection{Effect of magnetic anisotropies}

In Sect.~\ref{sec:CEF_4f}, we have revealed that the magnetic anisotropy of the CEF ground state for 4f electron at Ce changes drastically along three different directions depending on the values of the screened valences of ligand ions $\alpha=Z_{\rm SM}/Z_{\rm Au}$ [see Figs.~\ref{fig:E4f_CEF}(b)-(d)]. 
In this subsection, we discuss the relevance of these results to the experiments where spin-glass like behavior was observed in the 1/1 AC Au-SM-Ce~\cite{Imura,Sawano} and Ag-SM-Ce~\cite{Muro}. 

As shown in Figs.~\ref{fig:Tsai_cluster}(a), (b), (e), (f), there exist SM/Au mixed sites (green sites) in the Ce-based 1/1 AC. For instance, in the 1/1 AC Au-Sn-Ce, Sn of Ce atom exists $50\%$ at the mixed site~\cite{Kenzari}.  Although in the present study we treated as the SM/Au mixed sites are all occupied by SM atoms as a first step of analysis, in reality, it is occupied by the SM or Au atom site by site. Then, depending on the atomic configurations of the mixed sites [see Fig.~\ref{fig:Tsai_cluster}(f)], the threshold values of the magnetic anisotropy, i.e., $\alpha_{\rm c1}$ and $\alpha_{\rm c2}$ in Fig.~\ref{fig:PD_AC}(a), are considered to be different at each Ce site. 
Hence, it is likely that the magnetic easy axis points to different directions among the three easy-axis directions site by site. As shown in Figs.~\ref{fig:PD_AC}(b)-(d), the easy axis charges drastically toward three completely different directions. Hence, the effect of three directions of magnetic easy axis and the SM/Au mixed sites can be the origin of the spin glass behavior observed in the Ce-based 1/1 ACs.

\subsection{Tuning 4f and 5d energies so as to approach and valence crossover/transition}

In Sect.~\ref{sec:MT}, {\it resonance} between the atomic MT dipole moment and cluster MT moment in the vortex order in the 1/1 AC was discussed. Here we point out that such a situation can occur when the Ce-valence crossover (transition) temperature exists in the vortex order in the temperature $T$-pressure $P$ phase diagram. 

\begin{figure}[tb]
\centering
\includegraphics[width=7cm]{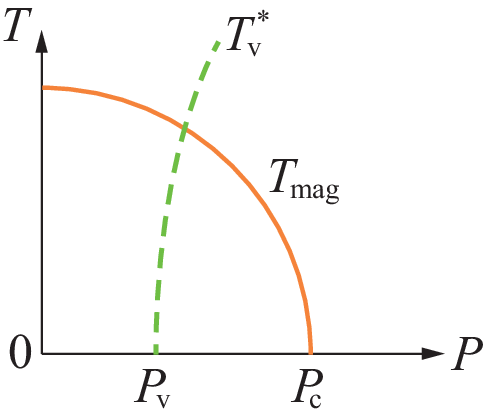}
\caption{(Color online) 
Temperature-pressure phase diagram with the magnetic transition temperature (solid line) and the valence crossover (transition) temperature (dashed line). $P_{\rm c}$ is the magnetic transition and $P_{\rm v}$ is the valence crossover or the valence transition at $T=0$. 
}
\label{fig:TP_PD}
\end{figure}

On the basis of the extended Anderson lattice model taking into account the Coulomb repulsions between f and conduction electrons as well as between f electrons, the $T$-$P$ phase diagram was discussed theoretically~\cite{WMbook}.  In the case of small hybridization between the f and conduction electrons, the $T$-$P$ phase diagram where the temperature of valence-crossover or valence-transition $T_{\rm v}$ crosses the magnetically ordered temperature $T_{\rm mag}$ as shown in Fig.~\ref{fig:TP_PD}. 

The charge transfer effect under the odd-parity CEF was theoretically studied on the basis of the extended Anderson lattice model for $\beta$-YbAlB$_4$~\cite{WM2019}. It is noted that similarly to the QC and AC shown in Fig.~\ref{fig:Tsai_cluster}(f), the local inversion symmetry is broken at the Yb site which seven B rings surround. It was theoretically shown that at the quantum critical point of the valence transition, the valence fluctuation, i.e., charge transfer fluctuation between the 4f-5d states at the Yb site diverges. Interestingly, electric dipole as well as MT dipole fluctuation is also enhanced~\cite{WM2019}. This result is the consequence of the many-body effect driven by the onsite Coulomb repulsion between the 4f and 5d states at the Yb site, which is considered to correspond to the situation where $\varepsilon_{\rm f}$ and $\varepsilon_{\rm d}$ approaches closely in the present study. 

In Sect. 5.3, it was discussed that the lattice-constant dependences of the 4f level and 5d level are different from both viewpoints of the point charge and hybridization. This argument is not restricted to the Ce-based systems but is also applied to the other rare-earth-based systems. For instance, in the Yb-based systems, Eq. (\ref{eq:Ef}) is replaced with 
\begin{eqnarray}
\varepsilon_{\rm f}\propto B_2^0\langle O_2^0\rangle+B_4^0\langle O_4^0\rangle+B_6^0\langle O_6^0\rangle, 
\label{eq:Ef_4f_Yb}
\end{eqnarray}
where $O_{\ell}^0$ is the Stevens operators for $J=7/2$ and $B_{\ell}^0$ is the coefficient with $B_{\ell}^m\propto R_i^{-\ell-1}$. As for the 5d level, Eq.~(\ref{eq:Ed}) is applied to the Yb-based systems. Hence, within the point charge model, the lattice constant dependences of $\varepsilon_{\rm f}$ and $\varepsilon_{\rm d}$ are different. 
  On the other hand, within the hybridization picture, Eq.~(\ref{eq:Ef2}) and Eq.~(\ref{eq:Ed2}) are applied to the 4f and 5d level in the hole picture, which also indicates the different lattice-constant dependence. 
  Therefore, applying positive or negative pressure to the Yb-based systems, the 4f level and 5d level can be tuned to approach in principle.

Hence, at $T_{\rm v}$ in the phase diagram shown in Fig.~\ref{fig:TP_PD}, {\it resonance} of the atomic MT dipole and the cluster MT moments is expected to occur. 

The valence transition or valence crossover is ubiquitous in rare-earth based systems as not only periodic crystals such as heavy-fermion systems and ACs but also QCs~\cite{WMbook}.  The $T$-$P$ phase diagram shown in Fig.~\ref{fig:TP_PD} was reported in CeRh$_{0.5}$Ir$_{0.5}$In$_5$~\cite{Kawasaki2020}. It was experimentally revealed that in the quantum critical QC Au$_{51}$Al$_{34}$Yb$_{15}$, the quantum critical point (QCP) of the valence transition is realized~\cite{Imura2020}. Moreover, in the 1/1 AC Au$_{51}$Al$_{35}$Yb$_{14}$ the quantum critical point of the valence transition was identified around $P\approx 2$~GPa~\cite{Imura2020}.   
In the sister compounds of the 1/1 AC Zn-Au-Yb, the first-order valence transition takes place at $P_{\rm v}\sim 4.5$~GPa and in the 2/1 AC Zn-Au-Yb, the valence anomaly was detected at $P_{\rm v}\sim 5.5$~GPa~\cite{Imura2023}. These observations suggest that it seems promising that the $T$-$P$ phase diagram shown in Fig.~\ref{fig:TP_PD} is realized in the rare-earth-based QCs and ACs. 

In the Ce-dilute system Ag-In-(Ce$_x$Y$_{1-x}$), it was observed that the magnetic susceptibility is expressed by a single scaling function of the ratio of temperature $T$ and magnetic field $B$, whose behavior is referred to as the $T/B$ scaling~\cite{Sawano}. It is noted that the $T/B$ scaling behavior was theoretically shown to arise from critical valence fluctuations~\cite{WM2016}. Actually, in the QC Au-Al-Yb and its 1/1 AC at $P\approx 2$~GPa, where the QCP of the valence transition exists, the $T/B$ scaling behavior was observed~\cite{NKSato2022,Matsukawa2016}. As discussed in Ref.~\cite{Sawano}, the observation of the $T/B$ scaling behavior in the Ce-dilute system suggests underlying proximity to the valence QCP in the Ce-based ACs.

Hence, applying positive or negative pressure to the vortex order [see Fig.~\ref{fig:PD_AC}(g)] is interesting to detect the {\it resonance}  between the atomic MT dipole moment and cluster MT moment giving rise to a sharp enhancement of the electro-magnetic effect at the valence crossover (transition) temperature $T_{\rm v}$ inside the magnetically ordered phase. 

\section{Summary}

We have developed general formulation of the CEF in rare-earth-based QCs and ACs not only for even parity but also for odd parity on the basis of the point charge model. 
By analyzing the CEF in the QC Au-SM-Ce and AC, we have discovered that a drastic change of three magnetic anisotropies emerges depending on the value of the ratio of the screened ligand ions $Z_{\rm SM}/Z_{\rm Au}$, which is sharp contrast to those in the Yb-, Tb-, and Dy-based systems. 

Then, we have constructed the magnetic model in the 1/1 AC taking into account the uniaxial magnetic anisotropy arising from the CEF.  By performing the numerical calculation, 
we have determined the ground-state phase diagram for the AF interactions. The result reveals that various noncollinear magnetic states are stabilized depending on the ratio of the N.N. and N.N.N. interactions and $Z_{\rm SM}/Z_{\rm Au}$. 
In the vortex order, the magnetic moments at the twelve vertices of each icosahedron swirl, which gives rise to the ferroic alignment of the cluster MT moment ${\bm T}_{\rm IC}$ along the (111) direction. Then, the electro-magnetic effect where the magnetic moment is induced in the perpendicular direction to the applied electric field. The whirling-anti-whirling order is characterized by the topological charge $n=+3$ and $-3$. 
 
By diagonalizing the CEF Hamiltonian for the 4f and 5d states at Ce, we have obtained the CEF energies. The eigenstates for the lowest and highest energies are attributed to the 5d state for $J=3/2$ and the middle three states are attributed to the 4f state for $J=5/2$. 

We have pointed out that applying positive or negative pressure to the 1/1 AC can make the 4f level and 5d level approach, which gives rise to the onsite 4f-5d hybridization. 
This causes the electric dipole at each Ce site, whose amount of 12 Ce sites per IC is cancelled in total. 
In the vortex order, the hybrid 4f-5d wave function gives rise to the MT dipole at each Ce site, whose amount of 12 Ce sites per IC ${\bm t}_{\rm IC}$ is pointed to the (111) direction. This gives rise to {\it resonance} between the cluster MT moment and the atomic MT dipole moment ${\bm T}_{\rm IC}\parallel{\bm t}_{\rm IC}$. 
The {\it resonance} is expected to occur at the valence crossover (transition) temperature $T_{\rm v}$ in the vortex order, where a sharp enhancement of the electro-magnetic effect emerges. 

By applying the 2-IC cluster mean-field method to the QC cluster, we have explored the ground state of the magnetic model with uniaxial anisotropy arising from the CEF for the AF N.N. and N.N.N. interactions.  Then, we have found the noncollinear magnetic states, which are characterized by the ferroic alignment of the MT dipole moment at each IC. This result can be interpreted as a mechanism for the amount of atomic MT dipole moment at each Ce per IC to be aligned ferroiclly works in the frustrated magnetic QC.  By applying positive or negative pressure to make the valence crossover (transition) in the magnetically ordered phase in the QC, the electro-magnetic effect due to the ferroic alignment of the MT dipole moments is expected to emerge. 

Recently, RKKY interactions between multipoles as well as magnetic moments have been derived from ab initio calculations for the Ce-based compounds~\cite{Yamada2024,Itokazu2025}. Microscopic derivation of the multipole interactions in the Ce-based AC and QC is interesting future subject.

We have formulated the odd-parity CEF in a general manner. 
Hence, our formulation is applicable to the rare-earth-based icosahedral QCs and ACs other than the Ce-based systems.  
As revealed recently in the analyses of the even-parity CEF in the QC Au-SM-R (R=Yb~\cite{WK2021}, Tb~\cite{WPNAS} and Dy~\cite{WI2023}) and the ACs, the tendency of emergence of the magnetic anisotropy is different depending on the rare-earth element. Hence, it is interesting to analyze the properties of the CEF and the multipoles not only for even-parity but also for odd-parity in the other rare-earth-based QCs and ACs, which are left for future studies. 

We have performed the density-functional calculation of $\langle r^n\rangle$ $(n=1,2,\dots,6)$ for radial parts of the $4f$ and $5d$ states in an isolated $R^{3+}$ ion as well as $R^{2+}$ ion ($R$=Ce, Pr, Nd, Pm, Sm, Eu, Gd, Tb, Dy, Ho, Er, Tm, and Yb). These results are useful to calculate the odd parity as well as even parity CEF not only in QCs and ACs but also in conventional periodic crystals composed of rare-earth element.

\begin{acknowledgment}
The authors thank K. Momma for expanding the functionality of VESTA in order to draw the magnetic moments and the atomic structures in the QC~\cite{Momma}. 
The author (S.W.) is grateful to H. Kusunose for valuable comments about the magnetic space group as well as cross-correlation effect of multipoles. 
This work was supported by JSPS KAKENHI Grant Numbers JP22H01170, JP23K17672, JP24H01675, JP19H05819, JP19H05818, and JP24K08041.
\end{acknowledgment}

\section{DFT calculation of the energy and $\langle r^n\rangle$ for rare-earth ions}
\label{sec:DFT}
In Sect.~\ref{sec:CEF_4f} and Sect.~\ref{sec:op_CEF}, we have employed the expectation values by the radial part of the $4f$ and $5d$ wave functions $\langle r^n\rangle$ for the $J=5/2$ and $J=3/2$ states respectively in Ce$^{3+}$, which appear in the coefficients of the CEF Hamiltonian matrix elements [see Eq.~(\ref{eq:B_def}) and Eq.~(\ref{eq:72_32_b})]. We have also calculated $\langle r^n\rangle$ $(n=1, 2,\dots,6)$ for the $4f$ state and the $5d$ state in an isolated $R^{3+}$ ion as well as $R^{2+}$ ion ($R$=Ce, Pr, Nd, Pm, Sm, Eu, Gd, Tb, Dy, Ho, Er, Tm, and Yb). 

We modified the atomic simulation code in the open-source \textit{ab initio} program package {\sc Quantum ESPRESSO}~\cite{Giannozzi} to compute the $n$th moment
\begin{align}
\langle r^n \rangle \equiv \int d^3r r^n |\varphi_{4f/5d,J}(\textbf{r})|^2,
\end{align}
where $\varphi_{4f/5d,J}(\textbf{r})$ is the 4$f$ or 5$d$ orbitals from the atomic Kohn--Sham--Dirac equation and $J$ is the total angular momentum.
The semi-empirical, three-parameter Becke--Lee--Yang--Parr (B3LYP) hybrid functional~\cite{Becke} was used to approximate the exchange-correlation functional. 
We set the occupations of the 5$d$ and 6$s$ orbitals to zero.
This occupation configuration is consistent with the resulting difference between the 4$f$ and 5$d$ Kohn--Sham--Dirac eigenvalue $\varepsilon$ (Table~\ref{tb:R4f} and Table~\ref{tb:R5d}, excepting Ce$^2+$).
In an isolated Ce$^{2+}$ ion, the valence electrons occupy 5$d$ orbitals first, while the 4$f$ is the valence orbital in other systems.

Table~\ref{tb:R4f} shows the energy and $\langle r^n\rangle$ for the $4f$ state with the total angular momentum $J=5/2$ and $7/2$ of $R^{2+}$ (upper panel) and $R^{3+}$ (lower panel). The units are eV and \AA, respectively. 


\begin{table}[!h] 
\begin{center}
\resizebox{\textwidth}{!}{%
\begin{tabular}{ccccccccccccccc}
\hline
 & \multicolumn{2}{c}{$\varepsilon$ (eV)} & \multicolumn{2}{c}{$\langle r\rangle~(\mathrm{\AA})$}
 & \multicolumn{2}{c}{$\langle r^2\rangle~(\mathrm{\AA}^2)$} & 
\multicolumn{2}{c}{$\langle r^3\rangle~(\mathrm{\AA}^3)$} & \multicolumn{2}{c}{$\langle r^4\rangle~(\mathrm{\AA}^4)$}
 & \multicolumn{2}{c}{$\langle r^5\rangle~(\mathrm{\AA}^5)$} & 
\multicolumn{2}{c}{$\langle r^6\rangle~(\mathrm{\AA}^6)$} \\
$J$ & 5/2 & 7/2 & 5/2 & 7/2 & 5/2 & 7/2 & 5/2 & 7/2 & 5/2 & 7/2 & 5/2 & 7/2 & 5/2 & 7/2 \\
\hline
2+ &  &  &  &  &  &  &  &  &  &  &  &  &  &  \\
Ce & -13.02 & -12.73 & 0.646 & 0.655 & 0.572 & 0.591 & 0.677 & 0.712 & 1.023 & 1.098 & 1.894 & 2.071 & 4.139 & 4.605 \\
Pr & -13.91 & -13.56 & 0.605 & 0.614 & 0.500 & 0.517 & 0.552 & 0.582 & 0.781 & 0.841 & 1.360 & 1.495 & 2.808 & 3.149 \\
Nd & -14.64 & -14.24 & 0.573 & 0.582 & 0.447 & 0.463 & 0.466 & 0.494 & 0.626 & 0.678 & 1.039 & 1.150 & 2.052 & 2.322 \\
Pm & -15.26 & -14.79 & 0.547 & 0.556 & 0.407 & 0.422 & 0.404 & 0.429 & 0.520 & 0.566 & 0.828 & 0.925 & 1.577 & 1.804 \\
Sm & -15.78 & -15.24 & 0.524 & 0.533 & 0.374 & 0.389 & 0.357 & 0.381 & 0.442 & 0.484 & 0.681 & 0.767 & 1.259 & 1.456 \\
Eu & -16.28 & -15.67 & 0.504 & 0.513 & 0.345 & 0.360 & 0.317 & 0.340 & 0.379 & 0.419 & 0.567 & 0.645 & 1.019 & 1.193 \\
Gd & -16.72 & -16.02 & 0.486 & 0.495 & 0.321 & 0.335 & 0.285 & 0.307 & 0.331 & 0.368 & 0.481 & 0.553 & 0.844 & 1.002 \\
Tb & -17.10 & -16.31 & 0.470 & 0.479 & 0.300 & 0.315 & 0.259 & 0.280 & 0.292 & 0.327 & 0.415 & 0.482 & 0.712 & 0.858 \\
Dy & -17.43 & -16.55 & 0.456 & 0.465 & 0.282 & 0.297 & 0.236 & 0.257 & 0.260 & 0.294 & 0.362 & 0.426 & 0.610 & 0.747 \\
Ho & -17.71 & -16.74 & 0.442 & 0.452 & 0.266 & 0.281 & 0.217 & 0.238 & 0.234 & 0.267 & 0.319 & 0.381 & 0.530 & 0.660 \\
Er & -17.96 & -16.88 & 0.430 & 0.440 & 0.252 & 0.267 & 0.201 & 0.221 & 0.212 & 0.245 & 0.284 & 0.344 & 0.465 & 0.590 \\
Tm & -18.18 & -16.98 & 0.419 & 0.429 & 0.239 & 0.254 & 0.186 & 0.207 & 0.193 & 0.225 & 0.255 & 0.314 & 0.412 & 0.534 \\
Yb & -18.36 & -17.04 & 0.407 & 0.419 & 0.227 & 0.243 & 0.174 & 0.194 & 0.177 & 0.209 & 0.231 & 0.289 & 0.368 & 0.488 \\
3+ &  &  &  &  &  &  &  &  &  &  &  &  &  &  \\
Ce & -28.74 & -28.40 & 0.559 & 0.564 & 0.402 & 0.409 & 0.362 & 0.372 & 0.399 & 0.414 & 0.522 & 0.548 & 0.794 & 0.840 \\
Pr & -29.88 & -29.48 & 0.534 & 0.539 & 0.367 & 0.374 & 0.317 & 0.327 & 0.335 & 0.349 & 0.423 & 0.446 & 0.622 & 0.661 \\
Nd & -30.87 & -30.41 & 0.513 & 0.518 & 0.339 & 0.346 & 0.282 & 0.291 & 0.288 & 0.301 & 0.352 & 0.372 & 0.502 & 0.537 \\
Pm & -31.74 & -31.21 & 0.494 & 0.500 & 0.315 & 0.322 & 0.253 & 0.262 & 0.251 & 0.263 & 0.298 & 0.317 & 0.415 & 0.446 \\
Sm & -32.51 & -31.91 & 0.478 & 0.483 & 0.295 & 0.302 & 0.230 & 0.239 & 0.221 & 0.233 & 0.257 & 0.274 & 0.349 & 0.378 \\
Eu & -33.19 & -32.51 & 0.463 & 0.468 & 0.277 & 0.284 & 0.210 & 0.219 & 0.197 & 0.209 & 0.224 & 0.241 & 0.299 & 0.326 \\
Gd & -33.85 & -33.09 & 0.449 & 0.455 & 0.261 & 0.268 & 0.193 & 0.201 & 0.177 & 0.188 & 0.196 & 0.212 & 0.257 & 0.283 \\
Tb & -34.44 & -33.59 & 0.436 & 0.442 & 0.246 & 0.254 & 0.178 & 0.186 & 0.159 & 0.170 & 0.174 & 0.189 & 0.223 & 0.248 \\
Dy & -34.98 & -34.03 & 0.424 & 0.431 & 0.234 & 0.241 & 0.164 & 0.173 & 0.145 & 0.155 & 0.155 & 0.170 & 0.196 & 0.219 \\
Ho & -35.46 & -34.41 & 0.413 & 0.420 & 0.222 & 0.230 & 0.153 & 0.162 & 0.132 & 0.143 & 0.139 & 0.154 & 0.173 & 0.196 \\
Er & -35.90 & -34.74 & 0.403 & 0.410 & 0.211 & 0.219 & 0.143 & 0.151 & 0.121 & 0.132 & 0.125 & 0.140 & 0.155 & 0.176 \\
Tm & -36.29 & -35.02 & 0.393 & 0.400 & 0.202 & 0.210 & 0.134 & 0.142 & 0.111 & 0.122 & 0.114 & 0.128 & 0.139 & 0.160 \\
Yb & -36.64 & -35.25 & 0.384 & 0.392 & 0.193 & 0.201 & 0.125 & 0.134 & 0.103 & 0.113 & 0.104 & 0.118 & 0.125 & 0.146 \\
\hline
\end{tabular}
}
\end{center}
\caption{Kohn--Sham--Dirac eigenvalue $\varepsilon$ and $n$th moment $\langle r^n \rangle$ for 4$f$ orbitals in lanthanides 2+ and 3+ ions.
The units are eV and \AA.
}
\label{tb:R4f}
\end{table}

Table~\ref{tb:R5d} shows the energy and $\langle r^n\rangle$ for the $5d$ state with the total angular momentum $J=3/2$ and $5/2$ of $R^{2+}$ (upper panel) and $R^{3+}$ (lower panel). The units are eV and \AA, respectively. 

\begin{table}[!h] 
\begin{center}
\resizebox{\textwidth}{!}{%
\begin{tabular}{ccccccccccccccc}
\hline
 & \multicolumn{2}{c}{$\varepsilon$ (eV)} & \multicolumn{2}{c}{$\langle r\rangle~(\mathrm{\AA})$}
 & \multicolumn{2}{c}{$\langle r^2\rangle~(\mathrm{\AA}^2)$} & 
\multicolumn{2}{c}{$\langle r^3\rangle~(\mathrm{\AA}^3)$} & \multicolumn{2}{c}{$\langle r^4\rangle~(\mathrm{\AA}^4)$}
 & \multicolumn{2}{c}{$\langle r^5\rangle~(\mathrm{\AA}^5)$} & 
\multicolumn{2}{c}{$\langle r^6\rangle~(\mathrm{\AA}^6)$} \\
$J$ & 3/2 & 5/2 & 3/2 & 5/2 & 3/2 & 5/2 & 3/2 & 5/2 & 3/2 & 5/2 & 3/2 & 5/2 & 3/2 & 5/2 \\
\hline
2+ &  &  &  &  &  &  &  &  &  &  &  &  &  &  \\
Ce & -13.57 & -13.36 & 1.426 & 1.447 & 2.355 & 2.423 & 4.412 & 4.600 & 9.296 & 9.817 & 21.84 & 23.35 & 56.76 & 61.40 \\
Pr & -13.69 & -13.46 & 1.403 & 1.425 & 2.282 & 2.352 & 4.223 & 4.414 & 8.799 & 9.324 & 20.47 & 21.98 & 52.74 & 57.32 \\
Nd & -13.77 & -13.53 & 1.383 & 1.406 & 2.223 & 2.295 & 4.071 & 4.267 & 8.410 & 8.941 & 19.42 & 20.93 & 49.70 & 54.27 \\
Pm & -13.83 & -13.57 & 1.367 & 1.390 & 2.174 & 2.248 & 3.947 & 4.147 & 8.096 & 8.636 & 18.59 & 20.11 & 47.32 & 51.92 \\
Sm & -13.86 & -13.59 & 1.352 & 1.377 & 2.131 & 2.209 & 3.842 & 4.048 & 7.838 & 8.389 & 17.91 & 19.46 & 45.43 & 50.08 \\
Eu & -13.88 & -13.60 & 1.339 & 1.365 & 2.093 & 2.173 & 3.750 & 3.962 & 7.612 & 8.176 & 17.33 & 18.91 & 43.81 & 48.53 \\
Gd & -13.89 & -13.60 & 1.327 & 1.354 & 2.060 & 2.143 & 3.671 & 3.889 & 7.423 & 8.001 & 16.84 & 18.46 & 42.48 & 47.30 \\
Tb & -13.89 & -13.59 & 1.316 & 1.345 & 2.031 & 2.117 & 3.604 & 3.828 & 7.262 & 7.856 & 16.44 & 18.10 & 41.40 & 46.33 \\
Dy & -13.89 & -13.57 & 1.307 & 1.336 & 2.006 & 2.095 & 3.545 & 3.777 & 7.126 & 7.738 & 16.11 & 17.81 & 40.51 & 45.57 \\
Ho & -13.87 & -13.54 & 1.299 & 1.329 & 1.983 & 2.076 & 3.494 & 3.734 & 7.011 & 7.642 & 15.83 & 17.58 & 39.80 & 44.99 \\
Er & -13.85 & -13.51 & 1.291 & 1.322 & 1.964 & 2.059 & 3.451 & 3.699 & 6.915 & 7.565 & 15.60 & 17.41 & 39.22 & 44.57 \\
Tm & -13.82 & -13.47 & 1.284 & 1.317 & 1.946 & 2.045 & 3.414 & 3.670 & 6.833 & 7.504 & 15.42 & 17.28 & 38.76 & 44.28 \\
Yb & -13.79 & -13.43 & 1.277 & 1.312 & 1.931 & 2.034 & 3.382 & 3.647 & 6.766 & 7.460 & 15.27 & 17.19 & 38.40 & 44.12 \\
3+ &  &  &  &  &  &  &  &  &  &  &  &  &  &  \\
Ce & -23.86 & -23.55 & 1.245 & 1.260 & 1.767 & 1.810 & 2.800 & 2.901 & 4.914 & 5.146 & 9.492 & 10.04 & 20.06 & 21.44 \\
Pr & -24.13 & -23.80 & 1.224 & 1.240 & 1.710 & 1.754 & 2.672 & 2.774 & 4.630 & 4.862 & 8.841 & 9.390 & 18.49 & 19.83 \\
Nd & -24.36 & -24.01 & 1.205 & 1.222 & 1.661 & 1.706 & 2.562 & 2.666 & 4.391 & 4.623 & 8.300 & 8.839 & 17.20 & 18.52 \\
Pm & -24.56 & -24.19 & 1.188 & 1.206 & 1.617 & 1.663 & 2.467 & 2.572 & 4.185 & 4.418 & 7.841 & 8.380 & 16.12 & 17.42 \\
Sm & -24.73 & -24.34 & 1.173 & 1.191 & 1.578 & 1.626 & 2.383 & 2.489 & 4.006 & 4.241 & 7.446 & 7.983 & 15.20 & 16.49 \\
Eu & -24.88 & -24.47 & 1.159 & 1.178 & 1.542 & 1.591 & 2.308 & 2.416 & 3.849 & 4.086 & 7.103 & 7.641 & 14.40 & 15.69 \\
Gd & -25.01 & -24.58 & 1.146 & 1.166 & 1.510 & 1.560 & 2.239 & 2.349 & 3.706 & 3.946 & 6.794 & 7.334 & 13.70 & 14.98 \\
Tb & -25.13 & -24.68 & 1.134 & 1.154 & 1.480 & 1.531 & 2.177 & 2.289 & 3.578 & 3.821 & 6.521 & 7.065 & 13.08 & 14.36 \\
Dy & -25.24 & -24.77 & 1.123 & 1.144 & 1.452 & 1.505 & 2.121 & 2.236 & 3.464 & 3.710 & 6.279 & 6.826 & 12.53 & 13.82 \\
Ho & -25.33 & -24.84 & 1.112 & 1.134 & 1.427 & 1.482 & 2.069 & 2.187 & 3.361 & 3.611 & 6.062 & 6.615 & 12.04 & 13.34 \\
Er & -25.41 & -24.90 & 1.102 & 1.125 & 1.403 & 1.460 & 2.022 & 2.142 & 3.267 & 3.521 & 5.867 & 6.426 & 11.61 & 12.92 \\
Tm & -25.48 & -24.95 & 1.093 & 1.117 & 1.382 & 1.440 & 1.979 & 2.102 & 3.182 & 3.440 & 5.691 & 6.257 & 11.22 & 12.54 \\
Yb & -25.54 & -24.99 & 1.085 & 1.109 & 1.361 & 1.421 & 1.940 & 2.065 & 3.105 & 3.368 & 5.532 & 6.106 & 10.87 & 12.21 \\
\hline
\end{tabular}
}
\end{center}
\caption{Kohn--Sham--Dirac eigenvalue $\varepsilon$ and $n$th moment $\langle r^n \rangle$ for 5$d$ orbitals in lanthanides 2+ and 3+ ions.
The units are eV and \AA.
}
\label{tb:R5d}
\end{table}


In this study, we have employed the results of Ce$^{3+}$ to analyze the CEF in the QC Au-SM-Ce and AC on the basis of the point-charge model. Our formulation of the CEF theory not only for the even parity but also for the odd parity can be generally applicable to the other rare-earth-based QCs and ACs. In such calculations, the results listed in Talbe~\ref{tb:R4f} and Table~\ref{tb:R5d} are available. 

The results shown in Talbe~\ref{tb:R4f} and Table~\ref{tb:R5d} are also useful to analyze the CEF in the rare-earth-based materials with conventional periodic-crystal structures. So far, in the calculation of the CEF for the even parity in the rare-earth-based compounds, the results of the Dirac-Fock calculation~\cite{Freeman} were often employed to evaluate $\langle r^n\rangle$ $(n=2, 4, 6)$. However, the expectation values of the radial part with odd powers $\langle r^n\rangle$ $(n=1,3,5)$ have not been reported to date, to our best knowledge. Our results presented in Table~\ref{tb:R4f} and Table~\ref{tb:R5d} provide $\langle r^n\rangle$ with $n$ being odd as well as even evaluated on an equal footing basis, which enable us to perform the full CEF calculation microscopically.

\section{Wave functions of 4f and 5d electrons at Ce}
\label{sec:psi_5d}

In Sec.~\ref{sec:op_CEF}, the odd-parity CEF is formulated on the basis of the point charge model. To obtain the off-diagonal matrix element of $H_{\rm CEF}$, the spherical parts of the $4f$ wavefunction and $5d$ wavefunction are used. In this Appendix, the explicit forms of the wavefunctions are given as follows.


The spherical part of the $4f$ wave function for $J=5/2$ at Ce $\psi^{4f}_{J,J_z}(\hat{\bm r})$, where  $\hat{\bm r}$ is the unit radial vector, is given by 
\begin{eqnarray}
\psi^{4f}_{\frac{5}{2},\frac{5}{2}}(\hat{\bm r})&=&-\sqrt{\frac{1}{7}}Y_3^2(\hat{\bm r})\psi_{\uparrow}+\sqrt{\frac{6}{7}}Y_3^3(\hat{\bm r})\psi_{\downarrow}, 
\label{eq:psi_4f_1}
\\
\psi^{4f}_{\frac{5}{2},\frac{3}{2}}(\hat{\bm r})&=&-\sqrt{\frac{2}{7}}Y_3^1(\hat{\bm r})\psi_{\uparrow}+\sqrt{\frac{5}{7}}Y_3^2(\hat{\bm r})\psi_{\downarrow}, 
\label{eq:psi_4f_2}
\\
\psi^{4f}_{\frac{5}{2},\frac{1}{2}}(\hat{\bm r})&=&-\sqrt{\frac{3}{7}}Y_3^0(\hat{\bm r})\psi_{\uparrow}+\sqrt{\frac{4}{7}}Y_3^1(\hat{\bm r})\psi_{\downarrow}, 
\label{eq:psi_4f_3}
\\
\psi^{4f}_{\frac{5}{2},-\frac{1}{2}}(\hat{\bm r})&=&-\sqrt{\frac{4}{7}}Y_3^{-1}(\hat{\bm r})\psi_{\uparrow}+\sqrt{\frac{3}{7}}Y_3^{-1}(\hat{\bm r})\psi_{\downarrow}, 
\label{eq:psi_4f_4}
\\
\psi^{4f}_{\frac{5}{2},-\frac{3}{2}}(\hat{\bm r})&=&-\sqrt{\frac{5}{7}}Y_3^{-2}(\hat{\bm r})\psi_{\uparrow}+\sqrt{\frac{2}{7}}Y_3^{-2}(\hat{\bm r})\psi_{\downarrow}, 
\label{eq:psi_4f_5}
\\
\psi^{4f}_{\frac{5}{2},-\frac{5}{2}}(\hat{\bm r})&=&-\sqrt{\frac{6}{7}}Y_3^{-3}(\hat{\bm r})\psi_{\uparrow}+\sqrt{\frac{1}{7}}Y_3^{-3}(\hat{\bm r})\psi_{\downarrow}, 
\label{eq:psi_4f_6}
\end{eqnarray}
where $Y_{\ell}^m(\hat{\bm r})$ is the spherical harmonics and $\psi_{\uparrow}$ and $\psi_{\downarrow}$ are the wavefunctions of $\uparrow$ spin and $\downarrow$ spin, respectively. 

The spherical part of the $5d$ wave function for $J=3/2$ at Ce $\psi^{5d}_{J,J_z}(\hat{\bm r})$ is given by 
\begin{eqnarray}
\psi^{5d}_{\frac{3}{2},\frac{3}{2}}(\hat{\bm r})&=&-\sqrt{\frac{1}{5}}Y_2^1(\hat{\bm r})\psi_{\uparrow}+\sqrt{\frac{4}{5}}Y_2^2(\hat{\bm r})\psi_{\downarrow}, 
\label{eq:psi_5d_1}
\\
\psi^{5d}_{\frac{3}{2},\frac{1}{2}}(\hat{\bm r})&=&-\sqrt{\frac{2}{5}}Y_2^0(\hat{\bm r})\psi_{\uparrow}+\sqrt{\frac{3}{5}}Y_2^1(\hat{\bm r})\psi_{\downarrow}, 
\label{eq:psi_5d_2}
\\
\psi^{5d}_{\frac{3}{2},-\frac{1}{2}}(\hat{\bm r})&=&-\sqrt{\frac{3}{5}}Y_2^{-1}(\hat{\bm r})\psi_{\uparrow}+\sqrt{\frac{2}{5}}Y_2^0(\hat{\bm r})\psi_{\downarrow}, 
\label{eq:psi_5d_3}
\\
\psi^{5d}_{\frac{3}{2},-\frac{3}{2}}(\hat{\bm r})&=&-\sqrt{\frac{4}{5}}Y_2^{-2}(\hat{\bm r})\psi_{\uparrow}+\sqrt{\frac{1}{5}}Y_2^{-1}(\hat{\bm r})\psi_{\downarrow}. 
\label{eq:psi_5d_4}
\label{eq:psi_5d}
\end{eqnarray}
%

\section{Matrix elements of the odd-parity CEF Hamiltonian}
\label{sec:H_CEF_odd}

In Sect.~\ref {sec:op_CEF}, the odd-parity CEF is formulated on the basis of the point-charge model. In this Appendix, the matrix elements of the odd-parity CEF Hamiltonian  Eq.~(\ref{eq:HCEF4f5d}) are explicitly given. 
Hereafter, the $4f$ state $\Ket{\frac{5}{2},J_z}_{4f}$ is denoted as $\Ket{\frac{5}{2},J_z}$ and the $5d$ state $\Ket{\frac{3}{2},J_z'}_{5d}$ is denoted as $\Ket{\frac{3}{2},J_z'}$.
\begin{eqnarray}
&&\Bra{\frac{5}{2},\frac{5}{2}}\mathcal{H}_{\rm CEF}\Ket{\frac{3}{2},\frac{3}{2}}\nonumber\\
&=&2\sqrt{\frac{\pi}{105}}\braket{r^1}(h_{1}^{-1}+ih_{1}^{-1}(s))c^1(3,2,2,1)+4\sqrt{\frac{2\pi}{35}}\braket{r^1}(h_{1}^{-1}+ih_{1}^{-1}(s))c^1(3,3,2,2)\nonumber\\
&+&\frac{2}{7}\sqrt{\frac{\pi}{5}}\braket{r^3}(h_{3}^{-1}+ih_{3}^{-1}(s))c^3(3,2,2,1)+\frac{4}{7}\sqrt{\frac{6\pi}{5}}\braket{r^3}(h_{3}^{-1}+ih_{3}^{-1}(s))c^3(3,3,2,2)\nonumber\\
&+&2\sqrt{\frac{\pi}{385}}\braket{r^5}(h_{5}^{-1}+ih_{5}^{-1}(s))c^5(3,2,2,1)+4\sqrt{\frac{6\pi}{385}}\braket{r^5}(h_{5}^{-1}+ih_{5}^{-1}(s))c^5(3,3,2,2), 
\end{eqnarray}
where $c^{k}({\ell},m,{\ell}',m')$ is defined by Eq.~(\ref{eq:c_func}) and $h_{\ell}^{m}$ and $h_{\ell}^{m}(s)$ are defined by Eq.~(\ref{eq:hlm}) and Eq.~(\ref{eq:hlm_s}), respectively.
\begin{eqnarray}
& &\Bra{\frac{5}{2},\frac{5}{2}}\mathcal{H}_{\rm CEF}\Ket{\frac{3}{2},\frac{1}{2}}\nonumber\\
&=&\frac{2}{7}\sqrt{\frac{2\pi}{5}}\braket{r^3}(h_{3}^{-2}+ih_{3}^{-2}(s))c^3(3,2,2,0)+\frac{6}{7}\sqrt{\frac{2\pi}{5}}\braket{r^3}(h_{3}^{-2}+ih_{3}^{-2}(s))c^3(3,3,2,1)\nonumber\\
&+&2\sqrt{\frac{2\pi}{385}}\braket{r^5}(h_{5}^{-2}+ih_{5}^{-2}(s))c^5(3,2,2,0)+6\sqrt{\frac{2\pi}{385}}\braket{r^5}(h_{5}^{-2}+ih_{5}^{-2}(s))c^5(3,3,2,1). 
\\
& &\Bra{\frac{5}{2},\frac{5}{2}}\mathcal{H}_{\rm CEF}\Ket{\frac{3}{2},-\frac{1}{2}}\nonumber\\
&=&\frac{2}{7}\sqrt{\frac{3\pi}{5}}\braket{r^3}(h_{3}^{-3}+ih_{3}^{-3}(s))c^3(3,2,2,-1)+\frac{4}{7}\sqrt{\frac{3\pi}{5}}\braket{r^3}(h_{3}^{-3}+ih_{3}^{-3}(s))c^3(3,3,2,0)\nonumber\\
&+&2\sqrt{\frac{3\pi}{385}}\braket{r^5}(h_{5}^{-3}+ih_{5}^{-3}(s))c^5(3,2,2,-1)+4\sqrt{\frac{3\pi}{385}}\braket{r^5}(h_{5}^{-3}+ih_{5}^{-3}(s))c^5(3,3,2,0). 
\\
& &\Bra{\frac{5}{2},\frac{5}{2}}\mathcal{H}_{\rm CEF}\Ket{\frac{3}{2},-\frac{3}{2}}\nonumber\\
&=&4\sqrt{\frac{\pi}{385}}\braket{r^5}(h_{5}^{-4}+ih_{5}^{-4}(s))c^5(3,2,2,-2)+2\sqrt{\frac{6\pi}{385}}\braket{r^5}(h_{5}^{-4}+ih_{5}^{-4}(s))c^5(3,3,2,-1).
\\
%
%
& &\Bra{\frac{5}{2},\frac{3}{2}}\mathcal{H}_{\rm CEF}\Ket{\frac{3}{2},\frac{3}{2}}\nonumber\\
&=&2\sqrt{\frac{2\pi}{105}}\braket{r^1}(h_{1}^{0}+ih_{1}^{0}(s))c^1(3,1,2,1)+4\sqrt{\frac{\pi}{21}}\braket{r^1}(h_{1}^{0}+ih_{1}^{0}(s))c^1(3,2,2,2)\nonumber\\
&+&\frac{2}{7}\sqrt{\frac{2\pi}{5}}\braket{r^3}(h_{3}^{0}+ih_{3}^{0}(s))c^3(3,1,2,1)+\frac{4\sqrt{\pi}}{7}\braket{r^3}(h_{3}^{0}+ih_{3}^{0}(s))c^3(3,2,2,2)\nonumber\\
&+&2\sqrt{\frac{2\pi}{385}}\braket{r^5}(h_{5}^{0}+ih_{5}^{0}(s))c^5(3,1,2,1)+4\sqrt{\frac{\pi}{77}}\braket{r^5}(h_{5}^{0}+ih_{5}^{0}(s))c^5(3,2,2,2). 
\\
&&\Bra{\frac{5}{2},\frac{3}{2}}\mathcal{H}_{\rm CEF}\Ket{\frac{3}{2},\frac{1}{2}}\nonumber\\
&=&4\sqrt{\frac{\pi}{105}}\braket{r^1}(h_{1}^{-1}+ih_{1}^{-1}(s))c^1(3,1,2,0)+2\sqrt{\frac{\pi}{7}}\braket{r^1}(h_{1}^{-1}+ih_{1}^{-1}(s))c^1(3,2,2,1)\nonumber\\
&+&\frac{4}{7}\sqrt{\frac{\pi}{5}}\braket{r^3}(h_{3}^{-1}+ih_{3}^{-1}(s))c^3(3,1,2,0)+\frac{2\sqrt{3\pi}}{7}\braket{r^3}(h_{3}^{-1}+ih_{3}^{-1}(s))c^3(3,2,2,1)\nonumber\\
&+&4\sqrt{\frac{\pi}{385}}\braket{r^5}(h_{5}^{-1}+ih_{5}^{-1}(s))c^5(3,1,2,0)+2\sqrt{\frac{3\pi}{77}}\braket{r^5}(h_{5}^{-1}+ih_{5}^{-1}(s))c^5(3,2,2,1). 
\\
&&\Bra{\frac{5}{2},\frac{3}{2}}\mathcal{H}_{\rm CEF}\Ket{\frac{3}{2},-\frac{1}{2}}\nonumber\\
&=&\frac{6}{7}\sqrt{\frac{6\pi}{5}}\braket{r^3}(h_{3}^{-2}+ih_{3}^{-2}(s))c^3(3,1,2,-1)+\frac{2\sqrt{2\pi}}{7}\braket{r^3}(h_{3}^{-2}+ih_{3}^{-2}(s))c^3(3,2,2,0)\nonumber\\
&+&\frac{2\sqrt{6\pi}}{385}\braket{r^5}(h_{5}^{-2}+ih_{5}^{-2}(s))c^5(3,1,2,-1)+\frac{2\sqrt{2\pi}}{77}\braket{r^5}(h_{5}^{-2}+ih_{5}^{-2}(s))c^5(3,2,2,0). 
\\
&&\Bra{\frac{5}{2},\frac{3}{2}}\mathcal{H}_{\rm CEF}\Ket{\frac{3}{2},-\frac{3}{2}}\nonumber\\
&=&\frac{4}{7}\sqrt{\frac{2\pi}{5}}\braket{r^3}(h_{3}^{-3}+ih_{3}^{-3}(s))c^3(3,1,2,-2)+\frac{2\sqrt{\pi}}{7}\braket{r^3}(h_{3}^{-3}+ih_{3}^{-3}(s))c^3(3,2,2,-1)\nonumber\\
&+&4\sqrt{\frac{2\pi}{385}}\braket{r^5}(h_{5}^{-3}+ih_{5}^{-3}(s))c^5(3,1,2,-2)+2\sqrt{\frac{\pi}{77}}\braket{r^5}(h_{5}^{-3}+ih_{5}^{-3}(s))c^5(3,2,2,-1). 
\end{eqnarray}
\begin{eqnarray}
&&\Bra{\frac{5}{2},\frac{1}{2}}\mathcal{H}_{\rm CEF}\Ket{\frac{3}{2},\frac{3}{2}}\nonumber\\
&=&2\sqrt{\frac{\pi}{35}}\braket{r^1}(h_{1}^{1}+ih_{1}^{1}(s))c^1(3,0,2,1)+8\sqrt{\frac{\pi}{105}}\braket{r^1}(h_{1}^{1}+ih_{1}^{1}(s))c^1(3,1,2,2)\nonumber\\
&+&\frac{2}{7}\sqrt{3\pi}{5}\braket{r^3}(h_{3}^{1}+ih_{3}^{1}(s))c^3(3,0,2,1)+\frac{8}{7}\sqrt{\pi}{5}\braket{r^3}(h_{3}^{1}+ih_{3}^{1}(s))c^3(3,1,2,2)\nonumber\\
&+&2\sqrt{\frac{3\pi}{385}}\braket{r^5}(h_{5}^{1}+ih_{5}^{1}(s))c^5(3,0,2,1)+8\sqrt{\frac{\pi}{385}}\braket{r^5}(h_{5}^{1}+ih_{5}^{1}(s))c^5(3,1,2,2). 
\\
&&\Bra{\frac{5}{2},\frac{1}{2}}\mathcal{H}_{\rm CEF}\Ket{\frac{3}{2},\frac{1}{2}}\nonumber\\
&=&2\sqrt{\frac{2\pi}{35}}\braket{r^1}(h_{1}^{0}+ih_{1}^{0}(s))c^1(3,0,2,0)+4\sqrt{\frac{\pi}{35}}\braket{r^1}(h_{1}^{0}+ih_{1}^{0}(s))c^1(3,1,2,1)\nonumber\\
&+&\frac{2}{7}\sqrt{6\pi}{5}\braket{r^3}(h_{3}^{0}+ih_{3}^{0}(s))c^3(3,0,2,0)+\frac{4}{7}\sqrt{\frac{3\pi}{5}}\braket{r^3}(h_{3}^{0}+ih_{3}^{0}(s))c^3(3,1,2,1)\nonumber\\
&+&2\sqrt{\frac{6\pi}{385}}\braket{r^5}(h_{5}^{0}+ih_{5}^{0}(s))c^5(3,0,2,0)+4\sqrt{\frac{3\pi}{385}}\braket{r^5}(h_{5}^{0}+ih_{5}^{0}(s))c^5(3,1,2,1). 
\\
&&\Bra{\frac{5}{2},\frac{1}{2}}\mathcal{H}_{\rm CEF}\Ket{\frac{3}{2},-\frac{1}{2}}\nonumber\\
&=&2\sqrt{\frac{3\pi}{35}}\braket{r^1}(h_{1}^{-1}+ih_{1}^{-1}(s))c^1(3,0,2,-1)+4\sqrt{\frac{2\pi}{105}}\braket{r^1}(h_{1}^{-1}+ih_{1}^{-1}(s))c^1(3,1,2,0)\nonumber\\
&+&\frac{6}{7}\sqrt{\frac{\pi}{5}}\braket{r^3}(h_{3}^{-1}+ih_{3}^{-1}(s))c^3(3,0,2,-1)+\frac{4}{7}\sqrt{\frac{2\pi}{5}}\braket{r^3}(h_{3}^{-1}+ih_{3}^{-1}(s))c^3(3,1,2,0)\nonumber\\
&+&6\sqrt{\frac{\pi}{385}}\braket{r^5}(h_{5}^{-1}+ih_{5}^{-1}(s))c^5(3,0,2,-1)+2\sqrt{\frac{2\pi}{385}}\braket{r^5}(h_{5}^{-1}+ih_{5}^{-1}(s))c^5(3,1,2,0). 
\\
&&\Bra{\frac{5}{2},\frac{1}{2}}\mathcal{H}_{\rm CEF}\Ket{\frac{3}{2},-\frac{3}{2}}\nonumber\\
&=&\frac{4}{7}\sqrt{\frac{3\pi}{5}}\braket{r^3}(h_{3}^{-2}+ih_{3}^{-2}(s))c^3(3,0,2,-2)+\frac{4}{7}\sqrt{\frac{\pi}{5}}\braket{r^3}(h_{3}^{-2}+ih_{3}^{-2}(s))c^3(3,1,2,-1)\nonumber\\
&+&4\sqrt{\frac{3\pi}{385}}\braket{r^5}(h_{5}^{-2}+ih_{5}^{-2}(s))c^5(3,0,2,-2)+4\sqrt{\frac{\pi}{385}}\braket{r^5}(h_{5}^{-2}+ih_{5}^{-2}(s))c^5(3,1,2,-1). 
\\
&&\Bra{\frac{5}{2},-\frac{1}{2}}\mathcal{H}_{\rm CEF}\Ket{\frac{3}{2},\frac{3}{2}}\nonumber\\
&=&\frac{4}{7}\sqrt{\frac{\pi}{5}}\braket{r^3}(h_{3}^{2}+ih_{3}^{2}(s))c^3(3,-1,2,1)+\frac{4}{7}\sqrt{\frac{3\pi}{5}}\braket{r^3}(h_{3}^{2}+ih_{3}^{2}(s))c^3(3,0,2,2)\nonumber\\
&+&4\sqrt{\frac{\pi}{385}}\braket{r^5}(h_{5}^{2}+ih_{5}^{2}(s))c^5(3,-1,2,1)+4\sqrt{\frac{3\pi}{385}}\braket{r^5}(h_{5}^{2}+ih_{5}^{2}(s))c^5(3,0,2,2). 
\\
&&\Bra{\frac{5}{2},-\frac{1}{2}}\mathcal{H}_{\rm CEF}\Ket{\frac{3}{2},\frac{1}{2}}\nonumber\\
&=&4\sqrt{\frac{2\pi}{105}}\braket{r^1}(h_{1}^{1}+ih_{1}^{1}(s))c^1(3,-1,2,0)+2\sqrt{\frac{3\pi}{35}}\braket{r^1}(h_{1}^{1}+ih_{1}^{1}(s))c^1(3,0,2,1)\nonumber\\
&+&\frac{4}{7}\sqrt{\frac{2\pi}{5}}\braket{r^3}(h_{3}^{1}+ih_{3}^{1}(s))c^3(3,-1,2,0)+\frac{6}{7}\sqrt{\frac{\pi}{5}}\braket{r^3}(h_{3}^{1}+ih_{3}^{1}(s))c^3(3,0,2,1)\nonumber\\
&+&4\sqrt{\frac{2\pi}{385}}\braket{r^5}(h_{5}^{1}+ih_{5}^{1}(s))c^5(3,-1,2,0)+6\sqrt{\frac{\pi}{385}}\braket{r^5}(h_{5}^{1}+ih_{5}^{1}(s))c^5(3,0,2,1). 
\end{eqnarray}
\begin{eqnarray}
&&\Bra{\frac{5}{2},-\frac{1}{2}}\mathcal{H}_{\rm CEF}\Ket{\frac{3}{2},-\frac{1}{2}}\nonumber\\
&=&4\sqrt{\frac{\pi}{35}}\braket{r^1}(h_{1}^{0}+ih_{1}^{0}(s))c^1(3,-1,2,-1)+2\sqrt{\frac{2\pi}{35}}\braket{r^1}(h_{1}^{0}+ih_{1}^{0}(s))c^1(3,0,2,0)\nonumber\\
&+&\frac{4}{7}\sqrt{\frac{3\pi}{5}}\braket{r^3}(h_{3}^{0}+ih_{3}^{0}(s))c^3(3,-1,2,-1)+\frac{2}{7}\sqrt{\frac{6\pi}{5}}\braket{r^3}(h_{3}^{0}+ih_{3}^{0}(s))c^3(3,0,2,0)\nonumber\\
&+&4\sqrt{\frac{3\pi}{385}}\braket{r^5}(h_{5}^{0}+ih_{5}^{0}(s))c^5(3,-1,2,-1)+2\sqrt{\frac{6\pi}{385}}\braket{r^5}(h_{5}^{0}+ih_{5}^{0}(s))c^5(3,0,2,0). 
\\
&&\Bra{\frac{5}{2},-\frac{1}{2}}\mathcal{H}_{\rm CEF}\Ket{\frac{3}{2},-\frac{3}{2}}\nonumber\\
&=&8\sqrt{\frac{\pi}{105}}\braket{r^1}(h_{1}^{-1}+ih_{1}^{-1})c^1(3,-1,2,-2)+2\sqrt{\frac{\pi}{35}}\braket{r^1}(h_{1}^{-1}+ih_{1}^{-1}(s))c^1(3,0,2,-1)\nonumber\\
&+&\frac{8}{7}\sqrt{\frac{\pi}{5}}\braket{r^3}(h_{3}^{-1}+ih_{3}^{-1}(s))c^3(3,-1,2,-2)+\frac{2}{7}\sqrt{\frac{3\pi}{5}}\braket{r^3}(h_{3}^{-1}+ih_{3}^{-1}(s))c^3(3,0,2,-1)\nonumber\\
&+&8\sqrt{\frac{\pi}{385}}\braket{r^5}(h_{5}^{-1}+ih_{5}^{-1}(s))c^5(3,-1,2,-2)+2\sqrt{\frac{3\pi}{385}}\braket{r^5}(h_{5}^{-1}+ih_{5}^{-1}(s))c^5(3,0,2,-1). 
\\
&&\Bra{\frac{5}{2},-\frac{3}{2}}\mathcal{H}_{\rm CEF}\Ket{\frac{3}{2},\frac{3}{2}}\nonumber\\
&=&\frac{2\sqrt{\pi}}{7}\braket{r^3}(h_{3}^{3}+ih_{3}^{3}(s))c^3(3,-2,2,1)+\frac{4}{7}\sqrt{\frac{2\pi}{5}}\braket{r^3}(h_{3}^{3}+ih_{3}^{3}(s))c^3(3,-1,2,2)\nonumber\\
&+&2\sqrt{\frac{\pi}{77}}\braket{r^5}(h_{5}^{3}+ih_{5}^{3}(s))c^5(3,-2,2,1)+4\sqrt{\frac{2\pi}{385}}\braket{r^5}(h_{5}^{3}+ih_{5}^{3}(s))c^5(3,-1,2,2). 
\\
&&\Bra{\frac{5}{2},-\frac{3}{2}}\mathcal{H}_{\rm CEF}\Ket{\frac{3}{2},\frac{1}{2}}\nonumber\\
&=&\frac{2\sqrt{2\pi}}{7}\braket{r^3}(h_{3}^{2}+ih_{3}^{2}(s))c^3(3,-2,2,0)+\frac{2}{7}\sqrt{\frac{6\pi}{5}}\braket{r^3}(h_{3}^{2}+ih_{3}^{2}(s))c^3(3,-1,2,1)\nonumber\\
&+&2\sqrt{\frac{2\pi}{77}}\braket{r^5}(h_{5}^{2}+ih_{5}^{2}(s))c^5(3,-2,2,0)+2\sqrt{\frac{6\pi}{385}}\braket{r^5}(h_{5}^{2}+ih_{5}^{2}(s))c^5(3,-1,2,1). 
\\
&&\Bra{\frac{5}{2},-\frac{3}{2}}\mathcal{H}_{\rm CEF}\Ket{\frac{3}{2},-\frac{1}{2}}\nonumber\\
&=&2\sqrt{\frac{\pi}{7}}\braket{r^1}(h_{1}^{1}+ih_{1}^{1}(s))c^1(3,-2,2,-1)+4\sqrt{\frac{\pi}{105}}\braket{r^1}(h_{1}^{1}+ih_{1}^{1}(s))c^1(3,-1,2,0)\nonumber\\
&+&\frac{2\sqrt{3\pi}}{7}\braket{r^3}(h_{3}^{1}+ih_{3}^{1}(s))c^3(3,-2,2,-1)+\frac{2}{7}\sqrt{\frac{\pi}{5}}\braket{r^3}(h_{3}^{1}+ih_{3}^{1}(s))c^3(3,-1,2,0)\nonumber\\
&+&2\sqrt{\frac{3\pi}{77}}\braket{r^5}(h_{5}^{1}+ih_{5}^{1}(s))c^5(3,-2,2,-1)+4\sqrt{\frac{\pi}{385}}\braket{r^5}(h_{5}^{1}+ih_{5}^{1}(s))c^5(3,-1,2,0). 
\\
&&\Bra{\frac{5}{2},-\frac{3}{2}}\mathcal{H}_{\rm CEF}\Ket{\frac{3}{2},-\frac{3}{2}}\nonumber\\
&=&4\sqrt{\frac{\pi}{21}}\braket{r^1}(h_{1}^{0}+ih_{1}^{0}(s))c^1(3,-2,2,-2)+2\sqrt{\frac{2\pi}{105}}\braket{r^1}(h_{1}^{0}+ih_{1}^{0}(s))c^1(3,-1,2,-1)\nonumber\\
&+&\frac{4\sqrt{\pi}}{7}\braket{r^3}(h_{3}^{0}+ih_{3}^{0}(s))c^3(3,-2,2,-2)+\frac{2}{7}\sqrt{\frac{2\pi}{5}}\braket{r^3}(h_{3}^{0}+ih_{3}^{0}(s))c^3(3,-1,2,-1)\nonumber\\
&+&4\sqrt{\frac{\pi}{77}}\braket{r^5}(h_{5}^{0}+ih_{5}^{0}(s))c^5(3,-2,2,-2)+2\sqrt{\frac{2\pi}{385}}\braket{r^5}(h_{5}^{0}+ih_{5}^{0}(s))c^5(3,-1,2,-1). 
\end{eqnarray}
\begin{eqnarray}
&&\Bra{\frac{5}{2},-\frac{5}{2}}\mathcal{H}_{\rm CEF}\Ket{\frac{3}{2},\frac{3}{2}}\nonumber\\
&=&2\sqrt{\frac{6\pi}{385}}\braket{r^5}(h_{5}^{4}+ih_{5}^{4}(s))c^5(3,-3,2,1)+4\sqrt{\frac{\pi}{385}}\braket{r^5}(h_{5}^{4}+ih_{5}^{4}(s))c^5(3,-2,2,2). 
\\
&&\Bra{\frac{5}{2},-\frac{5}{2}}\mathcal{H}_{\rm CEF}\Ket{\frac{3}{2},\frac{1}{2}}\nonumber\\
&=&\frac{4}{7}\sqrt{\frac{3\pi}{5}}\braket{r^3}(h_{3}^{3}+ih_{3}^{3}(s))c^3(3,-3,2,0)+\frac{2}{7}\sqrt{\frac{3\pi}{5}}\braket{r^3}(h_{3}^{3}+ih_{3}^{3}(s))c^3(3,-2,2,1)\nonumber\\
&+&4\sqrt{\frac{3\pi}{385}}\braket{r^5}(h_{5}^{3}+ih_{5}^{3}(s))c^5(3,-3,2,0)+2\sqrt{\frac{3\pi}{385}}\braket{r^5}(h_{5}^{3}+ih_{5}^{3}(s))c^5(3,-2,2,1).
\\
&&\Bra{\frac{5}{2},-\frac{5}{2}}\mathcal{H}_{\rm CEF}\Ket{\frac{3}{2},-\frac{1}{2}}\nonumber\\
&=&\frac{6}{7}\sqrt{\frac{2\pi}{5}}\braket{r^3}(h_{3}^{2}+ih_{3}^{2}(s))c^3(3,-3,2,-1)+\frac{2}{7}\sqrt{\frac{2\pi}{5}}\braket{r^3}(h_{3}^{2}+ih_{3}^{2}(s))c^3(3,-2,2,0)\nonumber\\
&+&6\sqrt{\frac{2\pi}{385}}\braket{r^5}(h_{5}^{2}+ih_{5}^{2}(s))c^5(3,-3,2,-1)+2\sqrt{\frac{2\pi}{385}}\braket{r^5}(h_{5}^{2}+ih_{5}^{2}(s))c^5(3,-2,2,0). 
\\
&&\Bra{\frac{5}{2},-\frac{5}{2}}\mathcal{H}_{\rm CEF}\Ket{\frac{3}{2},-\frac{3}{2}}\nonumber\\
&=&4\sqrt{\frac{2\pi}{35}}\braket{r^1}(h_{1}^{1}+ih_{1}^{1}(s))c^1(3,-3,2,-2)+2\sqrt{\frac{\pi}{105}}\braket{r^1}(h_{1}^{1}+ih_{1}^{1}(s))c^1(3,-2,2,-1)\nonumber\\
&+&\frac{4}{7}\sqrt{\frac{6\pi}{5}}\braket{r^3}(h_{3}^{1}+ih_{3}^{1}(s))c^3(3,-3,2,-2)+\frac{2}{7}\sqrt{\frac{\pi}{5}}\braket{r^3}(h_{3}^{1}+ih_{3}^{1}(s))c^3(3,-2,2,-1)\nonumber\\
&+&4\sqrt{\frac{6\pi}{385}}\braket{r^5}(h_{5}^{1}+ih_{5}^{1}(s))c^5(3,-3,2,-2)+2\sqrt{\frac{\pi}{385}}\braket{r^5}(h_{5}^{1}+ih_{5}^{1}(s))c^5(3,-2,2,-1). 
\end{eqnarray}
%

\section{Two-icosahedron cluster mean-field method}
\label{sec:2ICMF}

\begin{figure}[tb]
\centering
\includegraphics[width=7cm]{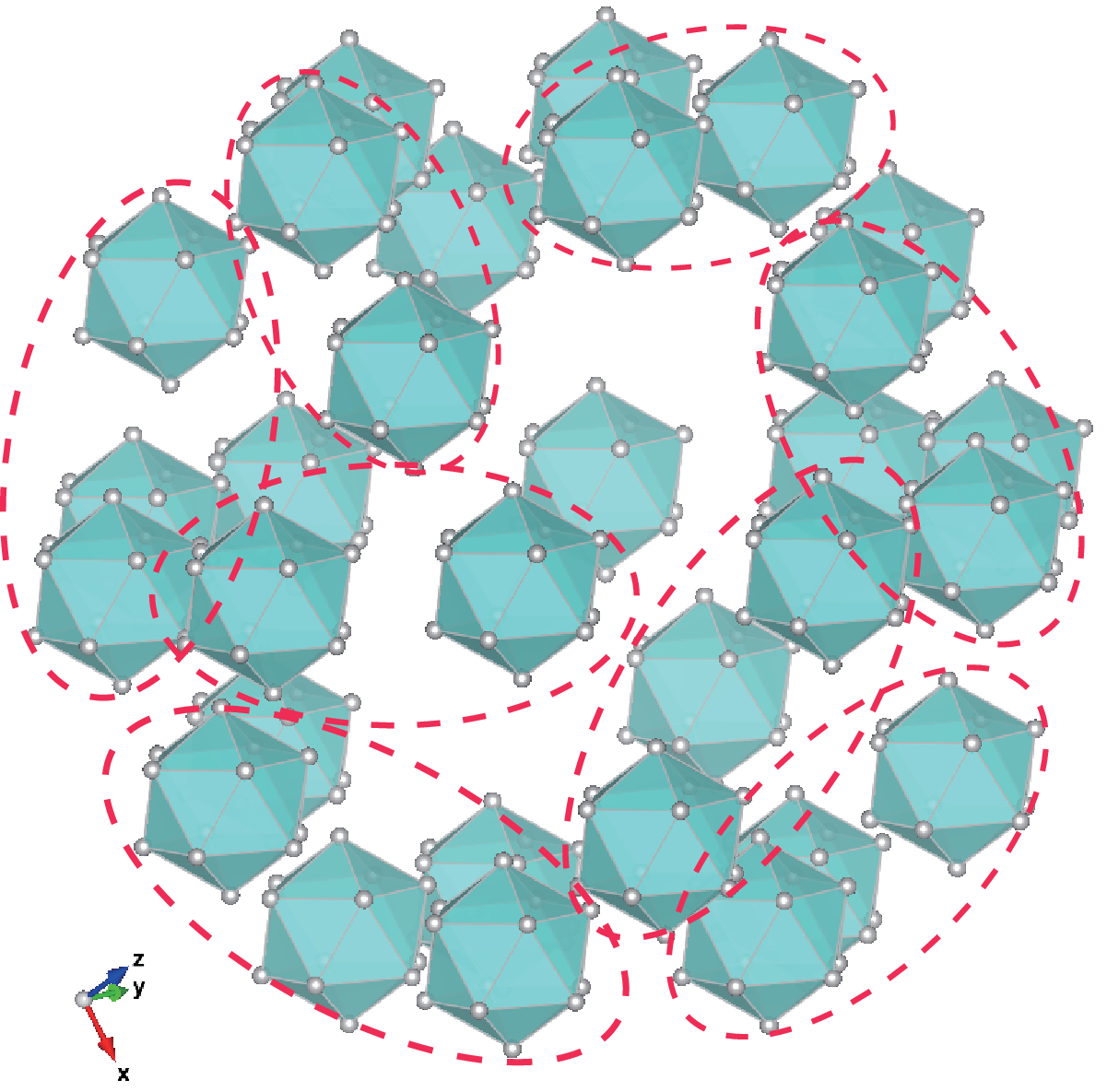}
\caption{(Color online) 
Schematic explanation for the two-IC mean-field approximation method. The magnetic model Eq.~(\ref{eq:H}) on each cluster composed of two ICs (frontmost two ICs) surrounded by a dashed line is solved by the exact-diagonalization method. The inter-cluster correlation is taken into account by the mean-field approximation (see text). 
}
\label{fig:QC_2IC_MF}
\end{figure}

To gain insight into the magnetic ground state in the Cd$_{5.7}$Yb-type QC~\cite{Takakura}, we have theoretically studied the magnetic model taking into account the magnetic anisotropy arising from the CEF, Eq.~(\ref{eq:H}),  on the QC cluster. 
Namely, we have considered the icosidodecahedron cluster at each vertex of whom the IC is located. 
In the present study, we have considered the cluster composed of 30 ICs under the open boundary condition, as shown in Fig.~\ref{fig:QC_2IC_MF}. Then, we have decomposed the 30 ICs into the 15 sets of the two ICs located closely each other (see two ICs enclosed by each dashed line in Fig.~\ref{fig:QC_2IC_MF}). 
The converted states obtained in this way are plotted by each symbol in the plane of $J_2/J_1$  and the anisotropy angle $\theta$ in Fig.~\ref{fig:QC_2IC_MF}.

First, we have obtained the ground state of the magnetic model $H$ on the cluster composed of the 2 ICs by performing the exact diagonalization method. 
This method enables us to calculate accurately all the eigenstates not only for the intra-IC interactions but also for the inter-IC interactions for the two ICs in the numerically-exact way at least inside the cluster. 
As for the outside of the cluster, the inter-cluster correlations are taken into account by applying the mean-field approximation as 
\begin{eqnarray}
\langle\hat{\bm J}_i\cdot\hat{\bm J}_j\rangle\approx\langle\hat{\bm J}_i\rangle\cdot\hat{\bm J}_j+\hat{\bm J}_i\cdot\langle\hat{\bm J}_j\rangle-\langle\hat{\bm J}_i\rangle\cdot\langle\hat{\bm J}_j\rangle.
\label{eq:JJ_MF}
\end{eqnarray}
These procedures are repeated iteratively until the ground state is converged. 

As for the initial state, we have inputted the several states such as the vortex state, the altanative state, and the whirling state obtained as the ground state in the 1/1 AC, as shown in Fig.~\ref{fig:PD_AC}. 
We have also checked whether the converged state shown in Fig.~\ref{fig:PD_QC}(a) is still obtained starting from the different initial state. 

\section{Electroic dipole moment}
\label{sec:EDapp}

In Sect.~\ref{sec:ED}, the definition of the electric dipole moment was given by Eq.~(\ref{eq:ED}). 
The expectation value for the CEF ground state $\bar{Q}_{\mu}\equiv\langle\psi_{0\eta}^{\rm tot}|\frac{Q_{\mu}}{r}|\psi_{0\eta}^{\rm tot}\rangle$ is calculated by Eq.~(\ref{eq:WFtot}) as 
\begin{eqnarray}
\bar{Q}_{\mu}=\sum_{J_z=-5/2}^{5/2}\sum_{J_z'=-3/2}^{3/2}a_{0\eta,J_z}^{\rm tot*}b_{0\eta,J_z'}^{\rm tot}\Bra{\frac{5}{2},J_z}\frac{Q_{\mu}}{r}\Ket{\frac{3}{2},J_z'}+h.c.
\label{eq:exp_Q}
\end{eqnarray}
In this Appendix, we show the explicit form of the matrix elements for $\Bra{J=5/2,J_z}\frac{Q_{\mu}}{r}\Ket{J'=3/2,J_z'}$ calculated analytically for $\mu=x, y, z$ on the basis of  
Eqs.~(\ref{eq:psi_4f_1})-(\ref{eq:psi_4f_6}) and Eqs.~(\ref{eq:psi_5d_1})-(\ref{eq:psi_5d_4}). The results for $\mu=x$ which are not zero are shown as follows. 
\begin{eqnarray}
\Bra{\frac{5}{2},\frac{5}{2}}\frac{x}{r}\Ket{\frac{3}{2},\frac{3}{2}}&=&-\frac{1}{\sqrt{70}}c^1(3,2,2,1)-2\sqrt{\frac{3}{35}}c^1(3,3,2,2),
\\
\Bra{\frac{5}{2},\frac{1}{2}}\frac{x}{r}\Ket{\frac{3}{2},\frac{3}{2}}&=&\sqrt{\frac{3}{70}}c^1(3,0,2,1)+2\sqrt{\frac{2}{35}}c^1(3,1,2,2),
\\
\Bra{\frac{5}{2},\frac{3}{2}}\frac{x}{r}\Ket{\frac{3}{2},\frac{1}{2}}&=&-\sqrt{\frac{2}{35}}c^1(3,1,2,0)-\sqrt{\frac{3}{14}}c^1(3,2,2,1),
\\
\Bra{\frac{5}{2},-\frac{1}{2}}\frac{x}{r}\Ket{\frac{3}{2},\frac{1}{2}}&=&\frac{2}{\sqrt{35}}c^1(3,-1,2,0)+\frac{3}{\sqrt{70}}c^1(3,0,2,1),
\\
\Bra{\frac{5}{2},\frac{1}{2}}\frac{x}{r}\Ket{\frac{3}{2},-\frac{1}{2}}&=&-\frac{3}{\sqrt{70}}c^1(3,0,2,-1)-\frac{2}{\sqrt{35}}c^1(3,1,2,0),
\\
\Bra{\frac{5}{2},-\frac{3}{2}}\frac{x}{r}\Ket{\frac{3}{2},-\frac{1}{2}}&=&\sqrt{\frac{3}{14}}c^1(3,-2,2,-1)+\sqrt{\frac{2}{35}}c^1(3,-1,2,0),
\\
\Bra{\frac{5}{2},-\frac{1}{2}}\frac{x}{r}\Ket{\frac{3}{2},-\frac{3}{2}}&=&-2\sqrt{\frac{2}{35}}c^1(3,-1,2,-2)-\sqrt{\frac{3}{70}}c^1(3,0,2,-1),
\\
\Bra{\frac{5}{2},-\frac{5}{2}}\frac{x}{r}\Ket{\frac{3}{2},-\frac{3}{2}}&=&2\sqrt{\frac{3}{35}}c^1(3,-3,2,-2)+\frac{1}{\sqrt{70}}c^1(3,-2,2,-1), 
\end{eqnarray}
where $c^k(\ell,m,\ell',m')$ is defined by Eq.~(\ref{eq:c_func}). 
The results for $\mu=y$  which are not zero are shown as follows. 
\begin{eqnarray}
\Bra{\frac{5}{2},\frac{5}{2}}\frac{y}{r}\Ket{\frac{3}{2},\frac{3}{2}}&=&i\frac{1}{\sqrt{70}}c^1(3,2,2,1)+i2\sqrt{\frac{3}{35}}c^1(3,3,2,2),
\\
\Bra{\frac{5}{2},\frac{1}{2}}\frac{y}{r}\Ket{\frac{3}{2},\frac{3}{2}}&=&i\sqrt{\frac{3}{70}}c^1(3,0,2,1)+i2\sqrt{\frac{2}{35}}c^1(3,1,2,2), 
\\
\Bra{\frac{5}{2},\frac{3}{2}}\frac{y}{r}\Ket{\frac{3}{2},\frac{1}{2}}&=&i\sqrt{\frac{2}{35}}c^1(3,1,2,0)+i\sqrt{\frac{15}{70}}c^1(3,2,2,1), 
\\
\Bra{\frac{5}{2},-\frac{1}{2}}\frac{y}{r}\Ket{\frac{3}{2},\frac{1}{2}}&=&i\frac{2}{\sqrt{35}}c^1(3,-1,2,0)+i\frac{3}{\sqrt{70}}c^1(3,0,2,1), 
\\
\Bra{\frac{5}{2},\frac{1}{2}}\frac{y}{r}\Ket{\frac{3}{2},-\frac{1}{2}}&=&i\frac{3}{\sqrt{70}}c^1(3,0,2,-1)+i\frac{2}{\sqrt{35}}c^1(3,1,2,0), 
\\
\Bra{\frac{5}{2},-\frac{3}{2}}\frac{y}{r}\Ket{\frac{3}{2},-\frac{1}{2}}&=&i\sqrt{\frac{15}{70}}c^1(3,-2,2,-1)+i\sqrt{\frac{2}{35}}c^1(3,-1,2,0), 
\\
\Bra{\frac{5}{2},-\frac{1}{2}}\frac{y}{r}\Ket{\frac{3}{2},-\frac{3}{2}}&=&i2\sqrt{\frac{2}{35}}c^1(3,-1,2,-2)+i\sqrt{\frac{3}{70}}c^1(3,0,2,-1), 
\\
\Bra{\frac{5}{2},-\frac{5}{2}}\frac{y}{r}\Ket{\frac{3}{2},-\frac{3}{2}}&=&i2\sqrt{\frac{3}{35}}c^1(3,-3,2,-2)+i\frac{1}{\sqrt{70}}c^1(3,-2,2,-1).
\end{eqnarray}
The results for $\mu=z$  which are not zero are shown as follows. 
\begin{eqnarray}
\Bra{\frac{5}{2},\frac{3}{2}}\frac{z}{r}\Ket{\frac{3}{2},\frac{3}{2}}&=&\sqrt{\frac{2}{35}}c^1(3,1,2,1)+\frac{2}{\sqrt{7}}c^1(3,2,2,2), 
\\
\Bra{\frac{5}{2},\frac{1}{2}}\frac{z}{r}\Ket{\frac{3}{2},\frac{1}{2}}&=&\sqrt{\frac{6}{35}}c^1(3,0,2,0)+2\sqrt{\frac{3}{35}}c^1(3,1,2,1), 
\\
\Bra{\frac{5}{2},-\frac{1}{2}}\frac{z}{r}\Ket{\frac{3}{2},-\frac{1}{2}}&=&2\sqrt{\frac{3}{35}}c^1(3,-1,2,-1)+\sqrt{\frac{6}{35}}c^1(3,0,2,0), 
\\
\Bra{\frac{5}{2},-\frac{3}{2}}\frac{z}{r}\Ket{\frac{3}{2},-\frac{3}{2}}&=&\frac{2}{\sqrt{7}}c^1(3,-2,2,-2)+\sqrt{\frac{2}{35}}c^1(3,-1,2,-1). 
\end{eqnarray}
%

\section{Magnetic toroidal dipole moment}
\label{sec:MTapp}

In Sect.~\ref{sec:MT}, the definition of the MT moment was given by Eq.~(\ref{eq:MT}). In this Appendix, the explicit form of the operators are given in the local coordinate [see Fig.~\ref{fig:Tsai_cluster}(f)] as 
\begin{eqnarray}
\tilde{t}_{{\ell}x}&=&\frac{2}{(\ell+1)(\ell+2)}\left\{\frac{y}{r}\ell_{z}-\frac{1}{2i}\frac{z}{r}(\ell_{+}-\ell_{-})-i\frac{x}{r}\right\}I
\nonumber
\\
&+&\frac{1}{\ell+1}\left\{\frac{y}{r}
\begin{pmatrix}
1 & 0 \\
0 & -1
\end{pmatrix}
-\frac{z}{r}
\begin{pmatrix}
0 & -i \\
i &  0
\end{pmatrix}
\right\}, 
\label{eq:t1x}
\\
\tilde{t}_{{\ell}y}&=&\frac{2}{(\ell+1)(\ell+2)}\left\{-\frac{x}{r}\ell_{z}+\frac{1}{2}\frac{z}{r}(\ell_{+}+\ell_{-})-i\frac{y}{r}\right\}I
\nonumber
\\
&+&\frac{1}{\ell+1}\left\{\frac{z}{r}
\begin{pmatrix}
0 & 1 \\
1 & 0
\end{pmatrix}
-\frac{x}{r}
\begin{pmatrix}
1 & 0 \\
0 & -1
\end{pmatrix}
\right\}, 
\label{eq:t1y}
\\
\tilde{t}_{{\ell}z}&=&\frac{2}{(\ell+1)(\ell+2)}\left\{\frac{1}{2i}\frac{x}{r}(\ell_{+}-\ell_{-})-\frac{1}{2}\frac{y}{r}(\ell_{+}+\ell_{-})-i\frac{z}{r}\right\}I
\nonumber
\\
&+&\frac{1}{\ell+1}\left\{\frac{x}{r}
\begin{pmatrix}
0 & -i \\
i & 0
\end{pmatrix}
-\frac{y}{r}
\begin{pmatrix}
0 & 1 \\
1 & 0
\end{pmatrix}
\right\}, 
\label{eq:t1z}
\end{eqnarray}
where ${\ell}_{+}$ $({\ell}_{-})$ is the raising (lowering) operator of the angular momentum 
and $I$ is the unit $2\times 2$ matrix.  
By substituting $x/r=-\sqrt{\frac{4\pi}{3}}\frac{1}{\sqrt{2}}\{Y_{1}^{1}(\hat{\bm r})-Y_{1}^{-1}(\hat{\bm r})\}$,  $y/r=-\sqrt{\frac{4\pi}{3}}\frac{1}{\sqrt{2}i}\{Y_{1}^{1}(\hat{\bm r})+Y_{1}^{-1}(\hat{\bm r})\}$, and $z/r=\sqrt{\frac{4\pi}{3}}Y_{1}^{0}(\hat{\bm r})$ into Eqs.~(\ref{eq:t1x}), (\ref{eq:t1y}), and (\ref{eq:t1z}), we obtain for the dipole $\ell=1$ 
%
\begin{eqnarray}
\tilde{t}_{1x}&=&\frac{i}{3\sqrt{2}}\sqrt{\frac{4\pi}{3}}\left[Y_1^1(\hat{\bm r})(\ell_z+1)+Y_1^{-1}(\hat{\bm r})(\ell_z-1)+\frac{1}{\sqrt{2}}Y_1^0(\hat{\bm r})(\ell_{+}-\ell_{-})
\right]I
\nonumber
\\
&+&\frac{i}{2}\sqrt{\frac{4\pi}{3}}
\begin{pmatrix}
\frac{1}{\sqrt{2}}\left\{Y_1^1(\hat{\bm r})+Y_1^{-1}(\hat{\bm r})\right\} & Y_1^0(\hat{\bm r}) \\
-Y_1^0(\hat{\bm r}) & -\frac{1}{\sqrt{2}}\left\{Y_1^1(\hat{\bm r})+Y_1^{-1}(\hat{\bm r})\right\}
\end{pmatrix},
\label{eq:t1x2}
\\
\tilde{t}_{1y}&=&\frac{1}{3\sqrt{2}}\sqrt{\frac{4\pi}{3}}\left[Y_1^1(\hat{\bm r})(\ell_z+1)-Y_1^{-1}(\hat{\bm r})(\ell_z-1)+\frac{1}{\sqrt{2}}Y_1^0(\hat{\bm r})(\ell_{+}+\ell_{-})
\right]I
\nonumber
\\
&+&\frac{1}{2}\sqrt{\frac{4\pi}{3}}
\begin{pmatrix}
\frac{1}{\sqrt{2}}\left\{Y_1^1(\hat{\bm r})-Y_1^{-1}(\hat{\bm r})\right\} & Y_1^0(\hat{\bm r}) \\
Y_1^0(\hat{\bm r}) & -\frac{1}{\sqrt{2}}\left\{Y_1^1(\hat{\bm r})-Y_1^{-1}(\hat{\bm r})\right\}
\end{pmatrix},
\label{eq:t1y2}
\\
\tilde{t}_{1z}&=&\frac{-i}{3\sqrt{2}}\sqrt{\frac{4\pi}{3}}\left[Y_1^{-1}(\hat{\bm r})\ell_{+}+Y_1^{1}(\hat{\bm r})\ell_{-}+\sqrt{2}Y_1^0(\hat{\bm r})
\right]I
\nonumber
\\
&-&\frac{i}{2}\sqrt{\frac{4\pi}{3}}
\begin{pmatrix}
0 & \sqrt{2}Y_1^{-1}(\hat{\bm r}) \\
\sqrt{2}Y_1^1(\hat{\bm r}) & 0
\end{pmatrix}. 
\label{eq:t1z2}
\end{eqnarray}
The expectation value of the MT dipole moment for the CEF ground state $\bar{T}_{1\mu\eta}\equiv-\mu_{\rm B}\langle\psi_{0\eta}^{\rm tot}|\tilde{t}_{1\mu}|\psi_{0\eta}^{\rm tot}\rangle$ is calculated by Eq.~(\ref{eq:WFtot}) as 
\begin{eqnarray}
\bar{T}_{1\mu\eta}=-\mu_{\rm B}\sum_{J_z=-5/2}^{5/2}\sum_{J_z'=-3/2}^{3/2}a_{0\eta,J_z}^{\rm tot*}b_{0\eta,J_z'}^{\rm tot}\Bra{\frac{5}{2},J_z}\tilde{t}_{1\mu}\Ket{\frac{3}{2},J_z'}+h.c.
\label{eq:exp_Q}
\end{eqnarray}
Then, by using Eqs.~(\ref{eq:psi_4f_1})-(\ref{eq:psi_4f_6}) and Eqs.~(\ref{eq:psi_5d_1})-(\ref{eq:psi_5d_4}), we have calculated $\Bra{J=5/2,J_z}\tilde{t}_{1\mu}\Ket{J'=3/2,J_z'}$ for $\mu=x, y, z$. The results for $\tilde{t}_{1x}$ which are not zero are shown as follows. 
\begin{eqnarray}
\Bra{\frac{5}{2},\frac{5}{2}}\tilde{t}_{1x}\Ket{\frac{3}{2},\frac{3}{2}}&=&i\frac{7}{6\sqrt{70}}c^1(3,2,2,1)-i\frac{2}{3\sqrt{35}}c^1(3,2,2,2)+i\sqrt{\frac{3}{35}}c^1(3,3,2,2), 
\\
\Bra{\frac{5}{2},\frac{1}{2}}\tilde{t}_{1x}\Ket{\frac{3}{2},\frac{3}{2}}&=&-i\frac{1}{\sqrt{70}}c^1(3,0,2,0)+i\frac{1}{2}\sqrt{\frac{3}{70}}c^1(3,0,2,1)
\nonumber
\\
&-&i\frac{1}{3\sqrt{35}}c^1(3,1,2,1)-i\frac{2}{3\sqrt{70}}c^1(3,1,2,2), 
\\
\Bra{\frac{5}{2},\frac{3}{2}}\tilde{t}_{1x}\Ket{\frac{3}{2},\frac{1}{2}}&=&i\frac{5}{3\sqrt{70}}c^1(3,1,2,0)-i\frac{1}{6}\sqrt{\frac{6}{35}}c^1(3,1,2,1)
\nonumber
\\
&+&i\frac{1}{6}\sqrt{\frac{3}{14}}c^1(3,2,2,1)+i\frac{1}{6}\sqrt{\frac{12}{7}}c^1(3,2,2,2), 
\\
\Bra{\frac{5}{2},-\frac{1}{2}}\tilde{t}_{1x}\Ket{\frac{3}{2},\frac{1}{2}}&=&-i\frac{2}{3}\sqrt{\frac{3}{35}}c^1(3,-1,2,-1)+i\frac{1}{3\sqrt{35}}c^1(3,-1,2,0)
\nonumber
\\
&-&i\frac{3}{2\sqrt{70}}c^1(3,0,2,1), 
\\
\Bra{\frac{5}{2},\frac{1}{2}}\tilde{t}_{1x}\Ket{\frac{3}{2},-\frac{1}{2}}&=&i\frac{3}{2\sqrt{70}}c^1(3,0,2,-1)-i\frac{1}{3\sqrt{35}}c^1(3,1,2,0)+i\frac{2}{3}\sqrt{\frac{3}{35}}c^1(3,1,2,1), 
\\
\Bra{\frac{5}{2},-\frac{3}{2}}\tilde{t}_{1x}\Ket{\frac{3}{2},-\frac{1}{2}}&=&-i\frac{1}{6}\sqrt{\frac{12}{7}}c^1(3,-2,2,-2)-i\frac{1}{6}\sqrt{\frac{3}{14}}c^1(3,-2,2,-1)
\nonumber
\\
&+&i\frac{1}{6}\sqrt{\frac{6}{35}}c^1(3,-1,2,-1)-i\frac{5}{3\sqrt{70}}c^1(3,-1,2,0),
\\
\Bra{\frac{5}{2},-\frac{1}{2}}\tilde{t}_{1x}\Ket{\frac{3}{2},-\frac{3}{2}}&=&i\frac{2}{3\sqrt{70}}c^1(3,-1,2,-2)+i\frac{1}{3\sqrt{35}}c^1(3,-1,2,-1)
\nonumber
\\
&-&i\frac{1}{2}\sqrt{\frac{3}{70}}c^1(3,0,2,-1)+i\frac{1}{\sqrt{70}}c^1(3,0,2,0), 
\\
\Bra{\frac{5}{2},-\frac{5}{2}}\tilde{t}_{1x}\Ket{\frac{3}{2},-\frac{3}{2}}&=&-i\sqrt{\frac{3}{35}}c^1(3,-3,2,-2)+i\frac{2}{3\sqrt{35}}c^1(3,-2,2,-2)
\nonumber
\\
&-&i\frac{7}{6\sqrt{70}}c^1(3,-2,2,-1). 
\end{eqnarray}
The results for $\tilde{t}_{1y}$ which are not zero are shown as follows. 
\begin{eqnarray}
\Bra{\frac{5}{2},\frac{5}{2}}\tilde{t}_{1y}\Ket{\frac{3}{2},\frac{3}{2}}&=&\frac{7}{6\sqrt{70}}c^1(3,2,2,1)-\frac{2}{3\sqrt{35}}c^1(3,2,2,2)+\sqrt{\frac{3}{35}}c^1(3,3,2,2),
\\
\Bra{\frac{5}{2},\frac{1}{2}}\tilde{t}_{1y}\Ket{\frac{3}{2},\frac{3}{2}}&=&\frac{1}{\sqrt{70}}c^1(3,0,2,0)-\frac{1}{2}\sqrt{\frac{3}{70}}c^1(3,0,2,1)
\nonumber
\\
&+&\frac{1}{3\sqrt{35}}c^1(3,1,2,1)+\frac{2}{3\sqrt{70}}c^1(3,1,2,2),
\\
\Bra{\frac{5}{2},\frac{3}{2}}\tilde{t}_{1y}\Ket{\frac{3}{2},\frac{1}{2}}&=&\frac{5}{3\sqrt{70}}c^1(3,1,2,0)-\frac{1}{6}\sqrt{\frac{6}{35}}c^1(3,1,2,1)
\nonumber
\\
&+&\frac{1}{6}\sqrt{\frac{3}{14}}c^1(3,2,2,1)+\frac{1}{6}\sqrt{\frac{12}{7}}c^1(3,2,2,2),
\\
\Bra{\frac{5}{2},-\frac{1}{2}}\tilde{t}_{1y}\Ket{\frac{3}{2},\frac{1}{2}}&=&\frac{2}{3}\sqrt{\frac{3}{35}}c^1(3,-1,2,-1)-\frac{1}{3\sqrt{35}}c^1(3,-1,2,0)
\nonumber
\\
&+&\frac{3}{2\sqrt{70}}c^1(3,0,2,1), 
\\
\Bra{\frac{5}{2},\frac{1}{2}}\tilde{t}_{1y}\Ket{\frac{3}{2},-\frac{1}{2}}&=&\frac{3}{2\sqrt{70}}c^1(3,0,2,-1)-\frac{1}{3\sqrt{35}}c^1(3,1,2,0)
\nonumber
\\
&+&\frac{2}{3}\sqrt{\frac{3}{35}}c^1(3,1,2,1), 
\\
\Bra{\frac{5}{2},-\frac{3}{2}}\tilde{t}_{1y}\Ket{\frac{3}{2},-\frac{1}{2}}&=&\frac{1}{6}\sqrt{\frac{12}{7}}c^1(3,-2,2,-2)+\frac{1}{6}\sqrt{\frac{3}{14}}c^1(3,-2,2,-1)
\nonumber
\\
&-&\frac{1}{6}\sqrt{\frac{6}{35}}c^1(3,-1,2,-1)+\frac{5}{3\sqrt{70}}c^1(3,-1,2,0),
\\
\Bra{\frac{5}{2},-\frac{1}{2}}\tilde{t}_{1y}\Ket{\frac{3}{2},-\frac{3}{2}}&=&\frac{2}{3\sqrt{70}}c^1(3,-1,2,-2)+\frac{1}{3\sqrt{35}}c^1(3,-1,2,-1)
\nonumber
\\
&-&\frac{1}{2}\sqrt{\frac{3}{70}}c^1(3,0,2,-1)+\frac{1}{\sqrt{70}}c^1(3,0,2,0), 
\\
\Bra{\frac{5}{2},-\frac{5}{2}}\tilde{t}_{1y}\Ket{\frac{3}{2},-\frac{3}{2}}&=&\sqrt{\frac{3}{35}}c^1(3,-3,2,-2)-\frac{2}{3\sqrt{35}}c^1(3,-2,2,-2)
\nonumber
\\
&+&\frac{7}{6\sqrt{70}}c^1(3,-2,2,-1). 
\end{eqnarray}
The results for $\tilde{t}_{1z}$ which are not zero are shown as follows.
\begin{eqnarray}
\Bra{\frac{5}{2},\frac{3}{2}}\tilde{t}_{1z}\Ket{\frac{3}{2},\frac{3}{2}}&=&-i\frac{1}{3}\sqrt{\frac{6}{35}}c^1(3,1,2,0)-i\frac{1}{3}\sqrt{\frac{2}{35}}c^1(3,1,2,1)
\nonumber
\\
&-&i\frac{8}{3\sqrt{35}}c^1(3,1,2,2)-i\frac{1}{6}\sqrt{\frac{2}{7}}c^1(3,2,2,1)-i\frac{2}{3\sqrt{7}}c^1(3,2,2,2),
\\
\Bra{\frac{5}{2},\frac{1}{2}}\tilde{t}_{1z}\Ket{\frac{3}{2},\frac{1}{2}}&=&-i\frac{1}{3}\sqrt{\frac{3}{35}}c^1(3,0,2,-1)-i\frac{1}{3}\sqrt{\frac{6}{35}}c^1(3,0,2,0)+i\frac{1}{\sqrt{70}}c^1(3,0,2,1)
\nonumber
\\
&-&i\frac{2}{3}\sqrt{\frac{3}{35}}c^1(3,1,2,1)-i\frac{2}{3}\sqrt{\frac{6}{35}}c^1(3,1,2,2), 
\\
\Bra{\frac{5}{2},-\frac{1}{2}}\tilde{t}_{1z}\Ket{\frac{3}{2},-\frac{1}{2}}&=&-i\frac{2}{3}\sqrt{\frac{6}{35}}c^1(3,-1,2,-2)-i\frac{2}{3}\sqrt{\frac{3}{35}}c^1(3,-1,2,-1)+i\frac{1}{\sqrt{70}}c^1(3,0,2,-1)
\nonumber
\\
&-&i\frac{1}{3}\sqrt{\frac{6}{35}}c^1(3,0,2,0)-i\frac{1}{3}\sqrt{\frac{3}{35}}c^1(3,0,2,1), 
\\
\Bra{\frac{5}{2},-\frac{3}{2}}\tilde{t}_{1z}\Ket{\frac{3}{2},-\frac{3}{2}}&=&-i\frac{2}{3\sqrt{7}}c^1(3,-2,2,-2)-i\frac{1}{6}\sqrt{\frac{2}{7}}c^1(3,-2,2,-1)-i\frac{8}{3\sqrt{35}}c^1(3,-1,2,-2)
\nonumber
\\
&-&i\frac{1}{3}\sqrt{\frac{2}{35}}c^1(3,-1,2,-1)-i\frac{1}{3}\sqrt{\frac{6}{35}}c^1(3,-1,2,0). 
\end{eqnarray}

* swata@mns.kyutech.ac.jp

\end{document}